# Perturbation theory of mesoscale plasmonic waveguide with an analytical treatment of nonclassical electromagnetic boundary condition

Jiling Xue[1,2] and Haitao Liu[1,2,*]
[1]Institute of Modern Optics, College of Electronic Information and Optical Engineering, Nankai University, Tianjin, 300350, China
[2]National Key Laboratory of Semiconductor Laser, Tianjin, 300350, China

The optical modes of mesoscale plasmonic waveguides (MPWs) are significantly affected by nonclassical quantum effects, which can be comprehensively described by the nonclassical electromagnetic boundary condition (NEBC) formulated with the surface-response Feibelman *d*-parameters. In this paper, a perturbation theory for the nonclassical waveguide modes (NWMs) supported by MPWs under the NEBC is proposed. In this theory, by adopting the classical waveguide modes (CWMs) under the classical electromagnetic boundary condition (CEBC) as the basis functions and treating the NEBC as a first-order perturbation, a general expression of the propagation constant of the NWM with an analytical dependence on the NEBC is derived. This theory transparently reveals the underlying general relation between the nonclassical effects and the propagation properties of the NWMs, thereby providing an effective tool for the understanding and design of MPW devices, as well as for the experimental measurement of the *d*-parameters.

Mesoscale plasmonic nanostructures (MPNs) with feature sizes (such as the size of nanogap) down to 1-20 nm can achieve extreme enhancement of light-matter interactions at the nanoscale, and attract intense research interests [1-6]. The optical response of MPNs is significantly affected by nonclassical quantum effects such as nonlocality, electron spill-in/out, and surface Landau damping, and cannot be accurately predicted by the classical electromagnetic theory due to the neglect of these nonclassical effects [7-11]. First-principles time-dependent density functional theory can rigorously account for these nonclassical effects, but its applicability is restricted by its prohibitive computational cost to few-atom clusters or highly symmetric ultrasmall systems, rendering it inapplicable to MPNs with overall sizes reaching hundreds of nanometers [7,8,11-14]. Alternatively, the first-principles surface-response Feibelman *d*-parameters ( $d_\perp$ and $d_\parallel$, representing the centroids of the induced charge and of the normal derivative of the tangential current density, respectively) can comprehensively describe the nonclassical effects in MPNs [8-10,15]. By modifying the classical electromagnetic boundary condition (CEBC) of Maxwell's equations to the nonclassical electromagnetic boundary condition (NEBC) formulated with the *d*-parameters [9], efficient computation of the optical response of MPNs can be realized [9,16-20].

As a typical MPN, the mesoscale plasmonic waveguide (MPW) supports waveguide modes that enable strong confinement and propagation of optical field at deep-nanometer scales [21-25]. This capability underpins significant applications, such as deep subwavelength interference lithography [26], ultracompact and ultrafast optical switching [27], ultracompact phase modulators [5], nanofocusing [4], and broadband enhancement and guiding of spontaneous emission [28] and Raman scattering [29]. Within the NEBC framework, theoretical investigations of MPWs have been conducted. For a planar metal-insulator-metal structure, the dispersion relation of the propagation constant of the gap surface-plasmon waveguide mode was obtained by numerically solving a transcendental equation, where only the $d_\perp$ parameter was considered (with $d_\parallel = 0$ ) [24]. For a planar graphene-dielectric-metal structure, the propagation constant of the acoustic graphene-plasmon waveguide mode was calculated, suggesting a method of inversely retrieving the *d*-parameters via the measurement of the propagation constant [25]. Through rigorous numerical calculations incorporating both $d_\perp$ and $d_\parallel$ parameters, the surface-plasmon-polariton waveguide modes supported by a metal nanowire on a metal substrate and a metal-dielectric planar interface were obtained [30]. Nevertheless, for MPWs of arbitrary geometries, general physically-transparent relations between the nonclassical effects and the propagation properties of waveguide modes are still absent.

In this paper, a perturbation theory for the nonclassical waveguide modes (NWMs) supported by MPWs under the NEBC is proposed. This theory yields a general expression of the propagation constant of the NWM with an analytical dependence on both $d_\perp$ and $d_\parallel$ parameters. Consequently, it transparently reveals the general relation between the nonclassical effects described by the NEBC and the propagation properties of NWMs, thereby providing a valuable tool for the understanding and design of MPW devices, as well as for the experimental measurement of the *d*-parameters which is urgently needed for building the database of *d*-parameters [9,23,25,31]. The theoretical framework begins with the construction of an expansion theory for the NWMs, which employs a complete set of classical waveguide modes (CWMs) under the CEBC as basis functions. The expansion theory can rigorously determine both the propagation constants and the electromagnetic field distributions of the NWMs. Building upon the expansion theory, with the $d_\perp$ and $d_\parallel$ parameters in the NEBC treated as a first-order perturbation, the perturbation theory for the NWMs is further developed. The validity of the proposed perturbation theory is verified by comparing its predictions with rigorous full-wave numerical calculations for representative MPW configurations.

*Expansion theory of NWMs under NEBC.* –A $z$-translationally invariant MPW is illustrated in Fig. 1. The electromagnetic field of the $n$th-order NWM supported by the MPW under the NEBC is denoted by $\tilde{\boldsymbol{\psi}}_n=[\tilde{\mathbf{E}}_n,\tilde{\mathbf{H}}_n]^{\mathrm{T}}$, satisfying $\tilde{\boldsymbol{\psi}}_n(\mathbf{r})=\tilde{\boldsymbol{\psi}}_n(\boldsymbol{\rho})\exp(i\tilde{k}_n z)$, where $\mathbf{r}=(x,y,z)$ are Cartesian coordinates, $\boldsymbol{\rho}=(x,y,0)$ denotes the Cartesian coordinates in the cross section of $z$=0, and $\tilde{k}_n$ is the propagation constant. $\tilde{\boldsymbol{\psi}}_n$ satisfies the NEBC [9] at the interface $\partial\Omega$ between the metal and dielectric, and satisfies the classical frequency-domain Maxwell's equations in the bulk regions away from the $\partial\Omega$. Here, reciprocal media are considered, that is, the permittivity and permeability tensors satisfying $\boldsymbol{\varepsilon}^{\mathrm{T}}=\boldsymbol{\varepsilon}$ and $\boldsymbol{\mu}^{\mathrm{T}}=\boldsymbol{\mu}$, respectively. In the cross section of $z$=0, by employing a complete set of CWMs as expansion basis functions [32], $\tilde{\boldsymbol{\psi}}_n$ can be expressed as

$$\tilde{\boldsymbol{\psi}}_n(\boldsymbol{\rho})=\boldsymbol{\Phi}\tilde{\mathbf{a}}_n, \tag{1}$$

where $\boldsymbol{\Phi}=[\boldsymbol{\Phi}^+,\boldsymbol{\Phi}^-]$, with $\boldsymbol{\Phi}^\pm$ being $2\times P$ matrices. The $p$th column elements of $\boldsymbol{\Phi}^+$ and $\boldsymbol{\Phi}^-$ are the electromagnetic fields $\boldsymbol{\psi}_p^+=[\mathbf{E}_p^+,\mathbf{H}_p^+]^{\mathrm{T}}$ and $\boldsymbol{\psi}_p^-=[\mathbf{E}_p^-,\mathbf{H}_p^-]^{\mathrm{T}}$ ($p$=1, 2, …, $P$) of the $p$th-order CWMs propagating along the positive and negative $z$-directions, respectively, satisfying $\boldsymbol{\psi}_p^\pm(\mathbf{r})=\boldsymbol{\psi}_p^\pm(\boldsymbol{\rho})\exp(\pm ik_p z)$, where $k_p$ is the propagation constant, satisfying Re($k_p$)+Im($k_p$)≥0. $\tilde{\mathbf{a}}_n=[\tilde{\mathbf{a}}_n^+;\tilde{\mathbf{a}}_n^-]$ is a column vector (where the semicolon denotes the row-wise concatenation of column vectors $\tilde{\mathbf{a}}_n^+$ and $\tilde{\mathbf{a}}_n^-$ into one column vector), and the $p$th element of $\tilde{\mathbf{a}}_n^\pm$ is the coefficient $c_p^\pm$ of $\boldsymbol{\psi}_p^\pm(\boldsymbol{\rho})$. The CWMs satisfy the source-free Maxwell's equations under the CEBC and the outgoing-wave boundary condition at infinity along the $x$ and $y$ directions.

Starting from the NEBC-modified Maxwell's equations [18], in which the NEBC is treated as a nonclassical surface source $\mathbf{J}_{\mathrm{ncl}}$, and based on the reciprocity theorem between the CWMs and $\mathbf{J}_{\mathrm{ncl}}$ [33], a set of CWM-coupling equations satisfied by $\tilde{\mathbf{a}}_n$ can be derived, expressed as the following matrix eigenvalue problem [derivation details are provided in Supplemental Material (SM), Sec. S1 [34]],

$$\begin{bmatrix}\mathbf{W}+i\mathbf{K}^{-,+} & i\mathbf{K}^{-,-}\\ -i\mathbf{K}^{+,+} & -\mathbf{W}-i\mathbf{K}^{+,-}\end{bmatrix}\tilde{\mathbf{a}}_n=\tilde{k}_n\tilde{\mathbf{a}}_n. \tag{2}$$

After obtaining the CWMs $\boldsymbol{\psi}_p^\pm$ ($p$=1, 2, …, $P$), solving the eigenvalue problem (2) yields the $n$th eigenvalue $\tilde{k}_n$ and its corresponding eigenvector $\tilde{\mathbf{a}}_n$, which respectively give the propagation constant and electromagnetic field distribution of the $n$th-order NWM. In Eq. (2), $\mathbf{W}$ is a diagonal matrix, with its $p$th diagonal element being the propagation constant $k_p$ of the $p$th-order CWM. The $p$th-row and $q$th-column element of matrix $\mathbf{K}^{\sigma,\tau}(\sigma,\tau\in\{+,-\})$ is given by

$$\kappa_{p,q}^{\sigma,\tau}=\frac{1}{F_p}\oint_L[i\omega d_\parallel[\![\varepsilon]\!]\mathbf{E}_{p,\parallel}^\sigma\cdot\mathbf{E}_{q,\parallel}^\tau-i\omega d_\perp[\![\varepsilon E_{p,\perp}^\sigma E_{q,\perp}^\tau]\!] \\ -i(\sigma k_p+\tau k_q)d_\perp H_{p,l}^\sigma[\![E_{q,\perp}^\tau]\!]]_{z=0}\,dl, \tag{3}$$

which represents the coefficient of the $p$th-order CWM excited by the nonclassical surface source $\mathbf{J}_{\mathrm{ncl}}$ arising from the $q$th-order CWM, and is referred to as the NEBC-induced CWM-coupling coefficient. In Eq. (3), a pseudo-energy flux $F_p$ is defined as $F_p=\iint_{-\infty}^{\infty}\mathbf{z}\cdot(\mathbf{E}_p^-\times\mathbf{H}_p^+-\mathbf{E}_p^+\times\mathbf{H}_p^-)_{z=0}\,dxdy$, where $(\hat{\mathbf{x}},\hat{\mathbf{y}},\hat{\mathbf{z}})$ are the unit vectors along the $(x,y,z)$ axes, respectively. $\omega$ is the angular frequency, $\varepsilon$ is the permittivity, and $d_\perp$, $d_\parallel$ are the Feibelman $d$-parameters. The $[\![\mathbf{f}]\!]=\mathbf{f}_d-\mathbf{f}_m$ denotes the discontinuity of the field $\mathbf{f}$ at $\partial\Omega$, with $\mathbf{f}_d$ and $\mathbf{f}_m$ denoting the values of $\mathbf{f}$ on the dielectric and the metal sides of $\partial\Omega$, respectively. The $f_\perp=\hat{\mathbf{n}}\cdot\mathbf{f}$ and $\mathbf{f}_\parallel=\mathbf{f}-\hat{\mathbf{n}}f_\perp$ denote the normal and tangential components of field $\mathbf{f}$ on $\partial\Omega$, respectively, where $\hat{\mathbf{n}}$ is the unit normal vector (pointing from the metal region $\Omega^-$ to dielectric region $\Omega^+$) on $\partial\Omega$. As shown in Fig. 1, the closed contour $L$ denotes the intersection curve of $\partial\Omega$ with the $z$=0 plane, $\hat{\mathbf{l}}=\hat{\mathbf{n}}\times\hat{\mathbf{z}}$ is the unit tangential vector of $L$, and $H_{p,l}^\sigma=\hat{\mathbf{l}}\cdot\mathbf{H}_p^\sigma$.

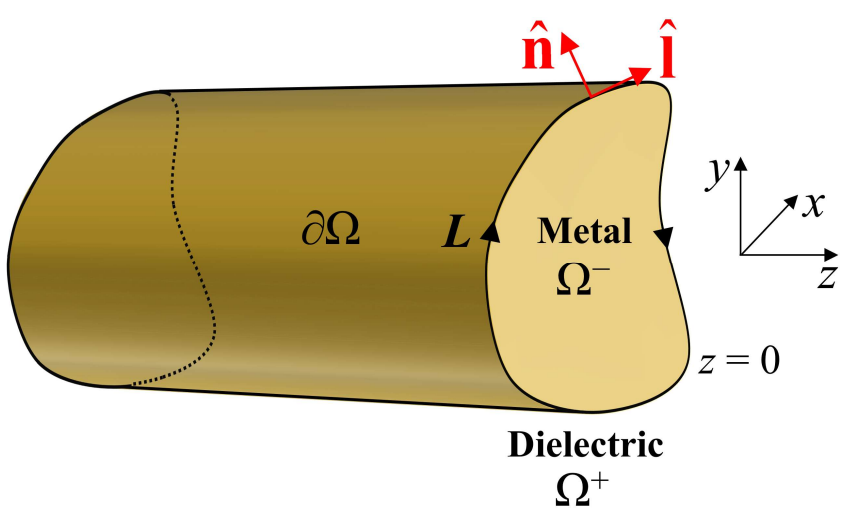

FIG. 1. Schematic of a $z$-translationally invariant MPW.

It can be proved that when the same number of $\boldsymbol{\psi}_p^+$ and $\boldsymbol{\psi}_p^-$ are considered in Eq. (2), if $\tilde{k}_n$ is the propagation constant of a NWM, then $-\tilde{k}_n$ must be the propagation constant of another NWM (proof details are provided in SM Sec. S2 [34]; numerical verifications are provided in Fig. S7 of Sec. S5B and Fig. S16 of Sec. S5C).

Consequently, we denote $\tilde{k}_{-n} = -\tilde{k}_n$ ($n$=1, 2, …), where $\tilde{k}_n$ and $\tilde{k}_{-n}$ are the propagation constants of the $n$th-order NWMs $\tilde{\boldsymbol{\psi}}_n$ and $\tilde{\boldsymbol{\psi}}_{-n}$ propagating along the positive and negative $z$-directions, respectively. It can also be proved that in the $z$=0 plane, $\tilde{\mathbf{E}}_n$ and $\tilde{\mathbf{E}}_{-n}$ satisfy a vectorial mirror symmetry, while $\tilde{\mathbf{H}}_n$ and $\tilde{\mathbf{H}}_{-n}$ satisfy a vectorial mirror antisymmetry (see SM Sec. S2 for the proof [34]). These conclusions are consistent with those of the CWMs [32], and are referred to as the symmetry relations for forward- and backward-propagating NWMs.

*Perturbation theory of NWMs under NEBC.* –Based on the established expansion theory of NWMs, the perturbation theory of NWMs is developed in the following. By treating the Feibelman $d$-parameters as a perturbation [9,10,16], it follows that

$$d_\perp = d_\perp^{(1)}\delta,\ d_\parallel = d_\parallel^{(1)}\delta, \tag{4}$$

where $\delta$ denotes the perturbation parameter (also referred to as the scaling factor of the $d$-parameters), with $\delta$=1 corresponding to the actual physical $d$-parameters. The propagation constant $\tilde{k}_n$ of NWM under the NEBC is a function of the $d$-parameters, i.e., $\tilde{k}_n = \tilde{k}_n(d_\perp, d_\parallel)$. Then the $\tilde{k}_n$ can be expressed as an asymptotic expansion with respect to $\delta$,

$$\tilde{k}_n = \tilde{k}_n(d_\perp^{(1)}\delta, d_\parallel^{(1)}\delta) = k^{(0)} + k^{(1)}\delta + O(\delta^2). \tag{5}$$

Applying the perturbation theory for eigenvalue problems (e.g., as applied in quantum mechanics [36]) to solve Eq. (2), we can obtain $k^{(0)} = \pm k_p$, $k^{(1)}\delta = \pm i\kappa_{p,p}^{-;+}$ (derivation details provided in SM, Sec. S3 [34]). Therefore,

$$\begin{aligned}\tilde{k}_{\pm p} = \pm\Bigg[ k_p + \frac{1}{F_p}\oint_L (&-\omega d_\parallel [\![\varepsilon]\!] \mathbf{E}_{p,\parallel}^- \cdot \mathbf{E}_{p,\parallel}^+ \\ &+\omega d_\perp [\![\varepsilon E_{p,\perp}^- E_{p,\perp}^+]\!])_{z=0}\, dl \Bigg] + O(\delta^2),\end{aligned} \tag{6}$$

where the subscript takes $n$=±$p$. Equation (6) constitutes the main result of this paper, i.e., the first-order asymptotic expansion of $\tilde{k}_{\pm p}$, *which is generally applicable to MPWs of arbitrary geometries*.

It is worth noting that in Eq. (2), if only one CWM is considered, i.e., either $\tilde{\mathbf{a}}_n = \tilde{c}_p^+$ or $\tilde{\mathbf{a}}_n = \tilde{c}_p^-$, then the corresponding eigenvalue is exactly the $\tilde{k}_{\pm p}$ given by Eq. (6).

To elucidate the physical implications of Eq. (6), by neglecting the contribution of $d_\parallel$ [9,15,18,37] and the metal loss, one can obtain from Eq. (6) (derivation details provided in SM Sec. S4 [34])

$$\tilde{k}_p - k_p \approx \oint_L \omega d_\perp \left[\!\left[ \mathrm{Re}(\varepsilon) \left|E_{p,\perp}^+\right|_{z=0}^2 \Big/ |F_p| \right]\!\right] dl + O(\delta^2), \tag{7}$$

where $|F_p|$≈4$\Phi_p$, with $\Phi_p$ being the energy flux of the CWM, and the permittivities of the metal and dielectric satisfy Re($\varepsilon$)<0 and Re($\varepsilon$)>0, respectively. Equation (7) indicates that:

i) Electron spill-in or spill-out [i.e., $\mathrm{Re}(d_\perp) < 0$ or $\mathrm{Re}(d_\perp) > 0$ ] [8,9,25,39] will respectively lead to $\mathrm{Re}(\tilde{k}_p) < \mathrm{Re}(k_p)$ or $\mathrm{Re}(\tilde{k}_p) > \mathrm{Re}(k_p)$. Here, an increase in $\mathrm{Re}(\tilde{k}_p)$ implies a decrease in the effective wavelength of the NWM, given by $\tilde{\lambda}_p = 2\pi / \mathrm{Re}(\tilde{k}_p)$ [40].

ii) Surface Landau damping [i.e., $\mathrm{Im}(d_\perp) > 0$ ] [8,37] gives rise to $\mathrm{Im}(\tilde{k}_p) > \mathrm{Im}(k_p)$. In this case, an increase in $\mathrm{Im}(\tilde{k}_p)$ implies an increase in propagation loss [40].

iii) The stronger the localization of the CWM electric field at the metal-dielectric interface (i.e., the larger $\left|E_{p,\perp}^+\right|_{z=0}^2 \Big/ |F_p|$), the larger the discrepancy between $\tilde{k}_p$ and $k_p$.

The above theoretical predictions will be verified by the numerical examples presented in the following section.

Furthermore, Eq. (6) or (7) transparently reveals the underlying relation between the Feibelman $d_\perp$, $d_\parallel$ parameters and the propagation constant $\tilde{k}_p$ of the NWM. For instance, an important application of this relation is for experimentally retrieving the $d$-parameters by measuring the $\tilde{k}_p$, analogously to the suggestions in Refs. [23,25].

*Verification.* –In this section, by considering several numerical examples, the proposed perturbation theory for the NWMs under the NEBC [see Eq. (6)] will be verified by comparing the theoretical predictions with the rigorous full-wave numerical results. The rigorous calculation of the NWM is carried out by using the pole-search approach [41] combined with the full-wave Fourier modal method (FMM) under the NEBC [42] (details provided in Supplement 1, Sec. S6 of Ref. [30]). The wavelength is set to $\lambda$=1 μm for all the numerical examples in this paper.

As illustrated in the inset of Fig. 2(b), the considered first numerical example is a single-nanowire MPW [47,48], which comprises a gold substrate, a polymethyl methacrylate (PMMA) nanogap, a gold nanowire, and the ambient air. The gold nanowire has a square cross section with a side length $D$=40 nm. The coordinate origin $O$ is set on the surface of the gold substrate and aligned with the center of the nanowire. The wavelength-dependent refractive index of gold is $n_{\mathrm{Au}}$=0.2549+6.8444i (at $\lambda$=1 μm) [49], and the refractive indices of PMMA and air are $n_{\mathrm{PMMA}}$=1.5 and $n_{\mathrm{Air}}$=1, respectively. Since the electric field at the Au-air interface is much weaker than that at the Au-PMMA interface, the NEBC is applied only at the Au-PMMA interface, where the Feibelman $d_\perp$ parameter is set to $d_\perp^{\text{Au-PMMA}} = (-0.4 + 0.2i)$ nm [9,18,25]. Generally, there is $|d_\parallel| \ll |d_\perp|$ [9,15,18,37]. To simultaneously reflect the contributions of both $d_{\perp}$ and $d_{\parallel}$ so as to provide a stringent numerical validation of the perturbation theory, the Feibelman $d_{\parallel}$ parameter is artificially set to $d_{\parallel}^{\text{Au-PMMA}} = (0.4 + 0.2i)$ nm. At this point, there is $|d_{\parallel}| = |d_{\perp}|$, and the signs of the real and imaginary parts of $d_{\parallel}$ are consistent with those given by the *s*-*d* polarization model [10,18,37,50]. To mitigate the computational errors caused by the electromagnetic field singularity at the two edges of the bottom surface of the nanowire, an apodization function is applied to the *d*-parameters on this surface (see SM Sec. S5A for details [34]).

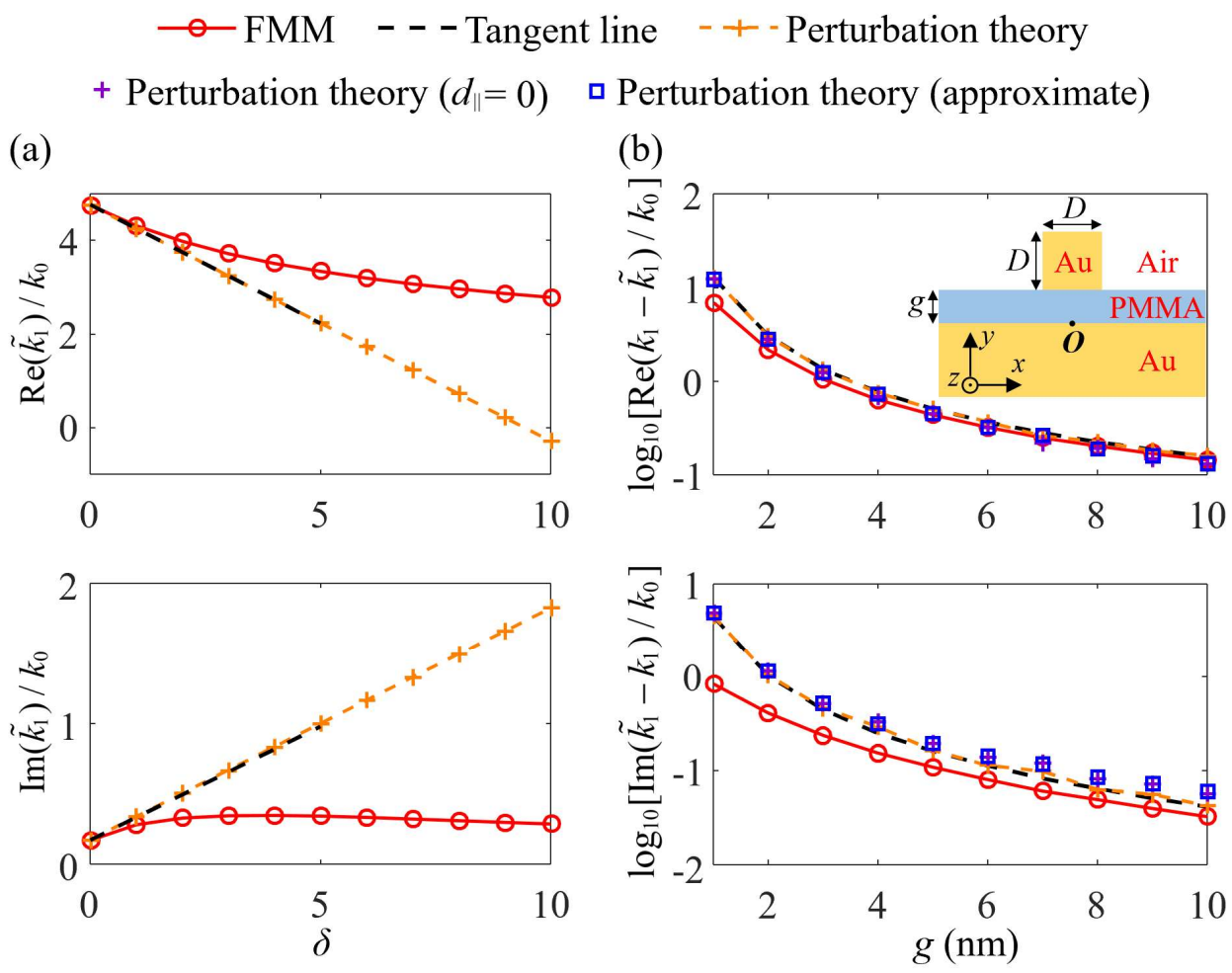

FIG. 2. (a) Propagation constant $\tilde{k}_1$ of the NWM supported by the single-nanowire MPW (as illustrated in the inset) as a function of the scaling factor $\delta$ of the Feibelman *d*-parameters, with the PMMA nanogap thickness *g*=5 nm. The plotted normalized quantity $\tilde{k}_1/k_0$ represents the complex effective index. (b) $\tilde{k}_1$ as a function of *g* for $\delta$=1. In (a) and (b), the red solid lines with circles represent the full-wave FMM numerical results, the black dashed lines represent the results of tangent lines to the FMM results at $\delta$=0, the orange dashed lines with pluses show the predictions of the perturbation theory, the purple pluses show the predictions of the perturbation theory considering only the contribution of $d_{\perp}$ (i.e., $d_{\parallel}=0$), and the blue squares represent the results from Eq. (7).

For this single-nanowire MPW, it supports only one bound and propagative mode (i.e., the field decays to zero at infinity along the transverse *x* and *y* directions, and the propagation constant is almost a real number [51]), whose nonclassical and classical propagation constants are denoted by $\tilde{k}_1$ and $k_1$, respectively, with the corresponding field distributions shown in Fig. S3 of SM Sec. S5B [34]. Figure 2(a) plots the variation of the $\tilde{k}_1$ with the scaling factor $\delta$ of the Feibelman *d*-parameters for a fixed PMMA nanogap thickness *g*=5 nm, where $d_{\perp} = d_{\perp}^{\text{Au-PMMA}}\delta$, $d_{\parallel} = d_{\parallel}^{\text{Au-PMMA}}\delta$ [i.e., $d_{\perp}^{(1)} = d_{\perp}^{\text{Au-PMMA}}$, $d_{\parallel}^{(1)} = d_{\parallel}^{\text{Au-PMMA}}$ in Eq. (4)]. The normalized quantity $\tilde{k}_1/k_0$ represents the complex effective index ($k_0$=2$\pi/\lambda$ being the wavenumber in the vacuum), and $\tilde{k}_1 = k_1$ if $\delta$=0. As shown in Fig. 2(a), the predictions of the perturbation theory [Eq. (6), orange dashed lines with pluses] are consistent with the tangent lines to the full-wave FMM results at $\delta$=0 [i.e., the first two terms on the right-hand side of Eq. (5) calculated via the FMM, black dashed lines], which verifies the validity of the perturbation theory. For $\delta$=1 (corresponding to the actual physical *d*-parameters), the predictions of the perturbation theory are quite close to the full-wave FMM results (red solid lines with circles). The results show that there are $\mathrm{Re}(\tilde{k}_1) < \mathrm{Re}(k_1)$ and $\mathrm{Im}(\tilde{k}_1) > \mathrm{Im}(k_1)$, which are consistent with the theoretical predictions of Eq. (7), and are attributed to $\mathrm{Re}(d_{\perp}) < 0$ and $\mathrm{Im}(d_{\perp}) < 0$, respectively.

Figure 2(b) presents the variations of $\mathrm{Re}(k_1 - \tilde{k}_1)$ and $\mathrm{Im}(\tilde{k}_1 - k_1)$ (both quantities are positive) with the PMMA nanogap thickness *g* for $\delta$=1. The dependence of the CWM propagation constant $k_1$ on *g* is provided in Fig. S4 of SM Sec. S5B [34]. As illustrated in Fig. 2(b), for all the values of *g*, the predictions of the perturbation theory (orange dashed lines with pluses) are consistent with the results of tangent lines to the full-wave FMM results at $\delta$=0 (black dashed lines), which further confirms the validity of the perturbation theory. Additionally, the predictions of the perturbation theory are quite close to the full-wave FMM results (red solid lines with circles). Figure 2(b) also reveals that the discrepancy between $\tilde{k}_1$ and $k_1$ gradually increases as *g* decreases, which is consistent with the theoretical predictions of Eq. (7) (blue squares), and is attributed to the enhanced confinement of the CWM electric field at the Au-PMMA interface within the nanogap.

Figure 2(b) also presents the predictions of the perturbation theory with $d_{\parallel} = 0$ (purple pluses), which are close to the predictions of the perturbation theory with $d_{\perp}d_{\parallel} \neq 0$ (orange dashed lines with pluses), indicating that the contribution of $d_{\perp}$ is dominant. To further verify the contributions of both $d_{\perp}$ and $d_{\parallel}$ in the perturbation theory, Fig. S5 in SM Sec. S5B [34] shows the results with $d_{\parallel}$ artificially magnified by a factor of 10 (i.e., $d_{\perp} = d_{\perp}^{\text{Au-PMMA}}$, $d_{\parallel} = 10 d_{\parallel}^{\text{Au-PMMA}}$). In this case, both $d_{\perp}$ and $d_{\parallel}$ contribute significantly, and the results demonstrate that the perturbation theory remains valid.

Figure 2(b) reveals that as the PMMA nanogap thickness *g* decreases, which leads to enhanced nonclassical effects within the nanogap [7,9,23,38], the discrepancy between the predictions of the perturbation theory and the full-wave

FMM results increases. This indicates that besides the first-order contributions of the $d$-parameters considered in the perturbation theory, it is necessary to include their higher-order contributions [i.e., the $O(\delta^2)$ term on the right-hand side of Eq. (6)]. To this end, all the bound and propagative CWMs (including one forward- and one backward-propagating CWMs) are incorporated into the expansion theory [see Eq. (2)], and the predicted $\tilde{k}_1$ is presented in Fig. S6 of SM Sec. S5B [34]. However, the results indicate that compared with the perturbation theory [Eq. (6)], the expansion theory does not significantly improve the accuracy. Therefore, to ensure the accuracy of the expansion theory so as to fully account for the higher-order contributions of the $d$-parameters, it is necessary to incorporate not only bound and propagative CWMs but also a large number of unbound or non-propagative (i.e., radiative or evanescent) CWMs into the expansion theory, so that the considered CWMs can constitute a complete set of basis functions [32]. This, however, comes at the cost of increased computational cost and diminished physical insight, which falls beyond the scope of this work and is left for future investigations.

In addition, Fig. S7 in SM Sec. S5B [34] shows that when all bound and propagative CWMs are considered in the expansion theory, the predicted propagation constants of the forward- and backward-propagating NWMs, $\tilde{k}_1$ and $\tilde{k}_{-1}$, satisfy the symmetry relation $\tilde{k}_{-1} = -\tilde{k}_1$, which confirms the general conclusion derived from the expansion theory and also agrees with the full-wave FMM results.

The considered second numerical example is a double-nanowire MPW, which supports two bound and propagative modes. The corresponding results are presented in SM Sec. S5C [34], further verifying the validity of the perturbation theory. Moreover, the relevant conclusions are analogous to those for the single-nanowire MPW.

The third numerical example focuses on the NWM supported by a planar metal-dielectric interface. Analytical solution for the propagation constant $\tilde{k}_1$ is derived via the perturbation theory [Eq. (6)] (see SM Sec. S6A [34]), and is compared with that derived via the asymptotic expansion method (see SM Sec. S6B [34]). It is found that the two solutions are mathematically identical and fully consistent with the full-wave FMM numerical results (see SM Sec. S6C [34]), which once again confirms the validity of the perturbation theory.

*Conclusions*. −For MPWs of arbitrary geometries, the proposed perturbation theory transparently reveals the general relation between the NEBC-described nonclassical effects (e.g., electron spill-in or spill-out and surface Landau damping) and the propagation properties of the NWMs (e.g., effective wavelength and propagation loss), thus providing an effective tool for the understanding and design of MPW devices, as well as for the experimental measurement of the Feibelman $d$-parameters.

Financial support from the National Natural Science Foundation of China (62475120, 62535007) is acknowledged.

*Corresponding author: liuht@nankai.edu.cn

## Supplemental Material

## Perturbation theory of mesoscale plasmonic waveguide with an analytical treatment of nonclassical electromagnetic boundary condition

Jiling Xue[1,2] and Haitao Liu[1,2,*]
[1]*Institute of Modern Optics, College of Electronic Information and Optical Engineering, Nankai University, Tianjin, 300350, China*
[2]*National Key Laboratory of Semiconductor Laser, Tianjin, 300350, China*
[*]Corresponding author: liuht@nankai.edu.cn

CONTENTS

**S1. Derivation of Eqs. (2) and (3) in the main text**

**A. Derivation of the CWM-coupling equations for solving the NWMs**

Let $\boldsymbol{\psi}_p^{\pm} = [\mathbf{E}_p^{\pm}, \mathbf{H}_p^{\pm}]^{\mathrm{T}}$ represent the electromagnetic fields of the $p$th-order classical waveguide modes (CWMs) propagating along the positive and negative $z$-directions (corresponding to the superscripts + and −, respectively), which satisfy the source-free Maxwell's equations under the classical electromagnetic boundary condition (CEBC),

$$\nabla \times \mathbf{E}_p^{\pm}(\mathbf{r}) = i\omega\boldsymbol{\mu} \cdot \mathbf{H}_p^{\pm}(\mathbf{r}), \tag{S1.1a}$$

$$\nabla \times \mathbf{H}_p^{\pm}(\mathbf{r}) = -i\omega\boldsymbol{\varepsilon} \cdot \mathbf{E}_p^{\pm}(\mathbf{r}), \tag{S1.1b}$$

and satisfy the outgoing-wave condition at infinity in the $x$ and $y$ directions. Along the $z$-direction, the $\boldsymbol{\psi}_p^{\pm}$ satisfy

$$\boldsymbol{\psi}_p^{\pm}(\mathbf{r}) = \boldsymbol{\psi}_p^{\pm}(\boldsymbol{\rho})\exp(\pm ik_p z), \tag{S1.2}$$

where $k_p$ is the propagation constant satisfying Re($k_p$)+Im($k_p$)≥0, $\mathbf{r}$=($x$,$y$,$z$) are the Cartesian coordinates, and $\boldsymbol{\rho}$=($x$,$y$,0) are the Cartesian coordinates in the cross section of $z$=0. Here, we consider reciprocal media, that is, the permittivity and permeability tensors satisfy $\boldsymbol{\varepsilon}^{\mathrm{T}}=\boldsymbol{\varepsilon}$ and $\boldsymbol{\mu}^{\mathrm{T}}=\boldsymbol{\mu}$, respectively. $\omega$ is the angular frequency.

The electromagnetic field $\boldsymbol{\Psi}$=[$\mathbf{E}$,$\mathbf{H}$]$^{\mathrm{T}}$ under the nonclassical electromagnetic boundary condition (NEBC) satisfies the following nonclassically corrected Maxwell's equations [1],

$$\nabla \times \mathbf{E} = i\omega\boldsymbol{\mu} \cdot \mathbf{H} - \mathbf{J}_{\text{ncl,m}} - \mathbf{J}_{\text{ext,m}}, \tag{S1.3a}$$

$$\nabla \times \mathbf{H} = -i\omega\boldsymbol{\varepsilon} \cdot \mathbf{E} + \mathbf{J}_{\text{ncl,e}} + \mathbf{J}_{\text{ext,e}}, \tag{S1.3b}$$

where $\mathbf{J}_{\text{ext,e}}$ and $\mathbf{J}_{\text{ext,m}}$ represent the electric and magnetic current density vectors, respectively, corresponding to the prescribed excitation sources. $\mathbf{J}_{\text{ncl,e}}$ is the nonclassical tangential surface electric-current source, and $\mathbf{J}_{\text{ncl,m}}$ is the nonclassical tangential surface magnetic-current source, defined as [1],

$$\mathbf{J}_{\text{ncl,e}} = \mathbf{j}_{\text{ncl,e}}(\llbracket \varepsilon \mathbf{E}_{\parallel} \rrbracket)\delta_{\partial\Omega} = i\omega d_{\parallel} \llbracket \varepsilon \mathbf{E}_{\parallel} \rrbracket \delta_{\partial\Omega}, \tag{S1.4a}$$

$$\mathbf{J}_{\text{ncl,m}} = \mathbf{j}_{\text{ncl,m}}(\llbracket E_{\perp} \rrbracket)\delta_{\partial\Omega} = \hat{\mathbf{n}} \times [\nabla_{\parallel}(d_{\perp} \llbracket E_{\perp} \rrbracket)]\delta_{\partial\Omega}. \tag{S1.4b}$$

Note that both $\mathbf{J}_{\text{ncl,e}}$ and $\mathbf{J}_{\text{ncl,m}}$ depend on the unknown electric field $\mathbf{E}$. $\hat{\mathbf{n}}$ represents the unit normal vector on the metal-dielectric interface $\partial\Omega$, pointing from the metal region $\Omega^-$ to dielectric region $\Omega^+$. $f_\perp = \hat{\mathbf{n}}\cdot\mathbf{f}$ denotes the normal component of the field $\mathbf{f}$ along $\hat{\mathbf{n}}$, and $\mathbf{f}_\parallel = \mathbf{f} - \mathbf{n}f_\perp$ denotes the tangential component of the field $\mathbf{f}$. $[\![\mathbf{f}]\!] = \mathbf{f}_d - \mathbf{f}_m$ denotes the discontinuity of the field $\mathbf{f}$ at the interface $\partial\Omega$, where $\mathbf{f}_d$ and $\mathbf{f}_m$ denote the field values on the dielectric and metal sides, respectively. On both sides of $\partial\Omega$, the media are assumed to be isotropic and nonmagnetic, i.e., $\boldsymbol{\varepsilon}=\varepsilon$ is a scalar, and $\boldsymbol{\mu}=\mu_0$ is the vacuum permeability. $d_\perp$ and $d_\parallel$ are the Feibelman $d$-parameters. $\delta_{\partial\Omega}$ denotes the surface Dirac function on the surface $\partial\Omega$, which is defined to satisfy,

$$\iiint_{R^3}[\mathbf{f}(\mathbf{r})\delta_{\partial\Omega}]\phi(\mathbf{r})d^3\mathbf{r} = \iint_{\partial\Omega}\mathbf{f}(\mathbf{r})\phi(\mathbf{r})ds, \tag{S1.5}$$

where $\phi(\mathbf{r})$ is an arbitrary scalar complex-valued test function that is infinitely differentiable and satisfies $\phi(|\mathbf{r}|\to\infty)=0$.

Let $\boldsymbol{\Psi}$ represent the electromagnetic field of the nonclassical waveguide mode (NWM), which satisfies the nonclassically corrected source-free Maxwell's equations [i.e., Eq. (S1.3) with $\mathbf{J}_{\text{ext,e}}=\mathbf{J}_{\text{ext,m}}=\mathbf{0}$], and satisfies the outgoing-wave condition at infinity along the $x$ and $y$ directions. Along the $z$-direction, the $\boldsymbol{\Psi}$ satisfies

$$\boldsymbol{\psi}(\mathbf{r}) = \boldsymbol{\psi}(\boldsymbol{\rho})\exp(ik_s z), \tag{S1.6}$$

where $k_s$ is the propagation constant. From Eq. (S1.6), the nonclassical surface source $\mathbf{J}_{\text{ncl}}=[\mathbf{J}_{\text{ncl,e}},-\mathbf{J}_{\text{ncl,m}}]^{\text{T}}$ defined by Eq. (S1.4) satisfies,

$$\mathbf{J}_{\text{ncl}}(\mathbf{r}) = \mathbf{J}_{\text{ncl}}(\boldsymbol{\rho})\exp(ik_s z). \tag{S1.7}$$

To solve for the NWM, it is assumed that the CWMs have already been obtained. Then by using the CWMs $\boldsymbol{\psi}_p^\pm(\boldsymbol{\rho})$ ($p$=1, 2, …, $P$) as a complete set of basis functions [2], the NWM field $\boldsymbol{\Psi}(\boldsymbol{\rho})$ in the cross section $z$=0 can be expanded in terms of the $\boldsymbol{\psi}_p^\pm(\boldsymbol{\rho})$ as,

$$\boldsymbol{\psi}(\boldsymbol{\rho}) = \sum_{p=1}^{P} c_p^+\boldsymbol{\psi}_p^+(\boldsymbol{\rho}) + \sum_{p=1}^{P} c_p^-\boldsymbol{\psi}_p^-(\boldsymbol{\rho}) = \boldsymbol{\Phi}^+\mathbf{a}^+ + \boldsymbol{\Phi}^-\mathbf{a}^- = \boldsymbol{\Phi}\mathbf{a}, \tag{S1.8}$$

where $c_p^\pm$ are the expansion coefficients of the $\boldsymbol{\psi}_p^\pm$, and $P$ is the number of the considered CWMs. $\boldsymbol{\Phi}^\pm$ is a $2\times P$ matrix whose $p$th column is $\boldsymbol{\psi}_p^\pm$, $\mathbf{a}^\pm$ is a column vector whose $p$th element is $c_p^\pm$, $\boldsymbol{\Phi} = [\boldsymbol{\Phi}^+, \boldsymbol{\Phi}^-]$, and $\mathbf{a} = [\mathbf{a}^+; \mathbf{a}^-]$, where the semicolon denotes the row-wise concatenation of the column vectors $\mathbf{a}^+$ and $\mathbf{a}^-$ into one column vector.

Next, based on the reciprocity theorem between the excitation source and the CWM [3], we will derive a set of CWM-coupling equations [i.e., Eq. (2) in the main text] satisfied by the expansion coefficients $c_p^\pm$. Solving these equations yields both the $c_p^\pm$ and the propagation constant $k_s$ of the NWM.

For this purpose, in Eq. (S1.8), the $c_p^+\boldsymbol{\psi}_p^+(\boldsymbol{\rho})$ and $c_p^-\boldsymbol{\psi}_p^-(\boldsymbol{\rho})$ should be excited by the $\mathbf{J}_{\text{ncl}}$ in the regions $z$<0 and $z$>0, respectively, yielding,

$$\begin{aligned} c_p^+ &= \frac{1}{F_p}\iint_{-\infty}^{\infty} dxdy\int_{-\infty}^{0}[-\mathbf{E}_p^-(x,y,z)\cdot\mathbf{J}_{\text{ncl,e}}(x,y,z) + \mathbf{H}_p^-(x,y,z)\cdot\mathbf{J}_{\text{ncl,m}}(x,y,z)]dz \\ &= \frac{1}{F_p}\iint_{-\infty}^{\infty} dxdy\int_{-\infty}^{0}[-\mathbf{E}_p^-(x,y,0)\cdot\mathbf{J}_{\text{ncl,e}}(x,y,0) + \mathbf{H}_p^-(x,y,0)\cdot\mathbf{J}_{\text{ncl,m}}(x,y,0)]\exp[i(k_s-k_p)z]dz \\ &= \frac{1}{i(k_s-k_p)F_p}\iint_{-\infty}^{\infty}[-\mathbf{E}_p^-\cdot\mathbf{j}_{\text{ncl,e}}([\![\varepsilon\mathbf{E}_\parallel]\!])\delta_{\partial\Omega} + \mathbf{H}_p^-\cdot\mathbf{j}_{\text{ncl,m}}([\![E_\perp]\!])\delta_{\partial\Omega}]_{z=0}\,dxdy, \end{aligned} \tag{S1.9a}$$

$$\begin{aligned} c_p^- &= \frac{1}{F_p}\iint_{-\infty}^{\infty} dxdy\int_{0}^{\infty}[-\mathbf{E}_p^+(x,y,z)\cdot\mathbf{J}_{\text{ncl,e}}(x,y,z) + \mathbf{H}_p^+(x,y,z)\cdot\mathbf{J}_{\text{ncl,m}}(x,y,z)]dz \\ &= \frac{1}{F_p}\iint_{-\infty}^{\infty} dxdy\int_{0}^{\infty}[-\mathbf{E}_p^+(x,y,0)\cdot\mathbf{J}_{\text{ncl,e}}(x,y,0) + \mathbf{H}_p^+(x,y,0)\cdot\mathbf{J}_{\text{ncl,m}}(x,y,0)]\exp[i(k_s+k_p)z]dz \\ &= -\frac{1}{i(k_s+k_p)F_p}\iint_{-\infty}^{\infty}[-\mathbf{E}_p^+\cdot\mathbf{j}_{\text{ncl,e}}([\![\varepsilon\mathbf{E}_\parallel]\!])\delta_{\partial\Omega} + \mathbf{H}_p^+\cdot\mathbf{j}_{\text{ncl,m}}([\![E_\perp]\!])\delta_{\partial\Omega}]_{z=0}\,dxdy, \end{aligned} \tag{S1.9b}$$

where the first equality uses the reciprocity theorem between the excitation source and the CWM [3], the second equality uses Eqs. (S1.2) and (S1.7), and the third equality uses Eq. (S1.4). A pseudo-energy flux is defined as,

$$F_p = \iint_{-\infty}^{\infty}\mathbf{z}\cdot(\mathbf{E}_p^-\times\mathbf{H}_p^+ - \mathbf{E}_p^+\times\mathbf{H}_p^-)_{z=0}\,dxdy, \tag{S1.10}$$

where $(\hat{\mathbf{x}},\hat{\mathbf{y}},\hat{\mathbf{z}})$ are the unit vectors along the $(x,y,z)$ axes, respectively. On the right-hand side of Eq. (S1.9), it follows from Eq. (S1.8) that the normal and tangential components of the electric field in the $z$=0 plane are given by, respectively,

$$E_{\perp}=\mathbf{n}\cdot\mathbf{E}=\mathbf{n}\cdot\left(\sum_{q=1}^{P}c_q^{+}\mathbf{E}_q^{+}+\sum_{q=1}^{P}c_q^{-}\mathbf{E}_q^{-}\right)=\sum_{q=1}^{P}c_q^{+}E_{q,\perp}^{+}+\sum_{q=1}^{P}c_q^{-}E_{q,\perp}^{-}, \tag{S1.11a}$$

$$\mathbf{E}_{\parallel}=\mathbf{E}-\hat{\mathbf{n}}\hat{\mathbf{n}}\cdot\mathbf{E}=(\ddot{\mathbf{I}}-\hat{\mathbf{n}}\hat{\mathbf{n}})\cdot\mathbf{E}=(\ddot{\mathbf{I}}-\hat{\mathbf{n}}\hat{\mathbf{n}})\cdot\left(\sum_{q=1}^{P}c_q^{+}\mathbf{E}_q^{+}+\sum_{q=1}^{P}c_q^{-}\mathbf{E}_q^{-}\right)=\sum_{q=1}^{P}c_q^{+}\mathbf{E}_{q,\parallel}^{+}+\sum_{q=1}^{P}c_q^{-}\mathbf{E}_{q,\parallel}^{-}, \tag{S1.11b}$$

where $\ddot{\mathbf{I}}$ denotes the identity tensor. Substituting Eq. (S1.11) into the right-hand side of Eq. (S1.9) yields a set of coupling equations satisfied by $c_p^{\pm}$ ($p$=1, 2, …, $P$),

$$\begin{aligned}c_p^{+}=\frac{1}{i(k_s-k_p)F_p}\Bigg\{&-\sum_{q=1}^{P}c_q^{+}\iint_{-\infty}^{+\infty}[\mathbf{E}_p^{-}\cdot\mathbf{j}_{\text{ncl,e}}([\![\varepsilon\mathbf{E}_{q,\parallel}^{+}]\!])\delta_{\partial\Omega}]_{z=0}\,dxdy\\&-\sum_{q=1}^{P}c_q^{-}\iint_{-\infty}^{+\infty}[\mathbf{E}_p^{-}\cdot\mathbf{j}_{\text{ncl,e}}([\![\varepsilon\mathbf{E}_{q,\parallel}^{-}]\!])\delta_{\partial\Omega}]_{z=0}\,dxdy\\&+\sum_{q=1}^{P}c_q^{+}\iint_{-\infty}^{+\infty}[\mathbf{H}_p^{-}\cdot\mathbf{j}_{\text{ncl,m}}([\![E_{q,\perp}^{+}]\!])\delta_{\partial\Omega}]_{z=0}\,dxdy\\&+\sum_{q=1}^{P}c_q^{-}\iint_{-\infty}^{+\infty}[\mathbf{H}_p^{-}\cdot\mathbf{j}_{\text{ncl,m}}([\![E_{q,\perp}^{-}]\!])\delta_{\partial\Omega}]_{z=0}\,dxdy\Bigg\},\end{aligned} \tag{S1.12a}$$

$$\begin{aligned}c_p^{-}=-\frac{1}{i(k_s+k_p)F_p}\Bigg\{&-\sum_{q=1}^{P}c_q^{+}\iint_{-\infty}^{+\infty}[\mathbf{E}_p^{+}\cdot\mathbf{j}_{\text{ncl,e}}([\![\varepsilon\mathbf{E}_{q,\parallel}^{+}]\!])\delta_{\partial\Omega}]_{z=0}\,dxdy\\&-\sum_{q=1}^{P}c_q^{-}\iint_{-\infty}^{+\infty}[\mathbf{E}_p^{+}\cdot\mathbf{j}_{\text{ncl,e}}([\![\varepsilon\mathbf{E}_{q,\parallel}^{-}]\!])\delta_{\partial\Omega}]_{z=0}\,dxdy\\&+\sum_{q=1}^{P}c_q^{+}\iint_{-\infty}^{+\infty}[\mathbf{H}_p^{+}\cdot\mathbf{j}_{\text{ncl,m}}([\![E_{q,\perp}^{+}]\!])\delta_{\partial\Omega}]_{z=0}\,dxdy\\&+\sum_{q=1}^{P}c_q^{-}\iint_{-\infty}^{+\infty}[\mathbf{H}_p^{+}\cdot\mathbf{j}_{\text{ncl,m}}([\![E_{q,\perp}^{-}]\!])\delta_{\partial\Omega}]_{z=0}\,dxdy\Bigg\}.\end{aligned} \tag{S1.12b}$$

Equation (S1.12) can be rewritten as,

$$i(k_s-k_p)c_p^{+}+\sum_{q=1}^{P}\kappa_{p,q}^{-,+}c_q^{+}+\sum_{q=1}^{P}\kappa_{p,q}^{-,-}c_q^{-}=0, \tag{S1.13a}$$

$$-i(k_s+k_p)c_p^{-}+\sum_{q=1}^{P}\kappa_{p,q}^{+,+}c_q^{+}+\sum_{q=1}^{P}\kappa_{p,q}^{+,-}c_q^{-}=0, \tag{S1.13b}$$

where the NEBC-induced CWM-coupling coefficients are defined as,

$$\kappa_{p,q}^{-,+}=\frac{1}{F_p}\iint_{-\infty}^{+\infty}[\mathbf{E}_p^{-}\cdot\mathbf{j}_{\text{ncl,e}}([\![\varepsilon\mathbf{E}_{q,\parallel}^{+}]\!])\delta_{\partial\Omega}]_{z=0}\,dxdy-\frac{1}{F_p}\iint_{-\infty}^{+\infty}[\mathbf{H}_p^{-}\cdot\mathbf{j}_{\text{ncl,m}}([\![E_{q,\perp}^{+}]\!])\delta_{\partial\Omega}]_{z=0}\,dxdy, \tag{S1.14a}$$

$$\kappa_{p,q}^{-,-}=\frac{1}{F_p}\iint_{-\infty}^{+\infty}[\mathbf{E}_p^{-}\cdot\mathbf{j}_{\text{ncl,e}}([\![\varepsilon\mathbf{E}_{q,\parallel}^{-}]\!])\delta_{\partial\Omega}]_{z=0}\,dxdy-\frac{1}{F_p}\iint_{-\infty}^{+\infty}[\mathbf{H}_p^{-}\cdot\mathbf{j}_{\text{ncl,m}}([\![E_{q,\perp}^{-}]\!])\delta_{\partial\Omega}]_{z=0}\,dxdy, \tag{S1.14b}$$

$$\kappa_{p,q}^{+,+}=\frac{1}{F_p}\iint_{-\infty}^{+\infty}[\mathbf{E}_p^{+}\cdot\mathbf{j}_{\text{ncl,e}}([\![\varepsilon\mathbf{E}_{q,\parallel}^{+}]\!])\delta_{\partial\Omega}]_{z=0}\,dxdy-\frac{1}{F_p}\iint_{-\infty}^{+\infty}[\mathbf{H}_p^{+}\cdot\mathbf{j}_{\text{ncl,m}}([\![E_{q,\perp}^{+}]\!])\delta_{\partial\Omega}]_{z=0}\,dxdy, \tag{S1.14c}$$

$$\kappa_{p,q}^{+,-}=\frac{1}{F_p}\iint_{-\infty}^{+\infty}[\mathbf{E}_p^{+}\cdot\mathbf{j}_{\text{ncl,e}}([\![\varepsilon\mathbf{E}_{q,\parallel}^{-}]\!])\delta_{\partial\Omega}]_{z=0}\,dxdy-\frac{1}{F_p}\iint_{-\infty}^{+\infty}[\mathbf{H}_p^{+}\cdot\mathbf{j}_{\text{ncl,m}}([\![E_{q,\perp}^{-}]\!])\delta_{\partial\Omega}]_{z=0}\,dxdy. \tag{S1.14d}$$

Substituting the definition (S1.4) of nonclassical surface source into Eq. (S1.14) yields,

$$\kappa_{p,q}^{-,+}=\frac{1}{F_p}\iint_{-\infty}^{+\infty}(\mathbf{E}_p^{-}\cdot i\omega d_{\parallel}[\![\varepsilon\mathbf{E}_{q,\parallel}^{+}]\!]\delta_{\partial\Omega})_{z=0}\,dxdy-\frac{1}{F_p}\iint_{-\infty}^{+\infty}\{\mathbf{H}_p^{-}\cdot[\hat{\mathbf{n}}\times\nabla_{\parallel}(d_{\perp}[\![E_{q,\perp}^{+}]\!])\delta_{\partial\Omega}]\}_{z=0}\,dxdy, \tag{S1.15a}$$

$$\kappa_{p,q}^{-,-}=\frac{1}{F_p}\iint_{-\infty}^{+\infty}(\mathbf{E}_p^{-}\cdot i\omega d_{\parallel}[\![\varepsilon\mathbf{E}_{q,\parallel}^{-}]\!]\delta_{\partial\Omega})_{z=0}\,dxdy-\frac{1}{F_p}\iint_{-\infty}^{+\infty}\{\mathbf{H}_p^{-}\cdot[\hat{\mathbf{n}}\times\nabla_{\parallel}(d_{\perp}[\![E_{q,\perp}^{-}]\!])\delta_{\partial\Omega}]\}_{z=0}\,dxdy, \tag{S1.15b}$$

$$\kappa_{p,q}^{+,+}=\frac{1}{F_p}\iint_{-\infty}^{+\infty}(\mathbf{E}_p^{+}\cdot i\omega d_{\parallel}[\![\varepsilon\mathbf{E}_{q,\parallel}^{+}]\!]\delta_{\partial\Omega})_{z=0}dxdy-\frac{1}{F_p}\iint_{-\infty}^{+\infty}\{\mathbf{H}_p^{+}\cdot[\hat{\mathbf{n}}\times\nabla_{\parallel}(d_{\perp}[\![E_{q,\perp}^{+}]\!])\delta_{\partial\Omega}]\}_{z=0}dxdy, \tag{S1.15c}$$

$$\kappa_{p,q}^{+,-}=\frac{1}{F_p}\iint_{-\infty}^{+\infty}(\mathbf{E}_p^{+}\cdot i\omega d_{\parallel}[\![\varepsilon\mathbf{E}_{q,\parallel}^{-}]\!]\delta_{\partial\Omega})_{z=0}dxdy-\frac{1}{F_p}\iint_{-\infty}^{+\infty}\{\mathbf{H}_p^{+}\cdot[\hat{\mathbf{n}}\times\nabla_{\parallel}(d_{\perp}[\![E_{q,\perp}^{-}]\!])\delta_{\partial\Omega}]\}_{z=0}dxdy. \tag{S1.15d}$$

Equation (S1.13) constitutes a set of *CWM-coupling equations* satisfied by the expansion coefficients $c_p^{\pm}$. Solving for its nontrivial solutions and substituting them into Eq. (S1.8) yields the electromagnetic field of the NWM.

The procedure for solving Eq. (S1.13) is as follows. By expressing Eq. (S1.13) in matrix form, we obtain,

$$\mathbf{A}(k_s)\mathbf{a}=\mathbf{0}, \tag{S1.16}$$

with the definition,

$$\mathbf{A}(k_s)=\begin{bmatrix} i(k_s\mathbf{I}-\mathbf{W})+\mathbf{K}^{-,+} & \mathbf{K}^{-,-} \\ \mathbf{K}^{+,+} & -i(k_s\mathbf{I}+\mathbf{W})+\mathbf{K}^{+,-} \end{bmatrix}, \tag{S1.17}$$

where $\mathbf{I}$ is the identity matrix, and $\mathbf{W}$ is a diagonal matrix whose $p$th diagonal element is the propagation constant $k_p$ of the $p$th-order CWM. The elements in the $p$th row and $q$th column of matrices $\mathbf{K}^{-,+}$, $\mathbf{K}^{-,-}$, $\mathbf{K}^{+,+}$, and $\mathbf{K}^{+,-}$ are $\kappa_{p,q}^{-,+}$, $\kappa_{p,q}^{-,-}$, $\kappa_{p,q}^{+,+}$, and $\kappa_{p,q}^{+,-}$, respectively. It can be easily proved that Eq. (S1.16) is equivalent to the matrix eigenvalue problem given in Eq. (2) in the main text. Solving for its $n$th eigenvalue $k_s=\tilde{k}_n$ ($n$=1, 2, …, 2$P$) yields the propagation constant of the $n$th-order NWM. Solving for the eigenvector $\mathbf{a}=\tilde{\mathbf{a}}_n$ corresponding to $\tilde{k}_n$ and substituting it into Eq. (S1.8) then yields the electromagnetic field of the $n$th-order NWM, i.e., $\boldsymbol{\psi}(\boldsymbol{\rho})=\tilde{\boldsymbol{\psi}}_n(\boldsymbol{\rho})=\boldsymbol{\Phi}\tilde{\mathbf{a}}_n$.

**B. Simplification of the expressions of the NEBC-induced CWM-coupling coefficients**

In the following, we consider simplifying the expressions for the NEBC-induced CWM-coupling coefficients, given on the right-hand side of Eq. (S1.15), so as to obtain the form of Eq. (3) in the main text. For Eq. (S1.15a), the first integral term on the right-hand side can be simplified as,

$$\begin{aligned}
&\iint_{-\infty}^{+\infty}(\mathbf{E}_p^{-}\cdot i\omega d_{\parallel}[\![\varepsilon\mathbf{E}_{q,\parallel}^{+}]\!]\delta_{\partial\Omega})_{z=0}dxdy\\
&=\iint_{-\infty}^{+\infty}(\mathbf{E}_p^{-}\cdot i\omega d_{\parallel}[\![\varepsilon\mathbf{E}_{q,\parallel}^{+}]\!])_{z=0}\delta_{\partial\Omega}dxdy\\
&=\iint_{-\infty}^{+\infty}dxdy\int_{-\infty}^{+\infty}(i\omega d_{\parallel}[\![\varepsilon]\!]\mathbf{E}_{p,\parallel}^{-}\cdot\mathbf{E}_{q,\parallel}^{+})_{z=0}\delta_{\partial\Omega}\mathrm{rect}(z/z_0)/z_0dz\\
&=\iint_{\partial\Omega}(i\omega d_{\parallel}[\![\varepsilon]\!]\mathbf{E}_{p,\parallel}^{-}\cdot\mathbf{E}_{q,\parallel}^{+})_{z=0}\mathrm{rect}(z/z_0)/z_0ds\\
&=\oint_L dl\int_{-\infty}^{+\infty}(i\omega d_{\parallel}[\![\varepsilon]\!]\mathbf{E}_{p,\parallel}^{-}\cdot\mathbf{E}_{q,\parallel}^{+})_{z=0}\mathrm{rect}(z/z_0)/z_0dz\\
&=\oint_L(i\omega d_{\parallel}[\![\varepsilon]\!]\mathbf{E}_{p,\parallel}^{-}\cdot\mathbf{E}_{q,\parallel}^{+})_{z=0}dl,
\end{aligned} \tag{S1.18}$$

where the closed contour $L$ is the intersection curve of $\partial\Omega$ with the $z$=0 plane [as shown in Fig. S1(a)], with definition,

$$\mathrm{rect}(z/z_0)=\begin{cases}1, & \text{for } |z|\le z_0/2,\\ 0, & \text{else.}\end{cases} \tag{S1.19}$$

In Eq. (S1.18), the first equality uses the fact that $\delta_{\partial\Omega}$ is independent of $z$, which can be proven as follows. There is,

$$\begin{aligned}
&\iiint_{R^3}\frac{\partial\delta_{\partial\Omega}}{\partial z}\phi d^3\mathbf{r}\\
&=\iiint_{R^3}\frac{\partial(\delta_{\partial\Omega}\phi)}{\partial z}d^3\mathbf{r}-\iiint_{R^3}\delta_{\partial\Omega}\frac{\partial\phi}{\partial z}d^3\mathbf{r}\\
&=\iint_{R^2}dxdy\int_{-\infty}^{+\infty}\frac{\partial(\delta_{\partial\Omega}\phi)}{\partial z}dz-\iint_{\partial\Omega}\frac{\partial\phi}{\partial z}ds\\
&=\iint_{R^2}dxdy\int_{-\infty}^{+\infty}\frac{\partial(\delta_{\partial\Omega}\phi)}{\partial z}dz-\oint_L dl\int_{-\infty}^{+\infty}\frac{\partial\phi}{\partial z}dz\\
&=\iint_{R^2}(\delta_{\partial\Omega}\phi)\Big|_{z=-\infty}^{+\infty}dxdy-\oint_L\phi\Big|_{z=-\infty}^{+\infty}dl\\
&=0,
\end{aligned} \tag{S1.20}$$

where $\phi(\mathbf{r})$ is an arbitrary test function as specified following Eq. (S1.5), the second equality uses the definition (S1.5) of the surface Dirac function, the third equality uses the translational invariance of the $\partial\Omega$ along the $z$ direction, and the last equality uses $\phi(|\mathbf{r}|\rightarrow\infty)=0$. Because Eq. (S1.20) holds for an arbitrary $\phi(\mathbf{r})$, it follows that $\frac{\partial\delta_{\partial\Omega}}{\partial z}=0$ , i.e., the $\delta_{\partial\Omega}$ is independent of $z$.

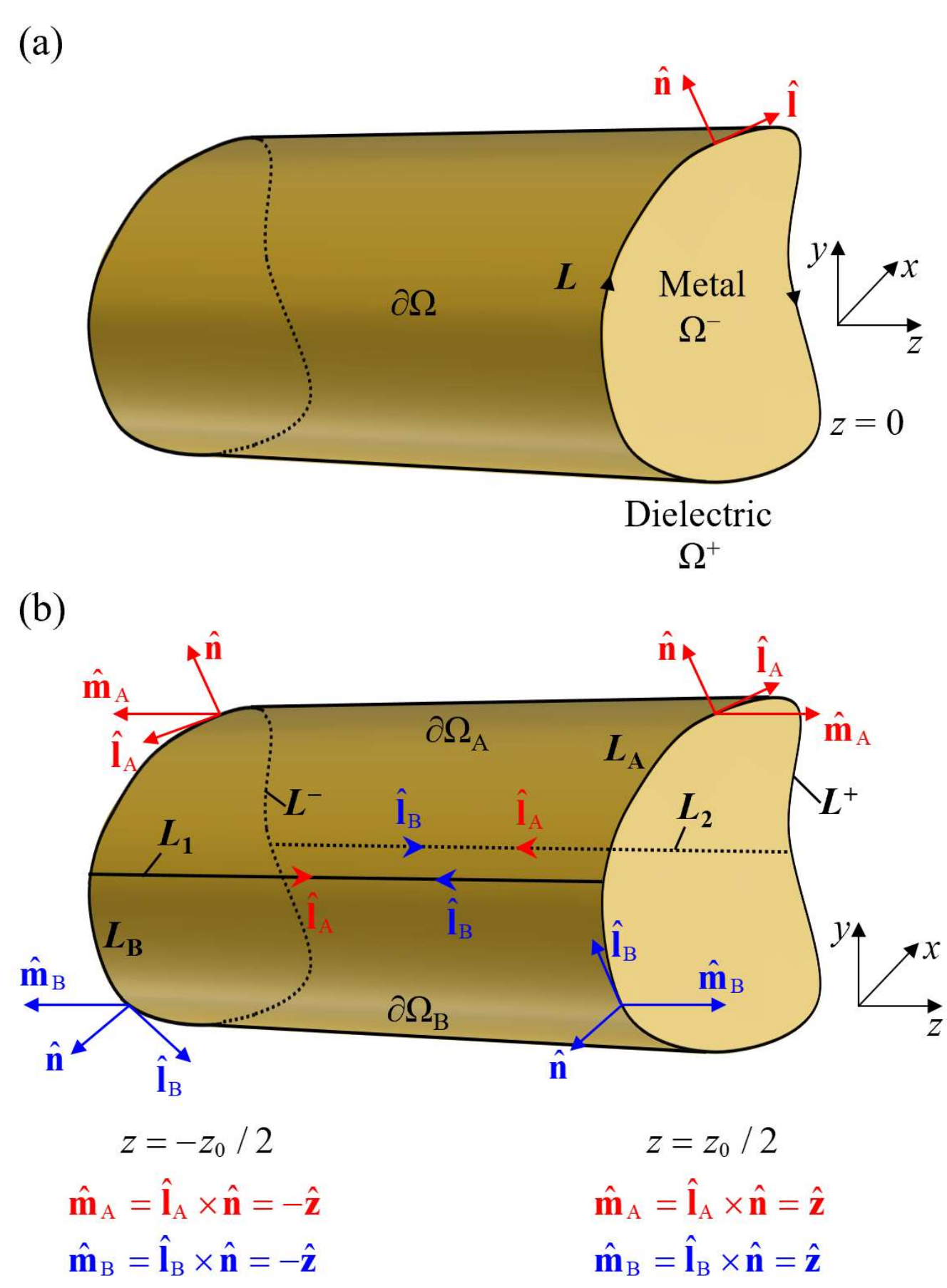

FIG. S1 (a) Schematic of a $z$-translationally invariant mesoscale plasmonic waveguide (MPW). (b) Schematic for the proof of Eq. (S1.25).

The second integral on the right-hand side of Eq. (S1.15a) can be rewritten as,

$$\begin{aligned}
&\iint_{-\infty}^{+\infty}\{\mathbf{H}_p^-\cdot[\hat{\mathbf{n}}\times\nabla_{\parallel}(d_\perp [\![E_{q,\perp}^+]\!])\delta_{\partial\Omega}]\}_{z=0}\,dxdy\\
&=\iint_{-\infty}^{+\infty}\{\mathbf{H}_p^-\cdot[\hat{\mathbf{n}}\times\nabla_{\parallel}(d_\perp [\![E_{q,\perp}^+]\!])]\}_{z=0}\,\delta_{\partial\Omega}dxdy\\
&=\frac{1}{N_{p,q}^{-,+}(z_0)}\iint_{-\infty}^{+\infty}dxdy\int_{-\infty}^{+\infty}\{\mathbf{H}_p^-\cdot[\hat{\mathbf{n}}\times\nabla_{\parallel}(d_\perp [\![E_{q,\perp}^+]\!])]\}_{z=0}\,\delta_{\partial\Omega}\exp[i(-k_p+k_q)z]\mathrm{rect}(z/z_0)/z_0dz\\
&=\frac{1}{z_0N_{p,q}^{-,+}(z_0)}\iint_{-\infty}^{+\infty}dxdy\int_{-\infty}^{+\infty}\mathbf{H}_p^-\cdot[\hat{\mathbf{n}}\times\nabla_{\parallel}(d_\perp [\![E_{q,\perp}^+]\!])\delta_{\partial\Omega'}]dz,
\end{aligned}\tag{S1.21}$$

where we define,

$$\begin{aligned}
N_{p,q}^{-,+}(z_0)&=\int_{-\infty}^{+\infty}\exp[i(-k_p+k_q)z]\mathrm{rect}(z/z_0)/z_0dz\\
&=\begin{cases}\dfrac{\exp[i(-k_p+k_q)z_0/2]-\exp[-i(-k_p+k_q)z_0/2]}{iz_0(-k_p+k_q)}, & \text{for } p\neq q,\\ 1, & \text{for } p=q,\end{cases}
\end{aligned}\tag{S1.22}$$

and the following is used,

$$\delta_{\partial\Omega}\mathrm{rect}(z/z_0)=\delta_{\partial\Omega'},\tag{S1.23}$$

where $\partial\Omega'$ is defined as the portion of $\partial\Omega$ within the range of $-z_0/2\le z\le z_0/2$. Equation (S1.23) can be proven as follows. There is,

$$\begin{aligned}&\iiint_{R^3}\delta_{\partial\Omega}\mathrm{rect}(z/z_0)\phi d^3\mathbf{r}=\iint_{\partial\Omega}\mathrm{rect}(z/z_0)\phi ds=\oint_L dl\int_{-\infty}^{+\infty}\mathrm{rect}(z/z_0)\phi dz\\&=\oint_L dl\int_{-z_0/2}^{z_0/2}\phi dz=\iint_{\partial\Omega'}\phi ds=\iiint_{R^3}\delta_{\partial\Omega'}\phi d^3\mathbf{r},\end{aligned}\tag{S1.24}$$

where $\phi(\mathbf{r})$ is an arbitrary test function as specified following Eq. (S1.5), and the first and last equalities use the definition (S1.5) of the surface Dirac function. Because Eq. (S1.24) holds for an arbitrary $\phi(\mathbf{r})$, Eq. (S1.23) is thus obtained.

It can be proven that, in the right-hand side of Eq. (S1.21), there is,

$$\hat{\mathbf{n}}\times\nabla_{\|}(d_\perp[\![E_{q,\perp}^+]\!])\delta_{\partial\Omega'}=-\nabla\times(d_\perp[\![E_{q,\perp}^+]\!]\hat{\mathbf{n}}\delta_{\partial\Omega'})+(d_\perp[\![E_{q,\perp}^+]\!]_{z=z_0/2}\hat{\mathbf{l}}\delta_{L^+}-d_\perp[\![E_{q,\perp}^+]\!]_{z=-z_0/2}\hat{\mathbf{l}}\delta_{L^-}),\tag{S1.25}$$

where $\hat{\mathbf{l}}=\hat{\mathbf{n}}\times\hat{\mathbf{z}}$, with $\hat{\mathbf{z}}$ being the unit vector along the positive $z$-direction. Because $\hat{\mathbf{l}}$ is independent of $z$, $\hat{\mathbf{l}}$ can be the unit tangential vector along the intersection curve [denoted as $L(z_C)$] of $\partial\Omega$ with the plane $z=z_C$ for an arbitrary constant $z_C$. For instance, taking $z_C=0$ defines $L=L(0)$ in Fig. S1(a). $\delta_{L^\pm}$ is the line Dirac function defined to satisfy,

$$\iiint_{R^3}[\mathbf{f}(\mathbf{r})\delta_{L^\pm}]\phi(\mathbf{r})d^3\mathbf{r}=\oint_{L^\pm}\mathbf{f}(\mathbf{r})\phi(\mathbf{r})dl,\tag{S1.26}$$

where $\phi(\mathbf{r})$ is an arbitrary test function as specified following Eq. (S1.5), and $L^+$ and $L^-$ are the intersection curves of $\partial\Omega$ with the planes $z=z_0/2$ and $z=-z_0/2$, respectively, as illustrated in Fig. S1(b).

*Proof*: First, projecting $\nabla\times(d_\perp[\![E_{q,\perp}^+]\!]\hat{\mathbf{n}}\delta_{\partial\Omega'})$ onto the test function $\phi(\mathbf{r})$ yields,

$$\begin{aligned}&\iiint_{R^3}\nabla\times(d_\perp[\![E_{q,\perp}^+]\!]\hat{\mathbf{n}}\delta_{\partial\Omega'})\phi d^3\mathbf{r}\\&=\iiint_{R^3}[\nabla\times(d_\perp[\![E_{q,\perp}^+]\!]\hat{\mathbf{n}}\delta_{\partial\Omega'}\phi)-(\nabla\phi)\times d_\perp[\![E_{q,\perp}^+]\!]\hat{\mathbf{n}}\delta_{\partial\Omega'}]d^3\mathbf{r}\\&=\oiint_{\partial R^3}\hat{\mathbf{N}}\times(d_\perp[\![E_{q,\perp}^+]\!]\hat{\mathbf{n}}\delta_{\partial\Omega'}\phi)ds-\iint_{\partial\Omega'}(\nabla\phi)\times d_\perp[\![E_{q,\perp}^+]\!]\hat{\mathbf{n}}ds,\end{aligned}\tag{S1.27}$$

where $\partial R^3$ denotes the boundary surface of the whole space $R^3$ at infinity, $\hat{\mathbf{N}}$ is the unit outwards-pointing normal vector on $\partial R^3$, and $\phi(\mathbf{r})$ is an arbitrary scalar complex-valued test function that is infinitely differentiable and satisfies $\phi(|\mathbf{r}|\to\infty)=0$. Due to $\phi(|\mathbf{r}|\to\infty)=0$, the first surface integral on the right-hand side of Eq. (S1.27) vanishes. For the integrand of the second integral on the right-hand side of Eq. (S1.27), there is,

$$(\nabla\phi)\times\hat{\mathbf{n}}d_\perp[\![E_{q,\perp}^+]\!]=(\nabla_S\phi)\times\hat{\mathbf{n}}d_\perp[\![E_{q,\perp}^+]\!]=\nabla_S\times(\phi\hat{\mathbf{n}}d_\perp[\![E_{q,\perp}^+]\!])-\phi\nabla_S\times(d_\perp[\![E_{q,\perp}^+]\!]\hat{\mathbf{n}}),\tag{S1.28}$$

where the first equality uses the definitions of the volume and surface gradients [see Eq. (S1.39a)], $\nabla_S\phi$ represents the surface gradient of $\phi$, and the second equality uses the surface differential identity [4]. For the second term on the right-hand side of Eq. (S1.28), there is,

$$\begin{aligned}&\nabla_S\times(d_\perp[\![E_{q,\perp}^+]\!]\hat{\mathbf{n}})\\&\overset{a}{=}\nabla_S(d_\perp[\![E_{q,\perp}^+]\!])\times\hat{\mathbf{n}}-d_\perp[\![E_{q,\perp}^+]\!]\nabla_S\times\hat{\mathbf{n}}\\&\overset{b}{=}\nabla_S(d_\perp[\![E_{q,\perp}^+]\!])\times\hat{\mathbf{n}}\\&\overset{c}{=}\nabla_{\|}(d_\perp[\![E_{q,\perp}^+]\!])\times\hat{\mathbf{n}},\end{aligned}\tag{S1.29}$$

where the following steps are used:

*a*. Application of surface differential identity [4].

*b*. Application of $\nabla_S\times\hat{\mathbf{n}}=\mathbf{0}$.

*c*. Application of the relation between $\nabla_S f$ and $\nabla_{\|}f$ [see Eq. (S1.39a)].

Substituting Eq. (S1.29) into Eq. (S1.28), and subsequently into the right-hand side of Eq. (S1.27), yields,

$$\begin{aligned}&\iiint_{R^3}\nabla\times(d_\perp[\![E_{q,\perp}^+]\!]\hat{\mathbf{n}}\delta_{\partial\Omega'})\phi d^3\mathbf{r}\\&=-\iint_{\partial\Omega'}\nabla_S\times(\phi\hat{\mathbf{n}}d_\perp[\![E_{q,\perp}^+]\!])ds-\iint_{\partial\Omega'}\hat{\mathbf{n}}\times\nabla_{\|}(d_\perp[\![E_{q,\perp}^+]\!])\phi ds\\&=-\iint_{\partial\Omega'}\nabla_S\times(\phi\hat{\mathbf{n}}d_\perp[\![E_{q,\perp}^+]\!])ds-\iiint_{R^3}\hat{\mathbf{n}}\times\nabla_{\|}(d_\perp[\![E_{q,\perp}^+]\!])\delta_{\partial\Omega'}\phi d^3\mathbf{r},\end{aligned}\tag{S1.30}$$

where the second equality follows from the definition (S1.5) of the surface Dirac function.

Next, we consider calculating the first surface integral on the right-hand side of Eq. (S1.30). As illustrated in Fig. S1(b), two straight lines $L_1$ and $L_2$ parallel to the $z$-axis are introduced on $\partial\Omega'$, and partition $\partial\Omega'$ into two portions $\partial\Omega_A$ and $\partial\Omega_B$, i.e., $\partial\Omega' = \partial\Omega_A \cup \partial\Omega_B$. Consequently, the first surface-integral term on the right-hand side of Eq. (S1.30) becomes,

$$\begin{aligned}
&\iint_{\partial\Omega'} \nabla_S \times (\phi \hat{\mathbf{n}} d_\perp \llbracket E_{q,\perp}^{+} \rrbracket) ds \\
&= \iint_{\partial\Omega_A} \nabla_S \times (\phi \hat{\mathbf{n}} d_\perp \llbracket E_{q,\perp}^{+} \rrbracket) ds + \iint_{\partial\Omega_B} \nabla_S \times (\phi \hat{\mathbf{n}} d_\perp \llbracket E_{q,\perp}^{+} \rrbracket) ds \\
&\overset{a}{=} \oint_{L_A} \hat{\mathbf{m}}_A \times (\phi \hat{\mathbf{n}} d_\perp \llbracket E_{q,\perp}^{+} \rrbracket) dl + \oint_{L_B} \hat{\mathbf{m}}_B \times (\phi \hat{\mathbf{n}} d_\perp \llbracket E_{q,\perp}^{+} \rrbracket) dl \\
&\overset{b}{=} \oint_{L^+} \hat{\mathbf{z}} \times (\phi \hat{\mathbf{n}} d_\perp \llbracket E_{q,\perp}^{+} \rrbracket) dl + \oint_{L^-} (-\hat{\mathbf{z}}) \times (\phi \hat{\mathbf{n}} d_\perp \llbracket E_{q,\perp}^{+} \rrbracket) dl \\
&\overset{c}{=} -\oint_{L^+} \hat{\mathbf{l}} (\phi d_\perp \llbracket E_{q,\perp}^{+} \rrbracket) dl + \oint_{L^-} \hat{\mathbf{l}} (\phi d_\perp \llbracket E_{q,\perp}^{+} \rrbracket) dl \\
&\overset{d}{=} -\iiint_{R^3} d_\perp \llbracket E_{q,\perp}^{+} \rrbracket_{z=z_0/2} \hat{\mathbf{l}} \delta_{L^+} \phi d^3\mathbf{r} + \iiint_{R^3} d_\perp \llbracket E_{q,\perp}^{+} \rrbracket_{z=-z_0/2} \hat{\mathbf{l}} \delta_{L^-} \phi d^3\mathbf{r} \\
&= -\iiint_{R^3} (d_\perp \llbracket E_{q,\perp}^{+} \rrbracket_{z=z_0/2} \hat{\mathbf{l}} \delta_{L^+} - d_\perp \llbracket E_{q,\perp}^{+} \rrbracket_{z=-z_0/2} \hat{\mathbf{l}} \delta_{L^-}) \phi d^3\mathbf{r},
\end{aligned} \tag{S1.31}$$

where the following steps are used:

*a*. Application of surface differential identity [4], where $\hat{\mathbf{m}}_\gamma = \hat{\mathbf{l}}_\gamma \times \hat{\mathbf{n}}$ ($\gamma$=A, B), $\hat{\mathbf{l}}_\gamma$ is the unit tangential vector of $L_\gamma$ ($\hat{\mathbf{l}}_\gamma$ satisfies the right-hand screw rule with respect to $\hat{\mathbf{n}}$), and $L_\gamma$ is the boundary curve of $\partial\Omega_\gamma$.

*b*. Application of the cancellation of the line integrals along $L_1$ and $L_2$, as well as $\hat{\mathbf{m}}_A = \hat{\mathbf{m}}_B = \hat{\mathbf{z}}$ on $L^+$ and $\hat{\mathbf{m}}_A = \hat{\mathbf{m}}_B = -\hat{\mathbf{z}}$ on $L^-$.

*c*. Application of the definition $\hat{\mathbf{l}} = \hat{\mathbf{n}} \times \hat{\mathbf{z}}$.

*d*. Application of the definition (S1.26) of the line Dirac function.

Substituting Eq. (S1.31) into Eq. (S1.30) yields,

$$\begin{aligned}
&\iiint_{R^3} \nabla \times (d_\perp \llbracket E_{q,\perp}^{+} \rrbracket \hat{\mathbf{n}} \delta_{\partial\Omega'}) \phi d^3\mathbf{r} \\
&= \iiint_{R^3} [d_\perp \llbracket E_{q,\perp}^{+} \rrbracket_{z=z_0/2} \hat{\mathbf{l}} \delta_{L^+} - d_\perp \llbracket E_{q,\perp}^{+} \rrbracket_{z=-z_0/2} \hat{\mathbf{l}} \delta_{L^-} - \hat{\mathbf{n}} \times \nabla_{\parallel} (d_\perp \llbracket E_{q,\perp}^{+} \rrbracket) \delta_{\partial\Omega'}] \phi d^3\mathbf{r}.
\end{aligned} \tag{S1.32}$$

Because Eq. (S1.32) holds for any test function $\phi(\mathbf{r})$, we obtain,

$$\nabla \times (d_\perp \llbracket E_{q,\perp}^{+} \rrbracket \hat{\mathbf{n}} \delta_{\partial\Omega'}) = d_\perp \llbracket E_{q,\perp}^{+} \rrbracket_{z=z_0/2} \hat{\mathbf{l}} \delta_{L^+} - d_\perp \llbracket E_{q,\perp}^{+} \rrbracket_{z=-z_0/2} \hat{\mathbf{l}} \delta_{L^-} - \hat{\mathbf{n}} \times \nabla_{\parallel} (d_\perp \llbracket E_{q,\perp}^{+} \rrbracket) \delta_{\partial\Omega'}, \tag{S1.33}$$

which directly yields Eq. (S1.25).

*The proof is complete.*

Substituting Eq. (S1.25) into the right-hand side of Eq. (S1.21) yields,

$$\begin{aligned}
&\iint_{-\infty}^{+\infty} \{\mathbf{H}_p^- \cdot [\hat{\mathbf{n}} \times \nabla_{\parallel} (d_\perp \llbracket E_{q,\perp}^{+} \rrbracket) \delta_{\partial\Omega}]\}_{z=0} dxdy \\
&= -\frac{1}{z_0 N_{p,q}^{-,+}(z_0)} \iint_{-\infty}^{+\infty} dxdy \int_{-\infty}^{+\infty} \mathbf{H}_p^- \cdot [\nabla \times (d_\perp \llbracket E_{q,\perp}^{+} \rrbracket \hat{\mathbf{n}} \delta_{\partial\Omega'})] dz \\
&\quad + \frac{1}{z_0 N_{p,q}^{-,+}(z_0)} \Big[ \oint_{L^+} (d_\perp H_{p,l}^- \llbracket E_{q,\perp}^{+} \rrbracket)_{z=z_0/2} dl - \oint_{L^-} (d_\perp H_{p,l}^- \llbracket E_{q,\perp}^{+} \rrbracket)_{z=-z_0/2} dl \Big],
\end{aligned} \tag{S1.34}$$

where the definition (S1.26) of line Dirac function is used, and we define $H_{p,l}^- = \mathbf{H}_p^- \cdot \hat{\mathbf{l}}$. In the right-hand side of Eq. (S1.34), there is

$$
\begin{aligned}
&\iint_{-\infty}^{+\infty} dxdy \int_{-\infty}^{+\infty} \mathbf{H}_p^- \cdot [\nabla \times (d_\perp [\![ E_{q,\perp}^+ ]\!] \hat{\mathbf{n}} \delta_{\partial\Omega'})] dz \\
&= \iint_{-\infty}^{+\infty} dxdy \int_{-\infty}^{+\infty} \{(\nabla \times \mathbf{H}_p^-) \cdot (d_\perp [\![ E_{q,\perp}^+ ]\!] \hat{\mathbf{n}} \delta_{\partial\Omega'}) - \nabla \cdot [\mathbf{H}_p^- \times (d_\perp [\![ E_{q,\perp}^+ ]\!] \hat{\mathbf{n}} \delta_{\partial\Omega'})]\} dz \\
&\overset{a}{=} \iint_{-\infty}^{+\infty} dxdy \int_{-\infty}^{+\infty} (\nabla \times \mathbf{H}_p^-) \cdot (d_\perp [\![ E_{q,\perp}^+ ]\!] \hat{\mathbf{n}} \delta_{\partial\Omega'}) dz \\
&\overset{b}{=} \iint_{-\infty}^{+\infty} dxdy \int_{-\infty}^{+\infty} (-i\omega\varepsilon \mathbf{E}_p^-) \cdot (d_\perp [\![ E_{q,\perp}^+ ]\!] \hat{\mathbf{n}} \delta_{\partial\Omega'}) dz \\
&\overset{c}{=} \iint_{\partial\Omega'} (-i\omega d_\perp [\![ \varepsilon E_{p,\perp}^- E_{q,\perp}^+ ]\!]) ds \\
&\overset{d}{=} \oint_L dl \int_{-z_0/2}^{z_0/2} (-i\omega d_\perp [\![ \varepsilon E_{p,\perp}^- E_{q,\perp}^+ ]\!])_{z=0} \exp[i(-k_p + k_q) z] dz \\
&\overset{e}{=} z_0 N_{p,q}^{-,+}(z_0) \oint_L (-i\omega d_\perp [\![ \varepsilon E_{p,\perp}^- E_{q,\perp}^+ ]\!])_{z=0} dl,
\end{aligned}
\tag{S1.35}
$$

where the following steps are used:

*a*. Application of

$$
\iint_{-\infty}^{+\infty} dxdy \int_{-\infty}^{+\infty} \nabla \cdot [\mathbf{H}_p^- \times (d_\perp [\![ E_{q,\perp}^+ ]\!] \hat{\mathbf{n}} \delta_{\partial\Omega'})] dz = \oiint_{\partial R^3} \hat{\mathbf{N}} \cdot [\mathbf{H}_p^- \times (d_\perp [\![ E_{q,\perp}^+ ]\!] \hat{\mathbf{n}} \delta_{\partial\Omega'})] ds = 0, \tag{S1.36}
$$

where $\partial R^3$ denotes the boundary surface of the whole space $R^3$ at infinity, and $\hat{\mathbf{N}}$ is the outwards-pointing normal vector on $\partial R^3$.

*b*. Application of Maxwell's equation (S1.1b).

*c*. Application of the definition (S1.5) of surface Dirac function, and the continuity of $\varepsilon E_{p,\perp}^-$ across $\partial\Omega$.

*d*. Application of the definition of $\partial\Omega'$ given after Eq. (S1.23), and Eq. (S1.2).

*e*. Application of the definition (S1.22) of $N_{p,q}^{-,+}(z_0)$.

On the right-hand side of Eq. (S1.34), there is

$$
\begin{aligned}
&\oint_{L^+} (d_\perp H_{p,l}^- [\![ E_{q,\perp}^+ ]\!])_{z=z_0/2} dl - \oint_{L^-} (d_\perp H_{p,l}^- [\![ E_{q,\perp}^+ ]\!])_{z=-z_0/2} dl \\
&= \oint_L (d_\perp H_{p,l}^- [\![ E_{q,\perp}^+ ]\!])_{z=0} \exp[i(-k_p + k_q) z_0 / 2] dl \\
&\quad - \oint_L (d_\perp H_{p,l}^- [\![ E_{q,\perp}^+ ]\!])_{z=0} \exp[-i(-k_p + k_q) z_0 / 2] dl \\
&= i z_0 (-k_p + k_q) N_{p,q}^{-,+}(z_0) \oint_L (d_\perp H_{p,l}^- [\![ E_{q,\perp}^+ ]\!])_{z=0} dl,
\end{aligned}
\tag{S1.37}
$$

where the first equality uses Eq. (S1.2), and the second equality uses Eq. (S1.22). Substituting Eqs. (S1.35) and (S1.37) into Eq. (S1.34) yields,

$$
\begin{aligned}
&\iint_{-\infty}^{+\infty} \{\mathbf{H}_p^- \cdot [\hat{\mathbf{n}} \times \nabla_\parallel (d_\perp [\![ E_{q,\perp}^+ ]\!]) \delta_{\partial\Omega}]\}_{z=0} dxdy \\
&= -\oint_L (-i\omega d_\perp [\![ \varepsilon E_{p,\perp}^- E_{q,\perp}^+ ]\!])_{z=0} dl + i(-k_p + k_q) \oint_L (d_\perp H_{p,l}^- [\![ E_{q,\perp}^+ ]\!])_{z=0} dl.
\end{aligned}
\tag{S1.38}
$$

Substituting Eqs. (S1.18) and (S1.38) into Eq. (S1.15a) yields Eq. (3) in the main text for $(\sigma,\tau)=(-,+)$. Similarly, one can obtain Eq. (3) in the main text for the other cases of $(\sigma,\tau)$.

In the following, we will provide a brief introduction to the surface differential operations used in the above derivations. For the interface $\partial\Omega$, an orthogonal curvilinear coordinate system $(v_1,v_2,v_3)$ can be defined such that $\partial\Omega$ is expressed as the parametric equations $\mathbf{r}=\mathbf{r}(v_1,v_2)$, with $\partial\mathbf{r}/\partial v_3$ being the normal vector on $\partial\Omega$. The unit vectors are defined as $\hat{\mathbf{u}}_i = (\partial\mathbf{r} / \partial v_i) / h_i$ ($i$=1, 2, 3), in which $h_i=|\partial\mathbf{r}/\partial v_i|$ represent the Lamé coefficients. Here, the $\hat{\mathbf{u}}_1$ and $\hat{\mathbf{u}}_2$ are the orthogonal unit tangential vectors on $\partial\Omega$, and the $\hat{\mathbf{u}}_3$ is the unit normal vector on $\partial\Omega$. With the constraint $h_3=1$, this coordinate system is referred to as the Dupin orthogonal curvilinear coordinate system. In this framework, the surface gradient, surface divergence, and surface curl can be respectively expressed as [4],

$$
\nabla_S f = \sum_{i=1}^{2} \frac{\hat{\mathbf{u}}_i}{h_i} \frac{\partial f}{\partial v_i} + J \hat{\mathbf{u}}_3 f, \tag{S1.39a}
$$

$$
\nabla_S \cdot \mathbf{F} = \sum_{i=1}^{2} \frac{\hat{\mathbf{u}}_i}{h_i} \cdot \frac{\partial \mathbf{F}}{\partial v_i} + J \hat{\mathbf{u}}_3 \cdot \mathbf{F}, \tag{S1.39b}
$$

$$\nabla_S \times \mathbf{F} = \sum_{i=1}^{2} \frac{\hat{\mathbf{u}}_i}{h_i} \times \frac{\partial \mathbf{F}}{\partial v_i} + J\hat{\mathbf{u}}_3 \times \mathbf{F}, \tag{S1.39c}$$

where $J = \dfrac{-1}{h_1 h_2}\dfrac{\partial(h_1 h_2)}{\partial v_3}$ is the curvature of $\partial\Omega$. The first terms of summation in the right-hand side of Eqs. (S1.39a)-(S1.39c) are the expressions of $\nabla_{\|} f$, $\nabla_{\|} \cdot \mathbf{F}$, and $\nabla_{\|} \times \mathbf{F}$, respectively. If the second terms on the right-hand side of Eqs. (S1.39a)-(S1.39c) are removed, and the upper limit of the summation index $i$ is changed from 2 to 3, then Eqs. (S1.39a)-(S1.39c) become the expressions for the volume differential operations $\nabla f$, $\nabla \cdot \mathbf{F}$, and $\nabla \times \mathbf{F}$, respectively. It is worth pointing out that, similar to the volume differential operations, the results of the surface differential operations given by Eq. (S1.39) do not depend on the selection of the coordinate system [4].

**S2. Proof of the symmetry relations between the NWMs propagating along the positive and negative *z*-directions**

Step 1: Prove that $\mathbf{K}^{-,+}$, $\mathbf{K}^{-,-}$, $\mathbf{K}^{+,+}$, and $\mathbf{K}^{+,-}$ in Eq. (2) in the main text satisfy,

$$\mathbf{K}^{+,-} = \mathbf{K}^{-,+} \text{ (a)}, \ \mathbf{K}^{+,+} = \mathbf{K}^{-,-} \text{ (b)}. \tag{S2.1}$$

*Proof*: In the $z$=0 plane, the electromagnetic fields $\boldsymbol{\psi}_p^+(\boldsymbol{\rho}) = [\mathbf{E}_p^+, \mathbf{H}_p^+]^{\mathrm{T}}$ and $\boldsymbol{\psi}_p^-(\boldsymbol{\rho}) = [\mathbf{E}_p^-, \mathbf{H}_p^-]^{\mathrm{T}}$ of the $p$th-order CWMs propagating along the positive and negative $z$-directions satisfy the following symmetry relations [2],

$$\begin{aligned} &E_{p,x}^+ = E_{p,x}^- \text{ (a)}, \ E_{p,y}^+ = E_{p,y}^- \text{ (b)}, \ E_{p,z}^+ = -E_{p,z}^- \text{ (c)}, \\ &H_{p,x}^+ = -H_{p,x}^- \text{ (d)}, \ H_{p,y}^+ = -H_{p,y}^- \text{ (e)}, \ H_{p,z}^+ = H_{p,z}^- \text{ (f)}, \end{aligned} \tag{S2.2}$$

which mean that with respect to the $z$=0 plane, $\mathbf{E}_p^+$ and $\mathbf{E}_p^-$ satisfy vectorial mirror symmetry, while $\mathbf{H}_p^+$ and $\mathbf{H}_p^-$ satisfy vectorial mirror antisymmetry. From Eqs. (S2.2a), (S2.2b), (S2.2d), and (S2.2e), it follows that in the $z$=0 plane,

$$\begin{aligned} E_{p,\perp}^+ &= \hat{\mathbf{n}} \cdot \mathbf{E}_p^+ = (n_x\hat{\mathbf{x}} + n_y\hat{\mathbf{y}}) \cdot (E_{p,x}^+\hat{\mathbf{x}} + E_{p,y}^+\hat{\mathbf{y}} + E_{p,z}^+\hat{\mathbf{z}}) \\ &= n_x E_{p,x}^+ + n_y E_{p,y}^+ = n_x E_{p,x}^- + n_y E_{p,y}^- = E_{p,\perp}^-, \end{aligned} \tag{S2.3a}$$

$$\begin{aligned} E_{p,l}^+ &= \hat{\mathbf{l}} \cdot \mathbf{E}_p^+ = (l_x\hat{\mathbf{x}} + l_y\hat{\mathbf{y}}) \cdot (E_{p,x}^+\hat{\mathbf{x}} + E_{p,y}^+\hat{\mathbf{y}} + E_{p,z}^+\hat{\mathbf{z}}) \\ &= l_x E_{p,x}^+ + l_y E_{p,y}^+ = l_x E_{p,x}^- + l_y E_{p,y}^- = E_{p,l}^-, \end{aligned} \tag{S2.3b}$$

$$\begin{aligned} H_{p,\perp}^+ &= \hat{\mathbf{n}} \cdot \mathbf{H}_p^+ = (n_x\hat{\mathbf{x}} + n_y\hat{\mathbf{y}}) \cdot (H_{p,x}^+\hat{\mathbf{x}} + H_{p,y}^+\hat{\mathbf{y}} + H_{p,z}^+\hat{\mathbf{z}}) \\ &= n_x H_{p,x}^+ + n_y H_{p,y}^+ = -n_x H_{p,x}^- - n_y H_{p,y}^- = -H_{p,\perp}^-, \end{aligned} \tag{S2.3c}$$

$$\begin{aligned} H_{p,l}^+ &= \hat{\mathbf{l}} \cdot \mathbf{H}_p^+ = (l_x\hat{\mathbf{x}} + l_y\hat{\mathbf{y}}) \cdot (H_{p,x}^+\hat{\mathbf{x}} + H_{p,y}^+\hat{\mathbf{y}} + H_{p,z}^+\hat{\mathbf{z}}) \\ &= l_x H_{p,x}^+ + l_y H_{p,y}^+ = -l_x H_{p,x}^- - l_y H_{p,y}^- = -H_{p,l}^-, \end{aligned} \tag{S2.3d}$$

where $\hat{\mathbf{n}}$ is the unit normal vector on the metal-dielectric interface $\partial\Omega$, and $\hat{\mathbf{l}} = \hat{\mathbf{n}} \times \hat{\mathbf{z}}$, as illustrated in Fig. S1(a) in Sec. S1B. From Eqs. (S2.2c) and (S2.3b), it follows that in the $z$=0 plane,

$$\mathbf{E}_{p,\|}^- \cdot \mathbf{E}_{q,\|}^+ = (E_{p,l}^-\hat{\mathbf{l}} + E_{p,z}^-\hat{\mathbf{z}}) \cdot (E_{q,l}^+\hat{\mathbf{l}} + E_{q,z}^+\hat{\mathbf{z}}) = E_{p,l}^- E_{q,l}^+ + E_{p,z}^- E_{q,z}^+ = E_{p,l}^+ E_{q,l}^- + E_{p,z}^+ E_{q,z}^- = \mathbf{E}_{p,\|}^+ \cdot \mathbf{E}_{q,\|}^-, \tag{S2.4a}$$

$$\mathbf{E}_{p,\|}^- \cdot \mathbf{E}_{q,\|}^- = (E_{p,l}^-\hat{\mathbf{l}} + E_{p,z}^-\hat{\mathbf{z}}) \cdot (E_{q,l}^-\hat{\mathbf{l}} + E_{q,z}^-\hat{\mathbf{z}}) = E_{p,l}^- E_{q,l}^- + E_{p,z}^- E_{q,z}^- = E_{p,l}^+ E_{q,l}^+ + E_{p,z}^+ E_{q,z}^+ = \mathbf{E}_{p,\|}^+ \cdot \mathbf{E}_{q,\|}^+. \tag{S2.4b}$$

Substituting Eqs. (S2.3a), (S2.3d), and (S2.4) into Eq. (3) in the main text yields,

$$\kappa_{p,q}^{+,-} = \kappa_{p,q}^{-,+} \text{ (a)}, \ \kappa_{p,q}^{+,+} = \kappa_{p,q}^{-,-} \text{ (b)}. \tag{S2.5}$$

From Eq. (S2.5), Eq. (S2.1) is thus obtained.

*The proof is complete*.

Step 2: Prove that if Eq. (2) in the main text possesses an eigenvalue $\tilde{k}_n$ (i.e., the propagation constant of the NWM), with its eigenvector being,

$$\tilde{\mathbf{a}}_n = \begin{bmatrix} \tilde{\mathbf{a}}_n^+ \\ \tilde{\mathbf{a}}_n^- \end{bmatrix}, \tag{S2.6}$$

then Eq. (2) must possess another eigenvalue $-\tilde{k}_n$, denoted by,

$$\tilde{k}_{-n} = -\tilde{k}_n, \ n = 1, \ 2, \ \cdots, \tag{S2.7}$$

and its eigenvector (denoted by $\tilde{\mathbf{a}}_{-n} = [\tilde{\mathbf{a}}_{-n}^+; \tilde{\mathbf{a}}_{-n}^-]$) is,

$$\tilde{\mathbf{a}}_{-n} = \begin{bmatrix} \tilde{\mathbf{a}}_n^- \\ \tilde{\mathbf{a}}_n^+ \end{bmatrix}. \tag{S2.8}$$

Consequently, $\tilde{k}_n$ and $\tilde{k}_{-n} = -\tilde{k}_n$ ($n$=1, 2, …) are the propagation constants of the $n$th-order NWMs $\tilde{\boldsymbol{\psi}}_n$ and $\tilde{\boldsymbol{\psi}}_{-n}$ propagating along the positive and negative $z$-directions, respectively.

*Proof*: Substituting Eqs. (S2.1) and (S2.6) into Eq. (2) in the main text yields,

$$\begin{bmatrix} \mathbf{W} + i\mathbf{K}^{-,+} & i\mathbf{K}^{-,-} \\ -i\mathbf{K}^{-,-} & -\mathbf{W} - i\mathbf{K}^{-,+} \end{bmatrix} \begin{bmatrix} \tilde{\mathbf{a}}_n^+ \\ \tilde{\mathbf{a}}_n^- \end{bmatrix} = \tilde{k}_n \begin{bmatrix} \tilde{\mathbf{a}}_n^+ \\ \tilde{\mathbf{a}}_n^- \end{bmatrix}. \tag{S2.9}$$

From Eq. (S2.9), one obtains,

$$\left( \mathbf{Q} \begin{bmatrix} \mathbf{W} + i\mathbf{K}^{-,+} & i\mathbf{K}^{-,-} \\ -i\mathbf{K}^{-,-} & -\mathbf{W} - i\mathbf{K}^{-,+} \end{bmatrix} \mathbf{Q} \right) \left( \mathbf{Q} \begin{bmatrix} \tilde{\mathbf{a}}_n^+ \\ \tilde{\mathbf{a}}_n^- \end{bmatrix} \right) = \tilde{k}_n \mathbf{Q} \begin{bmatrix} \tilde{\mathbf{a}}_n^+ \\ \tilde{\mathbf{a}}_n^- \end{bmatrix}, \tag{S2.10}$$

where

$$\mathbf{Q} = \begin{bmatrix} \mathbf{0} & \mathbf{I} \\ \mathbf{I} & \mathbf{0} \end{bmatrix}, \tag{S2.11}$$

with $\mathbf{I}$ denoting the identity matrix. Equation (S2.10) is exactly,

$$\begin{bmatrix} -\mathbf{W} - i\mathbf{K}^{-,+} & -i\mathbf{K}^{-,-} \\ i\mathbf{K}^{-,-} & \mathbf{W} + i\mathbf{K}^{-,+} \end{bmatrix} \begin{bmatrix} \tilde{\mathbf{a}}_n^- \\ \tilde{\mathbf{a}}_n^+ \end{bmatrix} = \tilde{k}_n \begin{bmatrix} \tilde{\mathbf{a}}_n^- \\ \tilde{\mathbf{a}}_n^+ \end{bmatrix}. \tag{S2.12}$$

Multiplying both sides of Eq. (S2.12) by −1 yields,

$$\begin{bmatrix} \mathbf{W} + i\mathbf{K}^{-,+} & i\mathbf{K}^{-,-} \\ -i\mathbf{K}^{-,-} & -\mathbf{W} - i\mathbf{K}^{-,+} \end{bmatrix} \begin{bmatrix} \tilde{\mathbf{a}}_n^- \\ \tilde{\mathbf{a}}_n^+ \end{bmatrix} = -\tilde{k}_n \begin{bmatrix} \tilde{\mathbf{a}}_n^- \\ \tilde{\mathbf{a}}_n^+ \end{bmatrix}. \tag{S2.13}$$

By comparing Eq. (S2.13) with Eq. (S2.9), one can conclude that Eq. (2) in the main text must possess another eigenvalue $-\tilde{k}_n$, whose eigenvector is given by Eq. (S2.8).

*The proof is complete*.

Step 3: The symmetry relations (S2.2) satisfied by the electromagnetic fields $\boldsymbol{\psi}_p^+(\boldsymbol{\rho}) = [\mathbf{E}_p^+, \mathbf{H}_p^+]^{\mathrm{T}}$ and $\boldsymbol{\psi}_p^-(\boldsymbol{\rho}) = [\mathbf{E}_p^-, \mathbf{H}_p^-]^{\mathrm{T}}$ of the CWMs in the $z$=0 plane are denoted by,

$$\hat{M}\boldsymbol{\psi}_p^+(\boldsymbol{\rho}) = \boldsymbol{\psi}_p^-(\boldsymbol{\rho}), \tag{S2.14}$$

where $\hat{M}$ is referred to as the symmetry transformation operator. Below, we prove that in the $z$=0 plane, the electromagnetic fields of the $n$th-order NWMs propagating along the positive and negative $z$-directions, $\tilde{\boldsymbol{\psi}}_n(\boldsymbol{\rho}) = [\tilde{\mathbf{E}}_n, \tilde{\mathbf{H}}_n]^{\mathrm{T}}$ and $\tilde{\boldsymbol{\psi}}_{-n}(\boldsymbol{\rho}) = [\tilde{\mathbf{E}}_{-n}, \tilde{\mathbf{H}}_{-n}]^{\mathrm{T}}$, satisfy the same symmetry relations, that is,

$$\hat{M}\tilde{\boldsymbol{\psi}}_n(\boldsymbol{\rho}) = \tilde{\boldsymbol{\psi}}_{-n}(\boldsymbol{\rho}). \tag{S2.15}$$

*Proof*: Substituting Eqs. (S2.6) and (S2.8) into Eq. (S1.8) yields,

$$\tilde{\boldsymbol{\psi}}_n(\boldsymbol{\rho}) = \boldsymbol{\Phi}\tilde{\mathbf{a}}_n = \boldsymbol{\Phi}^+\tilde{\mathbf{a}}_n^+ + \boldsymbol{\Phi}^-\tilde{\mathbf{a}}_n^- = \sum_{p=1}^{P} c_{p,n}^+ \boldsymbol{\psi}_p^+(\boldsymbol{\rho}) + \sum_{p=1}^{P} c_{p,n}^- \boldsymbol{\psi}_p^-(\boldsymbol{\rho}), \tag{S2.16a}$$

$$\tilde{\boldsymbol{\psi}}_{-n}(\boldsymbol{\rho}) = \boldsymbol{\Phi}\tilde{\mathbf{a}}_{-n} = \boldsymbol{\Phi}^+\tilde{\mathbf{a}}_n^- + \boldsymbol{\Phi}^-\tilde{\mathbf{a}}_n^+ = \sum_{p=1}^{P} c_{p,n}^- \boldsymbol{\psi}_p^+(\boldsymbol{\rho}) + \sum_{p=1}^{P} c_{p,n}^+ \boldsymbol{\psi}_p^-(\boldsymbol{\rho}). \tag{S2.16b}$$

Then, one has,

$$\begin{aligned} \hat{M}\tilde{\boldsymbol{\psi}}_n(\boldsymbol{\rho}) &= \hat{M}[\sum_{p=1}^{P} c_{p,n}^+ \boldsymbol{\psi}_p^+(\boldsymbol{\rho}) + \sum_{p=1}^{P} c_{p,n}^- \boldsymbol{\psi}_p^-(\boldsymbol{\rho})] = \sum_{p=1}^{P} c_{p,n}^+ \hat{M}\boldsymbol{\psi}_p^+(\boldsymbol{\rho}) + \sum_{p=1}^{P} c_{p,n}^- \hat{M}\boldsymbol{\psi}_p^-(\boldsymbol{\rho}) \\ &= \sum_{p=1}^{P} c_{p,n}^+ \boldsymbol{\psi}_p^-(\boldsymbol{\rho}) + \sum_{p=1}^{P} c_{p,n}^- \boldsymbol{\psi}_p^+(\boldsymbol{\rho}) = \tilde{\boldsymbol{\psi}}_{-n}(\boldsymbol{\rho}), \end{aligned} \tag{S2.17}$$

where the first equality uses Eq. (S2.16a), the second equality uses the fact that $\hat{M}$ is a linear operator, the third equality uses Eq. (S2.14) and its corollary,

$$\hat{M}\boldsymbol{\psi}_p^-(\boldsymbol{\rho}) = \hat{M}\hat{M}\boldsymbol{\psi}_p^+(\boldsymbol{\rho}) = \boldsymbol{\psi}_p^+(\boldsymbol{\rho}), \tag{S2.18}$$

and the last equality in Eq. (S2.17) uses Eq. (S2.16b). The second equality in Eq. (S2.18) indicates that $\hat{M}\hat{M}=\hat{I}$ is the identity operator.

*The proof is complete*.

**S3. Derivation of Eq. (6) in the main text**

From Eqs. (3) and (4) in the main text, we have,

$$\kappa_{p,q}^{\sigma,\tau}=\kappa_{p,q}^{\sigma,\tau,(1)}\delta, \tag{S3.1}$$

where $\sigma,\tau\in\{+,-\}$, and we define,

$$\kappa_{p,q}^{\sigma,\tau,(1)}=\frac{1}{F_p}\oint_L[i\omega d_{\parallel}^{(1)}[\![\varepsilon]\!]\mathbf{E}_{p,\parallel}^{\sigma}\cdot\mathbf{E}_{q,\parallel}^{\tau}-i\omega d_{\perp}^{(1)}[\![\varepsilon E_{p,\perp}^{\sigma}E_{q,\perp}^{\tau}]\!]-i(\sigma k_p+\tau k_q)d_{\perp}^{(1)}H_{p,l}^{\sigma}[\![E_{q,\perp}^{\tau}]\!]]_{z=0}dl. \tag{S3.2}$$

Consequently, for Eq. (2) in the main text, we have,

$$\mathbf{K}:=\begin{bmatrix}\mathbf{K}^{-,+} & \mathbf{K}^{-,-}\\ \mathbf{K}^{+,+} & \mathbf{K}^{+,-}\end{bmatrix}=\begin{bmatrix}\mathbf{K}^{-,+,(1)}\delta & \mathbf{K}^{-,-,(1)}\delta\\ \mathbf{K}^{+,+,(1)}\delta & \mathbf{K}^{+,-,(1)}\delta\end{bmatrix}=\mathbf{K}^{(1)}\delta, \tag{S3.3}$$

where ":=" means that "the left quantity (which is unknown) is defined as the right quantity (which is known)", and we define,

$$\mathbf{K}^{(1)}=\begin{bmatrix}\mathbf{K}^{-,+,(1)} & \mathbf{K}^{-,-,(1)}\\ \mathbf{K}^{+,+,(1)} & \mathbf{K}^{+,-,(1)}\end{bmatrix}, \tag{S3.4}$$

where the element in the $p$th row and $q$th column of matrix $\mathbf{K}^{\sigma,\tau,(1)}$ is $\kappa_{p,q}^{\sigma,\tau,(1)}$. Equation (2) thus becomes,

$$(\mathbf{H}^{(0)}+\mathbf{H}^{(1)}\delta)\tilde{\mathbf{a}}_n=\tilde{k}_n\tilde{\mathbf{a}}_n, \tag{S3.5}$$

where the definitions are,

$$\mathbf{H}^{(0)}=\begin{bmatrix}\mathbf{W} & \mathbf{0}\\ \mathbf{0} & -\mathbf{W}\end{bmatrix},\ \mathbf{H}^{(1)}=i\begin{bmatrix}\mathbf{K}^{-,+,(1)} & \mathbf{K}^{-,-,(1)}\\ -\mathbf{K}^{+,+,(1)} & -\mathbf{K}^{+,-,(1)}\end{bmatrix}. \tag{S3.6}$$

The propagation constant $\tilde{k}_n$ of the NWM under the NEBC is a function of the $d$-parameters, that is,

$$\tilde{k}_n=\tilde{k}_n(d_\perp,d_\parallel)=\tilde{k}_n(d_\perp^{(1)}\delta,d_\parallel^{(1)}\delta), \tag{S3.7}$$

where the second equality uses Eq. (4) in the main text. Then the $\tilde{k}_n$ can be expressed as an asymptotic expansion in terms of $\delta$ [i.e., Eq. (5) in the main text],

$$\tilde{k}_n=k^{(0)}+k^{(1)}\delta+O(\delta^2), \tag{S3.8}$$

where

$$k^{(0)}=\tilde{k}_n(0,0)\ \text{(a)},\ k^{(1)}=\left.\frac{d\tilde{k}_n(d_\perp^{(1)}\delta,d_\parallel^{(1)}\delta)}{d\delta}\right|_{\delta=0}=\frac{\partial\tilde{k}_n(0,0)}{\partial d_\perp}d_\perp^{(1)}+\frac{\partial\tilde{k}_n(0,0)}{\partial d_\parallel}d_\parallel^{(1)}\ \text{(b)}. \tag{S3.9}$$

Similarly, the $\tilde{\mathbf{a}}_n$ in Eq. (S3.5) can be expressed as an asymptotic expansion in terms of $\delta$,

$$\tilde{\mathbf{a}}_n=\mathbf{a}^{(0)}+\mathbf{a}^{(1)}\delta+O(\delta^2). \tag{S3.10}$$

Next, we will apply the perturbation theory of eigenvalue problem (e.g., as applied in quantum mechanics [5]) to solve for the $k^{(0)}$, $k^{(1)}$ in Eq. (S3.8), and the $\mathbf{a}^{(0)}$, $\mathbf{a}^{(1)}$ in Eq. (S3.10).

First, we solve for $k^{(0)}$ and $\mathbf{a}^{(0)}$. To this end, substituting Eqs. (S3.8) and (S3.10) into Eq. (S3.5) yields,

$$\mathbf{H}^{(0)}\mathbf{a}^{(0)}+(\mathbf{H}^{(0)}\mathbf{a}^{(1)}+\mathbf{H}^{(1)}\mathbf{a}^{(0)})\delta+O(\delta^2)=k^{(0)}\mathbf{a}^{(0)}+(k^{(0)}\mathbf{a}^{(1)}+k^{(1)}\mathbf{a}^{(0)})\delta+O(\delta^2), \tag{S3.11}$$

which gives,

$$\mathbf{H}^{(0)}\mathbf{a}^{(0)}=k^{(0)}\mathbf{a}^{(0)}, \tag{S3.12a}$$

$$\mathbf{H}^{(0)}\mathbf{a}^{(1)}+\mathbf{H}^{(1)}\mathbf{a}^{(0)}=k^{(0)}\mathbf{a}^{(1)}+k^{(1)}\mathbf{a}^{(0)}. \tag{S3.12b}$$

The solution to the matrix eigenvalue problem (S3.12a) is,

$$k^{(0)}=k_p=:k_{+,p}^{(0)}\ \text{(a)},\ \mathbf{a}^{(0)}=[\boldsymbol{\delta}_p;\mathbf{0}]=:\mathbf{a}_{+,p}^{(0)}\ \text{(b)}, \tag{S3.13}$$

or,

$$k^{(0)}=-k_p=:k_{-,p}^{(0)}\ \text{(a)},\ \mathbf{a}^{(0)}=[\mathbf{0};\boldsymbol{\delta}_p]=:\mathbf{a}_{-,p}^{(0)}\ \text{(b)}, \tag{S3.14}$$

where "=:" means that "the right quantity (which is unknown) is defined as the left quantity (which is known)". The column vector $\boldsymbol{\delta}_p$ with $P$ elements is defined such that its $p$th element is 1 and the remaining elements are 0. Equations (S3.13) and (S3.14) can be collectively rewritten as,

$$k^{(0)} = k^{(0)}_{\sigma,p} \text{ (a)}, \ \mathbf{a}^{(0)} = \mathbf{a}^{(0)}_{\sigma,p} \text{ (b)}, \ \sigma \in \{+,-\}, \ p = 1,2,\cdots,P, \tag{S3.15}$$

where the eigenvectors satisfy orthogonality relations,

$$(\mathbf{a}^{(0)}_{\sigma,p})^{\mathrm{T}} \mathbf{a}^{(0)}_{\tau,q} = \delta_{\sigma,\tau}\delta_{p,q}, \tag{S3.16}$$

with the Kronecker delta defined to satisfy $\delta_{\alpha,\beta}$=0 for $\alpha\neq\beta$ and $\delta_{\alpha,\beta}$=1 for $\alpha=\beta$.

Second, we solve for $k^{(1)}$ and $\mathbf{a}^{(1)}$. The $\mathbf{a}^{(1)}$ can be expressed as an expansion upon the complete basis of $\mathbf{a}^{(0)}_{\tau,q}$,

$$\mathbf{a}^{(1)} = \sum_{\tau\in\{+,-\}} \sum_{q=1}^{P} c_{\tau,q} \mathbf{a}^{(0)}_{\tau,q}, \tag{S3.17}$$

where $c_{\tau,q}$ are the unknown expansion coefficients to be determined below. Substituting Eqs. (S3.15) and (S3.17) into Eq. (S3.12b) yields,

$$\sum_{\tau\in\{+,-\}} \sum_{q=1}^{P} c_{\tau,q} k^{(0)}_{\tau,q} \mathbf{a}^{(0)}_{\tau,q} + \mathbf{H}^{(1)} \mathbf{a}^{(0)}_{\sigma,p} = k^{(0)}_{\sigma,p} \sum_{\tau\in\{+,-\}} \sum_{q=1}^{P} c_{\tau,q} \mathbf{a}^{(0)}_{\tau,q} + k^{(1)} \mathbf{a}^{(0)}_{\sigma,p}, \tag{S3.18}$$

where the left-hand side uses Eq. (S3.12a), that is,

$$\mathbf{H}^{(0)} \mathbf{a}^{(0)}_{\tau,q} = k^{(0)}_{\tau,q} \mathbf{a}^{(0)}_{\tau,q}. \tag{S3.19}$$

By left-multiplying both sides of Eq. (S3.18) by $(\mathbf{a}^{(0)}_{\sigma',p'})^{\mathrm{T}}$, and using the orthogonality relations in Eq. (S3.16), we obtain,

$$c_{\sigma',p'} k^{(0)}_{\sigma',p'} + (\mathbf{a}^{(0)}_{\sigma',p'})^{\mathrm{T}} \mathbf{H}^{(1)} \mathbf{a}^{(0)}_{\sigma,p} = c_{\sigma',p'} k^{(0)}_{\sigma,p} + k^{(1)} \delta_{\sigma',\sigma} \delta_{p',p}, \tag{S3.20}$$

where

$$(\mathbf{a}^{(0)}_{\sigma',p'})^{\mathrm{T}} \mathbf{H}^{(1)} \mathbf{a}^{(0)}_{\sigma,p} = \begin{cases} i\kappa^{-,+,(1)}_{p',p}, \text{ for } (\sigma',\sigma) = (+,+), \\ i\kappa^{-,-,(1)}_{p',p}, \text{ for } (\sigma',\sigma) = (+,-), \\ -i\kappa^{+,+,(1)}_{p',p}, \text{ for } (\sigma',\sigma) = (-,+), \\ -i\kappa^{+,-,(1)}_{p',p}, \text{ for } (\sigma',\sigma) = (-,-). \end{cases} \tag{S3.21}$$

Taking $\sigma'=\sigma$, $p'=p$ in Eq. (S3.20) yields,

$$k^{(1)} = (\mathbf{a}^{(0)}_{\sigma,p})^{\mathrm{T}} \mathbf{H}^{(1)} \mathbf{a}^{(0)}_{\sigma,p} = \begin{cases} i\kappa^{-,+,(1)}_{p,p}, \text{ for } \sigma = +, \\ -i\kappa^{+,-,(1)}_{p,p}, \text{ for } \sigma = -. \end{cases} \tag{S3.22}$$

Taking $(\sigma',p')\neq(\sigma,p)$ (i.e., $\sigma'\neq\sigma$ or $p'\neq p$) in Eq. (S3.20) yields,

$$c_{\sigma',p'} = \frac{(\mathbf{a}^{(0)}_{\sigma',p'})^{\mathrm{T}} \mathbf{H}^{(1)} \mathbf{a}^{(0)}_{\sigma,p}}{k^{(0)}_{\sigma,p} - k^{(0)}_{\sigma',p'}}, \tag{S3.23}$$

which gives the value of $c_{\tau,q}$ in Eq. (S3.17) for $(\tau,q)\neq(\sigma,p)$. To determine the value of $c_{\tau,q}$ in Eq. (S3.17) for $(\tau,q)=(\sigma,p)$, we impose the normalization condition on $\tilde{\mathbf{a}}_n$,

$$\tilde{\mathbf{a}}_n^{\mathrm{T}} \tilde{\mathbf{a}}_n = N_n, \tag{S3.24}$$

where $N_n$ is a constant independent of $\delta$. Substituting Eq. (S3.10) into Eq. (S3.24) yields,

$$(\mathbf{a}^{(0)})^{\mathrm{T}} \mathbf{a}^{(0)} + 2(\mathbf{a}^{(0)})^{\mathrm{T}} \mathbf{a}^{(1)} \delta + O(\delta^2) = N_n, \tag{S3.25}$$

which gives,

$$(\mathbf{a}^{(0)})^{\mathrm{T}} \mathbf{a}^{(0)} = N_n \text{ (a)}, \ (\mathbf{a}^{(0)})^{\mathrm{T}} \mathbf{a}^{(1)} = 0 \text{ (b)}. \tag{S3.26}$$

Substituting Eq. (S3.15b) into Eq. (S3.26a) yields,

$$N_n = 1, \tag{S3.27}$$

where Eq. (S3.16) is used. Substituting Eqs. (S3.15b) and (S3.17) into Eq. (S3.26b) yields,

$$c_{\sigma,p} = 0, \tag{S3.28}$$

where Eq. (S3.16) is used. Equation (S3.28) provides the value of $c_{\tau,q}$ in Eq. (S3.17) for $(\tau,q)=(\sigma,p)$. Substituting Eqs. (S3.23) and (S3.28) into Eq. (S3.17) yields,

$$\mathbf{a}^{(1)} = \left[ \sum_{\tau\in\{+,-\}} \sum_{q=1}^{P} \frac{(\mathbf{a}^{(0)}_{\tau,q})^{\mathrm{T}} \mathbf{H}^{(1)} \mathbf{a}^{(0)}_{\sigma,p}}{k^{(0)}_{\sigma,p} - k^{(0)}_{\tau,q}} \mathbf{a}^{(0)}_{\tau,q} \right]_{(\sigma,p)}, \tag{S3.29}$$

where $[\ ]_{(\sigma,p)}$ means a summation over indices $(\tau,q)\neq(\sigma,p)$ (i.e., $\tau\neq\sigma$ or $q\neq p$).

In conclusion, from Eqs. (S3.15a) and (S3.22), we obtain,

$$k^{(0)} = k_p \text{ (a)}, \; k^{(1)} = i\kappa_{p,p}^{-,+,(1)} \text{ (b)}, \tag{S3.30}$$

or,

$$k^{(0)} = -k_p \text{ (a)}, \; k^{(1)} = -i\kappa_{p,p}^{+,-,(1)} \text{ (b)}. \tag{S3.31}$$

Substituting Eqs. (S3.30) and (S3.31) into Eq. (S3.8) (taking $n$=$p$, $-p$), and using Eq. (S3.1) along with $\kappa_{p,p}^{+,-} = \kappa_{p,p}^{-,+}$ [see Eq. (S2.5a)], we obtain Eq. (6) in the main text.

Corresponding to Eqs. (S3.30) and (S3.31), respectively, we have,

$$\mathbf{a}^{(0)} = [\boldsymbol{\delta}_p;\mathbf{0}] \text{ (a)}, \; \mathbf{a}^{(1)} = \left[\sum_{q=1}^{P} \frac{i\kappa_{q,p}^{-,+,(1)}}{k_p - k_q}[\boldsymbol{\delta}_q;\mathbf{0}]\right]_p + \sum_{q=1}^{P} \frac{-i\kappa_{q,p}^{+,+,(1)}}{k_p + k_q}[\mathbf{0};\boldsymbol{\delta}_q] \text{ (b)}, \tag{S3.32}$$

or,

$$\mathbf{a}^{(0)} = [\mathbf{0};\boldsymbol{\delta}_p] \text{ (a)}, \; \mathbf{a}^{(1)} = \sum_{q=1}^{P} \frac{i\kappa_{q,p}^{-,-,(1)}}{-k_p - k_q}[\boldsymbol{\delta}_q;\mathbf{0}] + \left[\sum_{q=1}^{P} \frac{-i\kappa_{q,p}^{+,-,(1)}}{-k_p + k_q}[\mathbf{0};\boldsymbol{\delta}_q]\right]_p \text{ (b)}, \tag{S3.33}$$

where Eqs. (S3.15b) and (S3.29) are used, and $[\;]_p$ means a summation over the index $q\neq p$. Substituting Eqs. (S3.32) and (S3.33) into Eq. (S3.10) (taking $n$=$p$, $-p$) yields,

$$\tilde{\mathbf{a}}_p = [\boldsymbol{\delta}_p;\mathbf{0}] + \left[\sum_{q=1}^{P} \frac{i\kappa_{q,p}^{-,+}}{k_p - k_q}[\boldsymbol{\delta}_q;\mathbf{0}]\right]_p + \sum_{q=1}^{P} \frac{-i\kappa_{q,p}^{+,+}}{k_p + k_q}[\mathbf{0};\boldsymbol{\delta}_q] + O(\delta^2), \tag{S3.34}$$

or,

$$\tilde{\mathbf{a}}_{-p} = [\mathbf{0};\boldsymbol{\delta}_p] + \sum_{q=1}^{P} \frac{i\kappa_{q,p}^{-,-}}{-k_p - k_q}[\boldsymbol{\delta}_q;\mathbf{0}] + \left[\sum_{q=1}^{P} \frac{-i\kappa_{q,p}^{+,-}}{-k_p + k_q}[\mathbf{0};\boldsymbol{\delta}_q]\right]_p + O(\delta^2), \tag{S3.35}$$

where Eq. (S3.1) is used.

Equation (6) in the main text indicates that the calculation of the first-order asymptotic expansion of $\tilde{k}_{\pm p}$ only requires the full-wave numerical calculation of the $p$th-order CWM (including the electromagnetic fields $\boldsymbol{\psi}_p^{\pm} = [\mathbf{E}_p^{\pm}, \mathbf{H}_p^{\pm}]^{\mathrm{T}}$ and the propagation constant $k_p$). In contrast, Eqs. (S3.34) and (S3.35) show that the calculation of the first-order asymptotic expansion of $\tilde{\mathbf{a}}_{\pm p}$ necessitates the full-wave numerical calculation of a complete set of CWMs (including the electromagnetic fields $\boldsymbol{\psi}_q^{\pm}$ and propagation constants $k_q$, $q$=1, 2, …, $P$). Compared with the former, the latter incurs a significantly higher computational cost and provides less physical insight.

### S4. Derivation of Eq. (7) in the main text

In general, since $|d_{\parallel}| \ll |d_{\perp}|$ [1,6-8], and the dominant component of the CWM electric field $\mathbf{E}_p^{\pm}$ is the normal component $E_{p,\perp}^{\pm}$ on the dielectric side, the first term of the integrand on the right-hand side of Eq. (6) in the main text (with subscript $n$=$p$) can be neglected. The electric fields of CWMs propagating along the positive and negative $z$-directions satisfy $E_{p,\perp}^{-} = E_{p,\perp}^{+}$ [see Eq. (S2.3a)]. Assuming a lossless dielectric and approximately neglecting the metal loss, we have $\varepsilon\approx\mathrm{Re}(\varepsilon)$, which renders $E_{p,\perp}^{\pm}$ and $F_p$ approximately real-valued, yielding $F_p\approx|F_p|\approx 4\Phi_p$, where $\Phi_p$ denotes the energy flux of the CWM. Consequently, Eq. (7) can be approximately obtained from Eq. (6) in the main text. It should be noted that since the imaginary and real parts of $d_{\perp}$ are of the same order of magnitude [6,9], the imaginary part of $d_{\perp}$ is retained in Eq. (7).

### S5. Additional results for the numerical examples of Au-nanowire MPWs

#### A. Apodization functions of Feibelman *d*-parameters for the Au-nanowire MPWs

First, an apodization function is defined as,

$$f(x;x_a,x_b) = \begin{cases} 1, \text{ for } |x| \le x_a, \\ c_0 + c_1|x| + c_2|x|^2 + c_3|x|^3, \text{ for } x_a < |x| \le x_b, \\ 0, \text{ for } |x| > x_b, \end{cases} \tag{S5.1}$$

where $0<x_a<x_b$, $[x_a, x_b]$ is the apodization interval, and $c_j$ ($j$=0, 1, 2, 3) are the coefficients to be determined. It is required that $f(x;x_a,x_b)$ and its derivatives are continuous at $x$=$x_a$, $x_b$, that is,

$$f(x_a^+;x_a,x_b)=f(x_a^-;x_a,x_b)=1, \tag{S5.2a}$$

$$f(x_b^-;x_a,x_b)=f(x_b^+;x_a,x_b)=0, \tag{S5.2b}$$

$$\left.\frac{df(x;x_a,x_b)}{dx}\right|_{x=x_a^+}=\left.\frac{df(x;x_a,x_b)}{dx}\right|_{x=x_a^-}=0, \tag{S5.2c}$$

$$\left.\frac{df(x;x_a,x_b)}{dx}\right|_{x=x_b^-}=\left.\frac{df(x;x_a,x_b)}{dx}\right|_{x=x_b^+}=0. \tag{S5.2d}$$

Substituting Eq. (S5.1) into Eq. (S5.2) yields,

$$\begin{aligned}&c_0+c_1x_a+c_2x_a^2+c_3x_a^3=1,\\&c_0+c_1x_b+c_2x_b^2+c_3x_b^3=0,\\&c_1+2c_2x_a+3c_3x_a^2=0,\\&c_1+2c_2x_b+3c_3x_b^2=0.\end{aligned} \tag{S5.3}$$

Equation (S5.3) constitutes a system of linear equations for $c_j$ ($j$=0, 1, 2, 3), whose solution determines the values of $c_j$. Substituting $c_j$ into Eq. (S5.1) yields the apodization function $f(x;x_a,x_b)$.

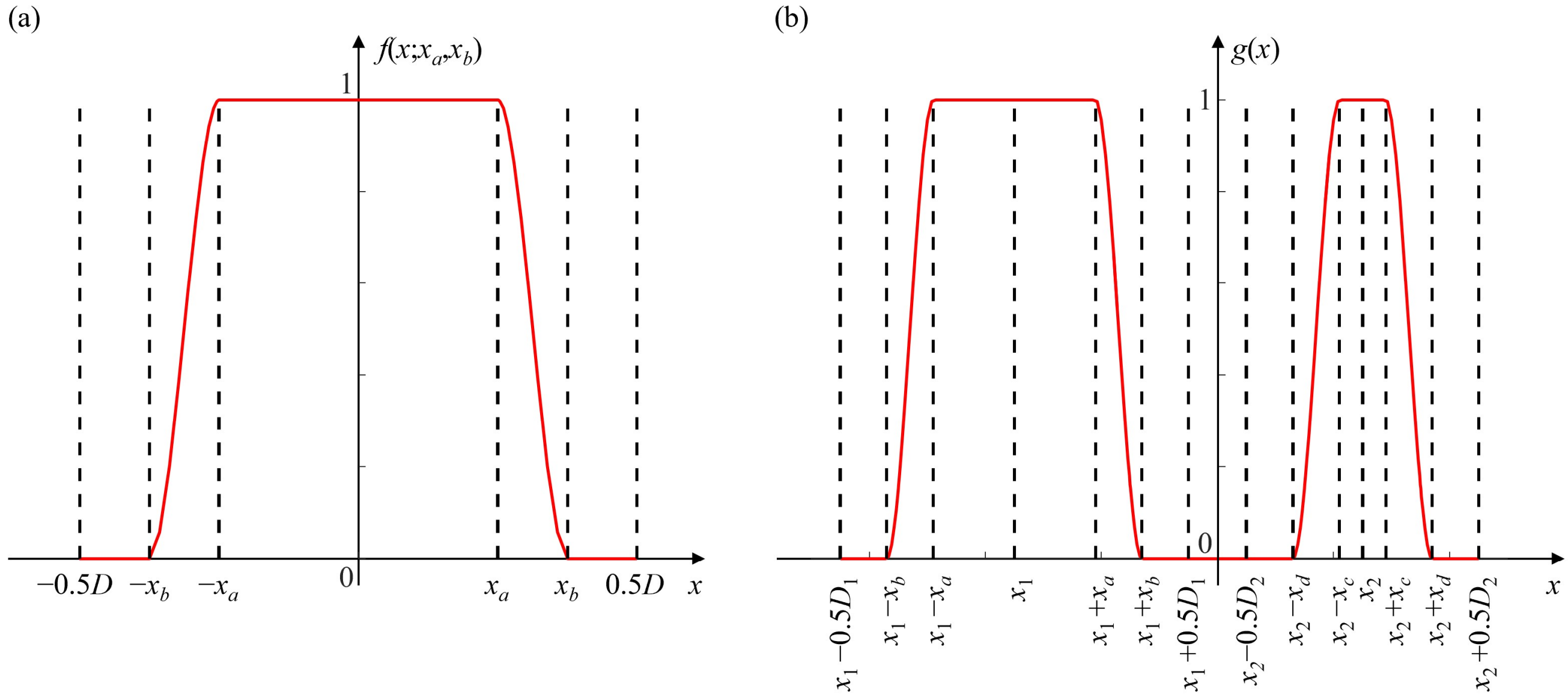

FIG. S2 (a) For the single-nanowire MPW, profile of the apodization function $f(x;x_a,x_b)$ of the $d$-parameters on the bottom surface of the nanowire as a function of the $x$ coordinate. (b) For the double-nanowire MPW, profile of the apodization function $g(x)=f(x-x_1;x_a,x_b)+f(x-x_2;x_c,x_d)$ of the $d$-parameters on the bottom surface of the nanowires as a function of the $x$ coordinate.

Next, for the single-nanowire MPW, the Feibelman $d$-parameters on the bottom surface of the Au nanowire are set to $d_{\parallel,\perp}=(d_{\parallel,\perp}^{\text{Au-PMMA}}\delta)f(x;x_a,x_b)$, where $x_a$=0.5$D$−2$w$, $x_b$=0.5$D$−$w$, $w$=1 nm (width of the apodization and zero intervals), and $D$=40 nm (side length of the nanowire cross section). The values of $d_{\parallel,\perp}^{\text{Au-PMMA}}$ are provided in the *Verification* Section of the main text, and $\delta$ is the scaling factor of the $d$-parameters. The profile of the apodization function $f(x;x_a,x_b)$ as a function of $x$ is shown in Fig. S2(a) ($w$ is magnified for a clear display of details).

For the double-nanowire MPW, the Feibelman $d$-parameters on the bottom surface of the Au nanowires are set to $d_{\parallel,\perp}=(d_{\parallel,\perp}^{\text{Au-PMMA}}\delta)[f(x-x_1;x_a,x_b)+f(x-x_2;x_c,x_d)]$, where $x_1$=−0.5$d$−0.5$D_1$ (coordinate of the center of the left nanowire), $x_a$=0.5$D_1$−2$w$, $x_b$=0.5$D_1$−$w$, $x_2$=0.5$d$+0.5$D_2$ (coordinate of the center of the right nanowire), $x_c$=0.5$D_2$−2$w$, $x_d$=0.5$D_2$−$w$, $w$=1 nm (width of the apodization and zero intervals), $D_1$=40 nm (side length of the cross section of the left nanowire), and $D_2$=60 nm (side length of the cross section of the right nanowire). The profile of the apodization function $g(x)=f(x-x_1;x_a,x_b)+f(x-x_2;x_c,x_d)$ as a function of $x$ is shown in Fig. S2(b) ($w$ is magnified for a clear display of details).

**B. Additional results for the single-nanowire MPW**

For the single bound and propagative mode supported by the single-nanowire MPW, Fig. S3 presents the electromagnetic field distributions of the CWM (row 1) and NWM (row 2) in the cross section of $z$=0, revealing a close resemblance between the two. These results are obtained via full-wave Fourier modal method (FMM) numerical calculations, with a polymethyl methacrylate (PMMA) nanogap thickness of $g$=5 nm. Figure S4 shows the propagation constant $k_1$ of the CWM as a function of $g$. Consistent with the main text, a wavelength of $\lambda$=1 μm is adopted in the calculations of this section.

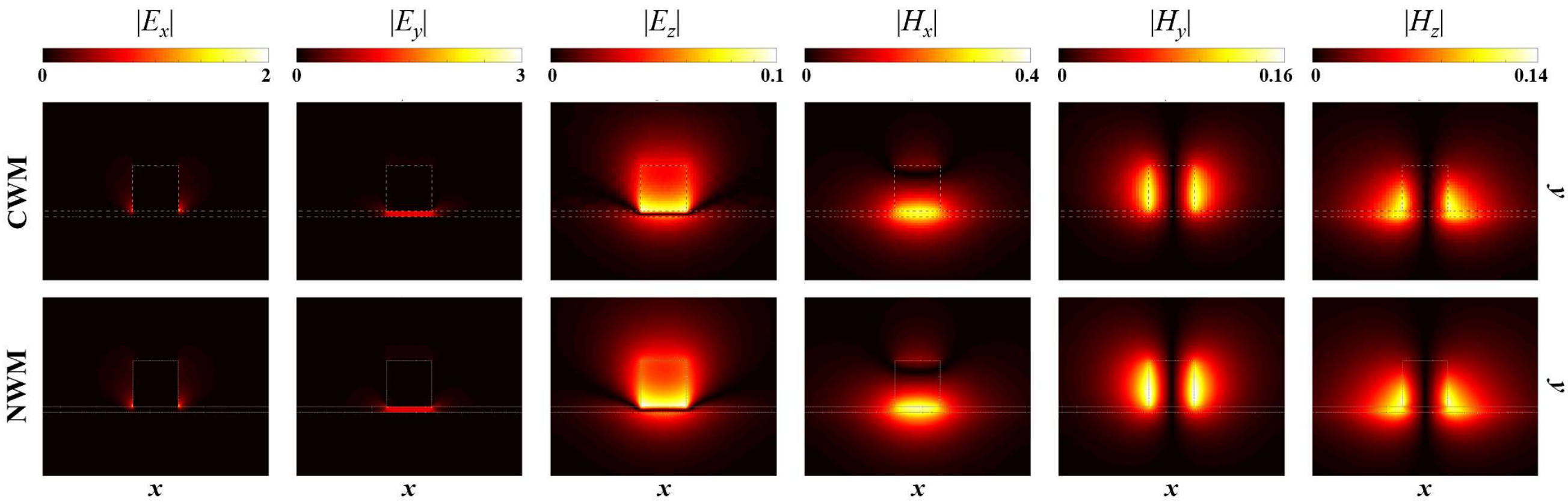

FIG. S3 For the single bound and propagative mode supported by the single-nanowire MPW, electromagnetic field distributions of the CWM (row 1) and NWM (row 2) in the cross section of $z$=0. The normalization is chosen such that $E_y$=1 at ($x$,$y$)=(0,$g$/2). The magnetic field **H** (in SI units) has been multiplied by the vacuum wave impedance $\eta_0$. The coordinate origin $O$ is set on the surface of the Au substrate and aligned with the center of the nanowire. The superimposed lines show the boundaries of the refractive index discontinuity. These results are obtained by full-wave FMM numerical calculations with a PMMA nanogap thickness of $g$=5 nm.

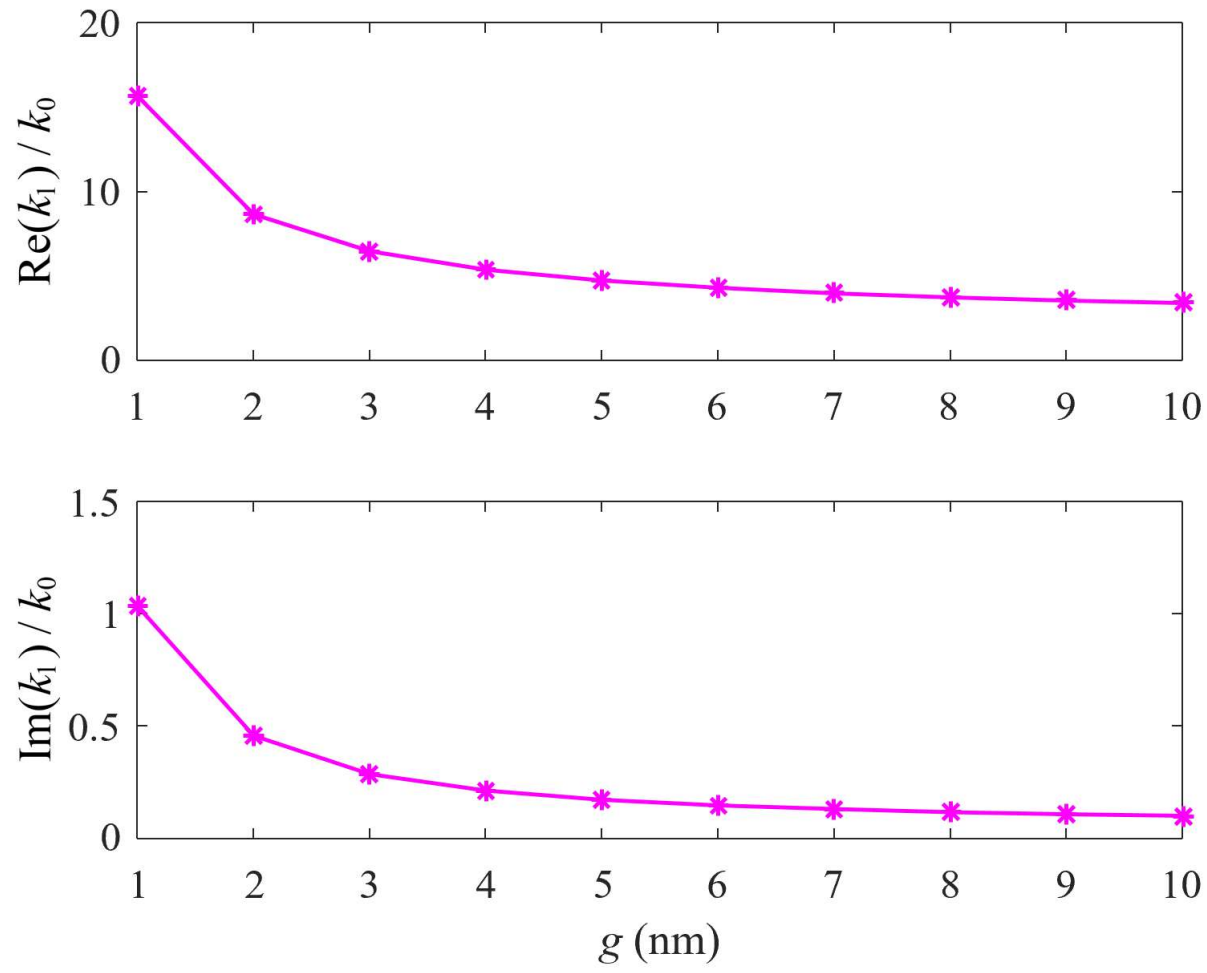

FIG. S4 For the CWM supported by the single-nanowire MPW, variation of the propagation constant $k_1$ with the PMMA nanogap thickness $g$. The normalized quantity $k_1/k_0$, i.e., the complex effective index, is plotted. The results are obtained via full-wave FMM numerical calculations.

In Fig. 2(b) of the main text, the predictions of the perturbation theory with $d_\parallel = 0$ (purple pluses) are close to the predictions with $d_\perp d_\parallel \neq 0$ (orange dashed lines with pluses), indicating that the contribution of $d_\perp$ is dominant. To further verify the contributions of $d_\perp$ and $d_\parallel$ in the perturbation theory, Fig. S5 displays the curves of $\mathrm{Re}(k_1 - \tilde{k}_1)$ and $\mathrm{Im}(\tilde{k}_1 - k_1)$ as functions of the PMMA nanogap thickness $g$ when $d_\parallel$ is artificially magnified by a factor of 10 (i.e., $d_\perp = d_\perp^{\text{Au-PMMA}}$, $d_\parallel = 10 d_\parallel^{\text{Au-PMMA}}$). Figure S5 shows that for all the values of $g$, the predictions of the perturbation theory (orange dashed lines with pluses) are consistent with the results of tangent lines to the full-wave FMM numerical results at $\delta$=0 [i.e., the first two terms on the right-hand side of Eq. (5) in the main text calculated using FMM, black dashed lines], which further verifies the validity of the perturbation theory. In addition, the predictions of the perturbation theory with $d_\parallel = 0$ (purple pluses) differ significantly from those with $d_\perp d_\parallel \neq 0$, indicating that both $d_\perp$ and $d_\parallel$ contribute significantly. Figure S5(b) shows that for all the values of $g$, there is $\mathrm{Re}(k_1 - \tilde{k}_1) > 0$ [which is consistent with Fig. 2(b) in the main text]; and for most values of $g$, there is $\mathrm{Im}(\tilde{k}_1 - k_1) < 0$ when $d_\perp d_\parallel \neq 0$ [which is different from Fig. 2(b) in the main text].

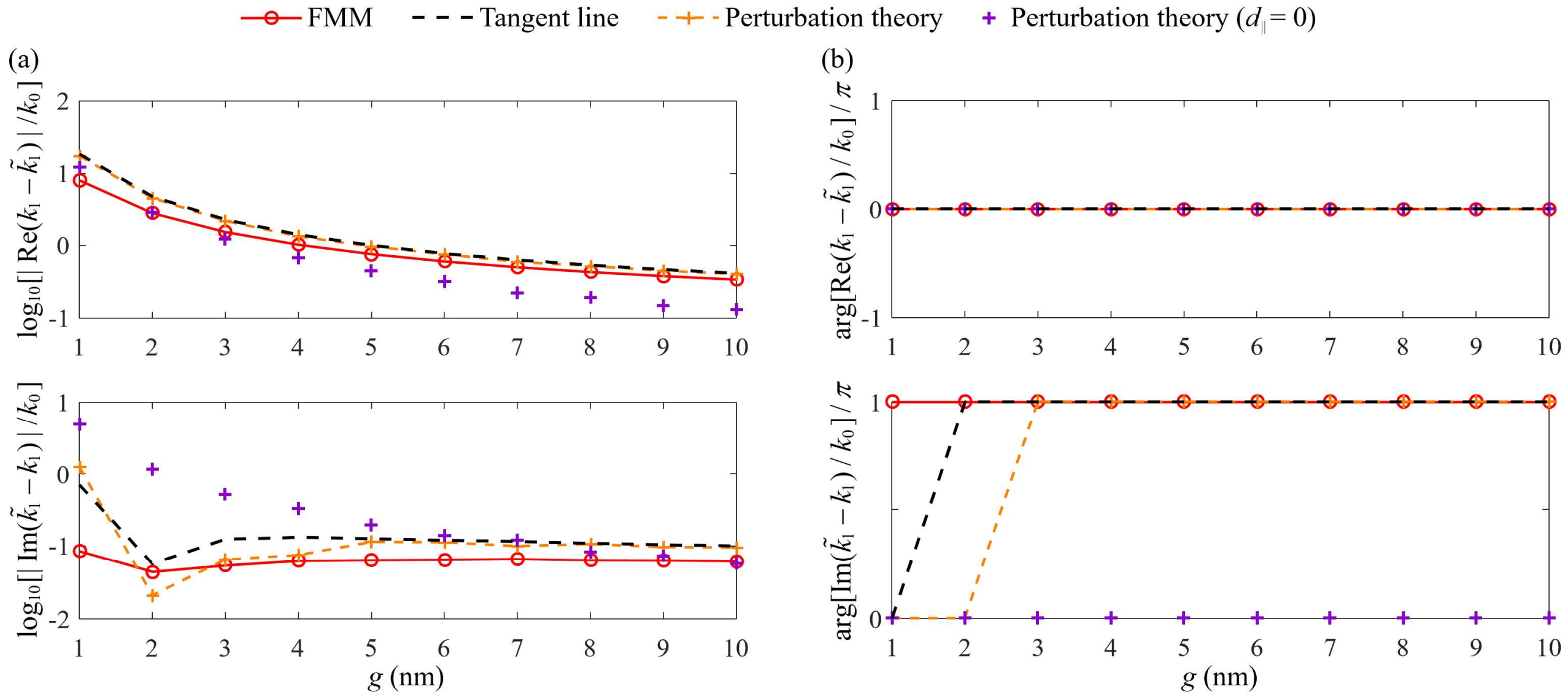

FIG. S5 Propagation constant $\tilde{k}_1$ of the NWM supported by the single-nanowire MPW as a function of the PMMA nanogap thickness $g$, with $d_\perp = d_\perp^{\text{Au-PMMA}}$, $d_\parallel = 10 d_\parallel^{\text{Au-PMMA}}$. In (a), the ordinates are $\log_{10}[|\mathrm{Re}(k_1 - \tilde{k}_1)|/k_0]$ and $\log_{10}[|\mathrm{Im}(\tilde{k}_1 - k_1)|/k_0]$. In (b), the ordinates are $\arg[\mathrm{Re}(k_1 - \tilde{k}_1)/k_0]/\pi$ and $\arg[\mathrm{Im}(\tilde{k}_1 - k_1)/k_0]/\pi$. In (a) and (b), the red solid lines with circles represent the full-wave FMM numerical results; the black dashed lines represent the results of tangent lines to the FMM results at $\delta$=0; the orange dashed lines with pluses show the predictions of the perturbation theory; and the purple pluses show the predictions of the perturbation theory considering only the contribution of $d_\perp$ (i.e., $d_\parallel = 0$).

Figure 2(b) in the main text shows that as the PMMA nanogap thickness $g$ decreases, which leads to enhanced nonclassical effects in the nanogap, the discrepancy between the predictions of the perturbation theory and the full-wave FMM results increases. This indicates that, in addition to the first-order contributions of the $d$-parameters considered in the perturbation theory, it is necessary to account for their higher-order contributions at this point [i.e., the $O(\delta^2)$ terms on the right-hand side of Eq. (6) in the main text]. To this end, we explore whether the expansion theory [i.e., Eq. (2) in the main text, which is capable of incorporating the higher-order contributions of the $d$-parameters] can be employed to improve the accuracy of the perturbation theory.

In the expansion theory, considering all the bound and propagative CWMs supported by the single-nanowire MPW (including one forward- and one backward-propagating CWMs), Eq. (2) becomes,

$$\begin{bmatrix} k_1 + i\kappa_{1,1}^{-,+} & i\kappa_{1,1}^{-,-} \\ -i\kappa_{1,1}^{-,-} & -k_1 - i\kappa_{1,1}^{-,+} \end{bmatrix} \tilde{\mathbf{a}}_n = \tilde{k}_n \tilde{\mathbf{a}}_n, \tag{S5.4}$$

where Eq. (S2.1) is used. Solving the matrix eigenvalue problem (S5.4) yields the eigenvalues,

$$\tilde{k}_{\pm 1} = \pm\sqrt{(k_1 + i\kappa_{1,1}^{-,+})^2 + (\kappa_{1,1}^{-,-})^2}, \tag{S5.5}$$

and the corresponding eigenvectors,

$$\tilde{\mathbf{a}}_1 = \begin{bmatrix} 1 \\ \dfrac{-i\kappa_{1,1}^{-,-}}{\tilde{k}_1 + k_1 + i\kappa_{1,1}^{-,+}} \end{bmatrix}, \ \tilde{\mathbf{a}}_{-1} = \begin{bmatrix} \dfrac{-i\kappa_{1,1}^{-,-}}{\tilde{k}_1 + k_1 + i\kappa_{1,1}^{-,+}} \\ 1 \end{bmatrix}. \tag{S5.6}$$

It can be seen that Eqs. (S5.5) and (S5.6) are consistent with the general conclusions of the expansion theory [see Eqs. (S2.6)-(S2.8)]. The above $\tilde{k}_{\pm 1}$ and $\tilde{\mathbf{a}}_{\pm 1}$ give the propagation constants and electromagnetic field distributions of the forward- (+) and backward-propagating (−) NWMs, respectively.

In Fig. S6(a), with the PMMA nanogap thickness $g$=5 nm and $d_\perp = d_\perp^{\text{Au-PMMA}}\delta$, $d_\parallel = d_\parallel^{\text{Au-PMMA}}\delta$, the propagation constant $\tilde{k}_1$ of the forward-propagating NWM is plotted as a function of the $d$-parameters scaling factor $\delta$. When $\delta$=0, there is $\tilde{k}_1 = k_1$, which is exactly the propagation constant of the CWM. In Fig. S6(b), with $\delta$=1, $\tilde{k}_1$ is plotted as a function of $g$, where both $\text{Re}(k_1 - \tilde{k}_1)$ and $\text{Im}(\tilde{k}_1 - k_1)$ are greater than 0. Figure S6 presents the results of tangent lines to the full-wave FMM numerical results at $\delta$=0 (black dashed lines) and the results of tangent lines to the results of expansion theory at $\delta$=0 (black squares), which represent the first two terms on the right-hand side of Eq. (5) in the main text calculated by the FMM and the expansion theory, respectively. The consistency between these two results, as well as their agreement with the results of perturbation theory [see Eq. (6) and Fig. 2 in the main text], verifies the correctness of the implementation of the expansion theory. However, Fig. S6 shows that compared with the perturbation theory, the expansion theory (orange dashed lines with pluses) does not significantly improve the accuracy in predicting the full-wave FMM numerical results (red solid lines with circles), especially when $\delta$ is large or $g$ is small. Therefore, to ensure the accuracy of the expansion theory so as to fully account for the higher-order contributions of the $d$-parameters, it is necessary to incorporate not only the bound and propagative CWMs but also a large number of unbound or non-propagative (i.e., radiative or evanescent) CWMs into the expansion theory, so that the considered CWMs can form a complete set of basis functions [2]. However, this comes at the cost of a drastic increase in computational effort and a degradation of physical insight.

In addition, Fig. S7 shows that the propagation constants of the forward- and backward-propagating NWMs predicted by the expansion theory, $\tilde{k}_1$ (orange dashed lines) and $\tilde{k}_{-1}$ (orange diamonds), satisfy the symmetry relation $\tilde{k}_{-1} = -\tilde{k}_1$ [see Eq. (S5.5)]. This conclusion is consistent with the general conclusion of the expansion theory [see Eq. (S2.7)] and also consistent with the full-wave FMM numerical results (red solid lines and red crosses).

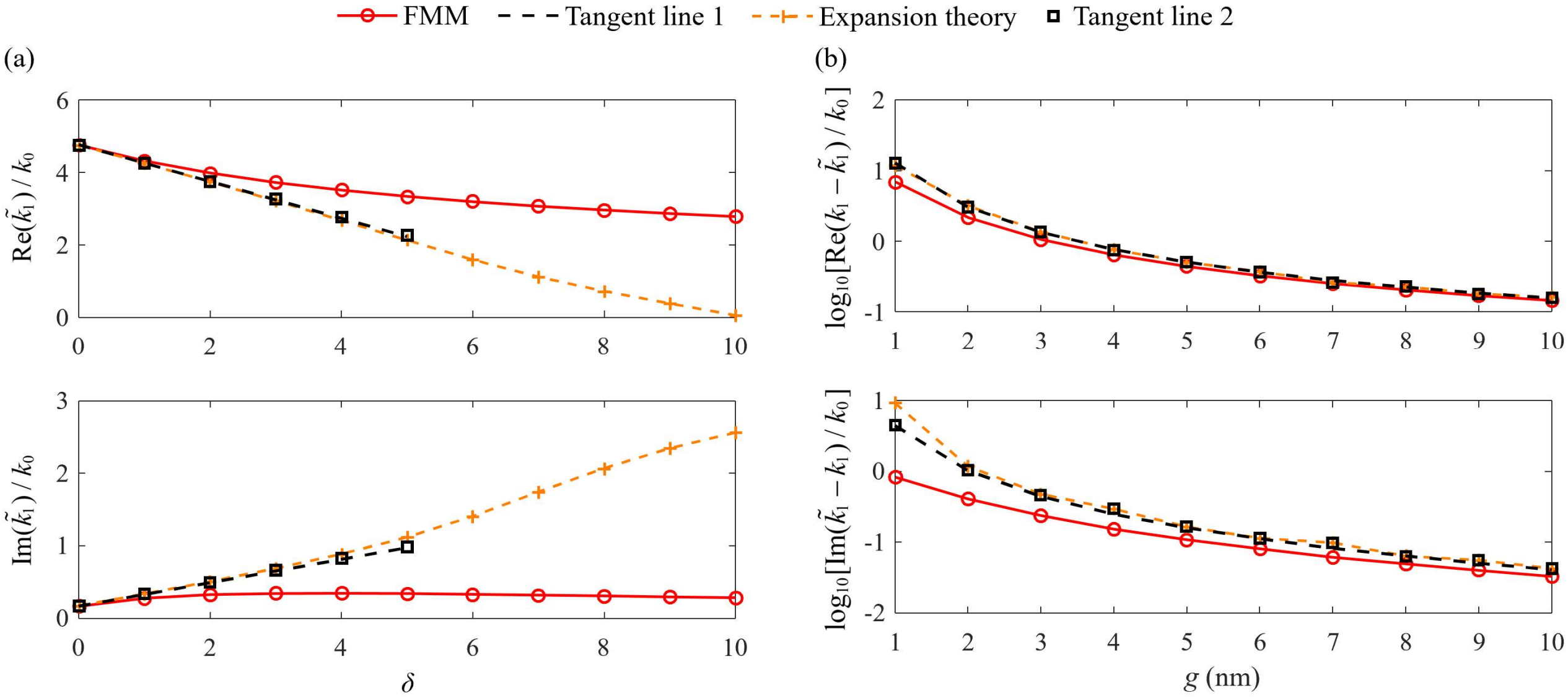

FIG. S6 (a) Propagation constant $\tilde{k}_1$ of the NWM supported by the single-nanowire MPW as a function of the scaling factor $\delta$ of the Feibelman $d$-parameters, with the PMMA nanogap thickness $g$=5 nm. The shown normalized quantity $\tilde{k}_1/k_0$ represents the complex effective index. (b) $\tilde{k}_1$ as a function of $g$ for $\delta$=1. In (a) and (b), the red solid lines with circles represent the full-wave FMM numerical results, and the orange dashed lines with pluses represent the predictions of the expansion theory. The black dashed lines and black squares show the results of tangent lines to the results of the FMM and the expansion theory at $\delta$=0, respectively.

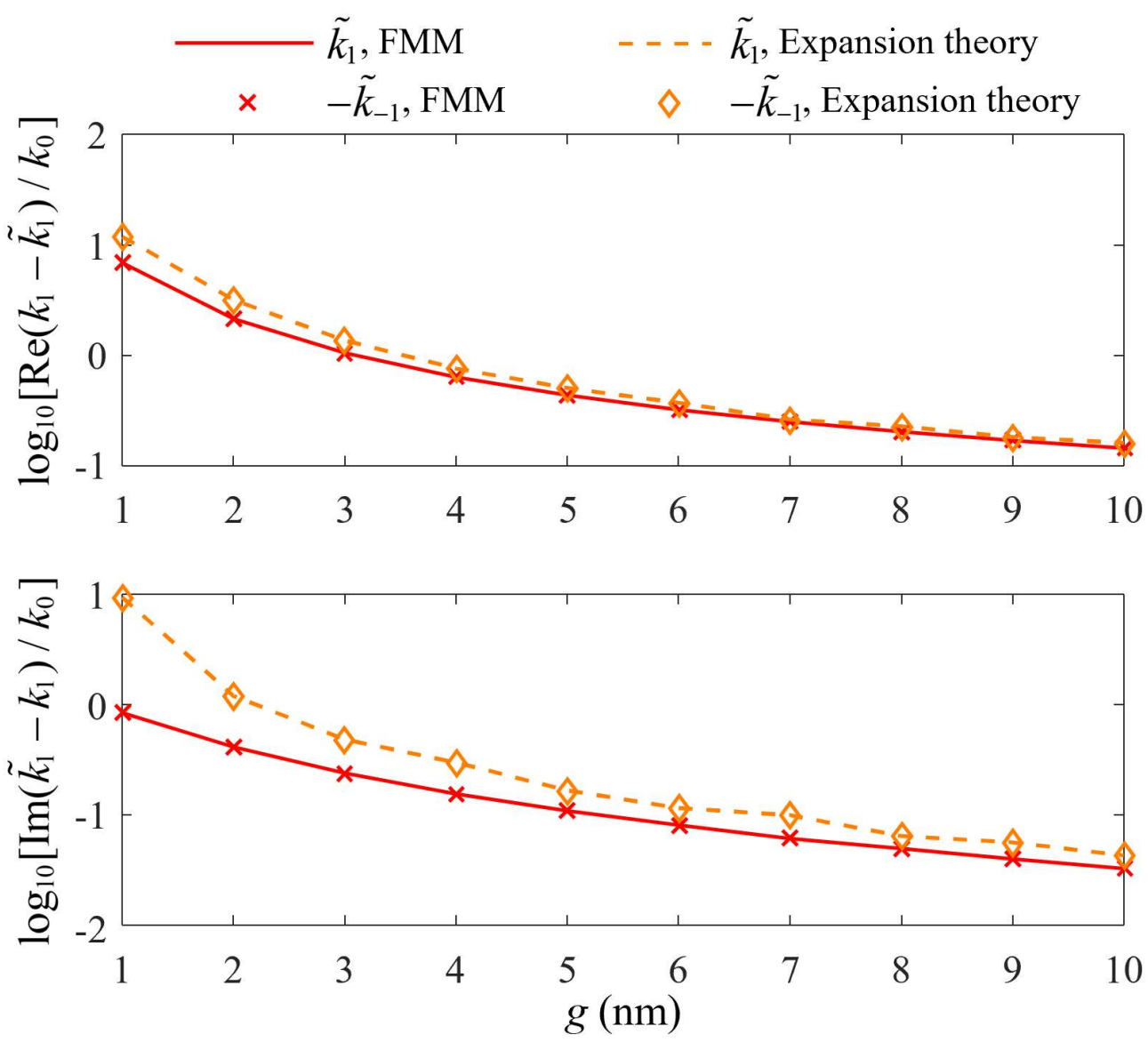

FIG. S7 Propagation constants $\tilde{k}_1$ and $\tilde{k}_{-1}$ of the forward- and backward-propagating NWMs supported by the single-nanowire MPW as functions of the PMMA nanogap thickness $g$. Here, $\mathrm{Re}(k_1 - \tilde{k}_1)$, $\mathrm{Im}(\tilde{k}_1 - k_1)$, $\mathrm{Re}[k_1 - (-\tilde{k}_{-1})]$, and $\mathrm{Im}[(-\tilde{k}_{-1}) - k_1]$ are all greater than 0. The scaling factor of Feibelman $d$-parameters is set to $\delta$=1, i.e., $d_\perp = d_\perp^{\text{Au-PMMA}}$ and $d_\parallel = d_\parallel^{\text{Au-PMMA}}$. The red solid lines and red crosses represent $\tilde{k}_1$ and $-\tilde{k}_{-1}$ calculated by the full-wave FMM, respectively, while the orange dashed lines and orange diamonds represent the corresponding predictions of the expansion theory.

## C. Results for the double-nanowire MPW

As shown in Fig. S8, the double-nanowire MPW consists of a gold substrate, a PMMA nanogap, two gold nanowires, and the air environment. The left and right nanowires have square cross sections with side lengths $D_1$=60 nm and $D_2$=40 nm, respectively, and the distance between them is $d$=10 nm. The refractive indices of the materials are set to be the same as those of the single-nanowire MPW (see the *Verification* Section of the main text). The coordinate origin $O$ is set on the surface of the Au substrate and is equidistant from the facing sidewalls of the two nanowires. Due to the same considerations as in the numerical example of the single-nanowire MPW, the NEBC is applied only at the Au-PMMA interface (with $d_\perp = d_\perp^{\text{Au-PMMA}}\delta$ and $d_\parallel = d_\parallel^{\text{Au-PMMA}}\delta$), and an apodization function is applied to the $d$-parameters on the bottom surfaces of the nanowires (see Sec. S5A for details). Consistent with the main text, a wavelength of $\lambda$=1 μm is adopted in the calculations of this section.

This double-nanowire MPW supports two bound and propagative modes, whose nonclassical and classical propagation constants are denoted by $\tilde{k}_p$ and $k_p$, respectively ($p$=1, 2 corresponding to the different modes). Figure S9 displays the electromagnetic field distributions of the CWM (row 1) and NWM (row 2) in the cross section of $z$=0, revealing a close resemblance between the two. These results are obtained via full-wave FMM numerical calculations with a PMMA nanogap thickness of $g$=5 nm. Figure S10 plots the dependence of the propagation constant $k_p$ of the CWM on $g$.

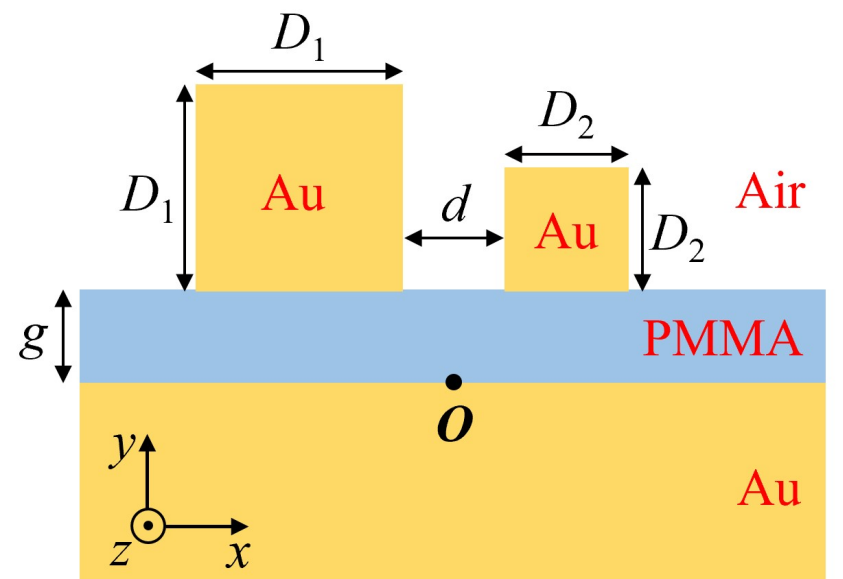

FIG. S8 Schematic of the double-nanowire MPW.

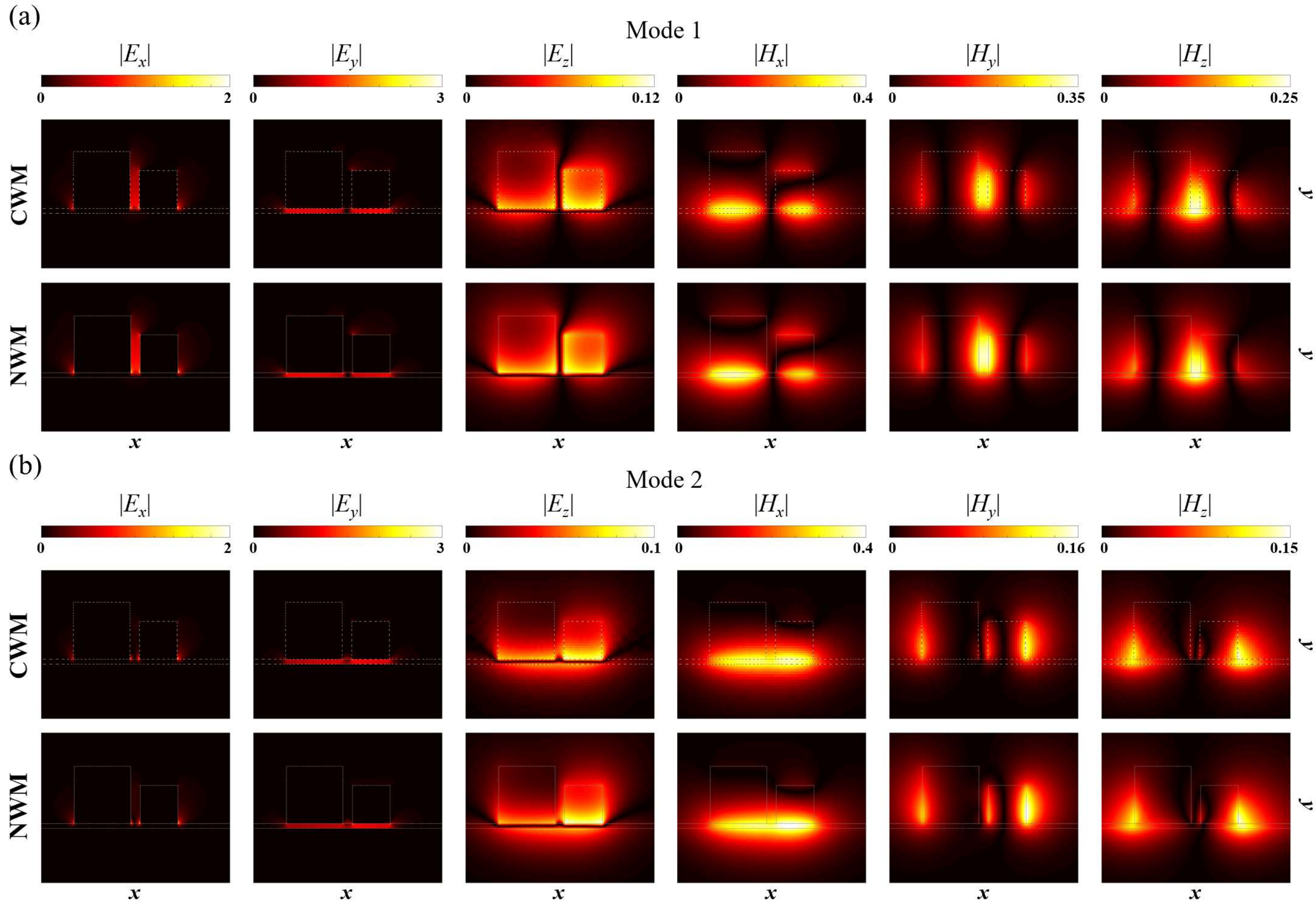

FIG. S9 Electromagnetic field distributions of the CWM (row 1) and NWM (row 2) in the cross section of $z$=0 for the two bound and propagative modes [referred to as modes 1 and 2, corresponding to (a) and (b), respectively] supported by the double-nanowire MPW. The normalizations of modes 1 and 2 satisfy $E_y$=1 at $(x,y)=(-d/2-D_1/2,g/2)$ and $(d/2+D_2/2,g/2)$, respectively. The magnetic field **H** (in SI units) has been multiplied by the vacuum wave impedance $\eta_0$. The coordinate origin $O$ is set on the surface of the Au substrate and is equidistant from the facing sidewalls of the two nanowires. The superimposed lines indicate the boundaries of the refractive index discontinuity. These results are obtained via full-wave FMM numerical calculations with a PMMA nanogap thickness of $g$=5 nm.

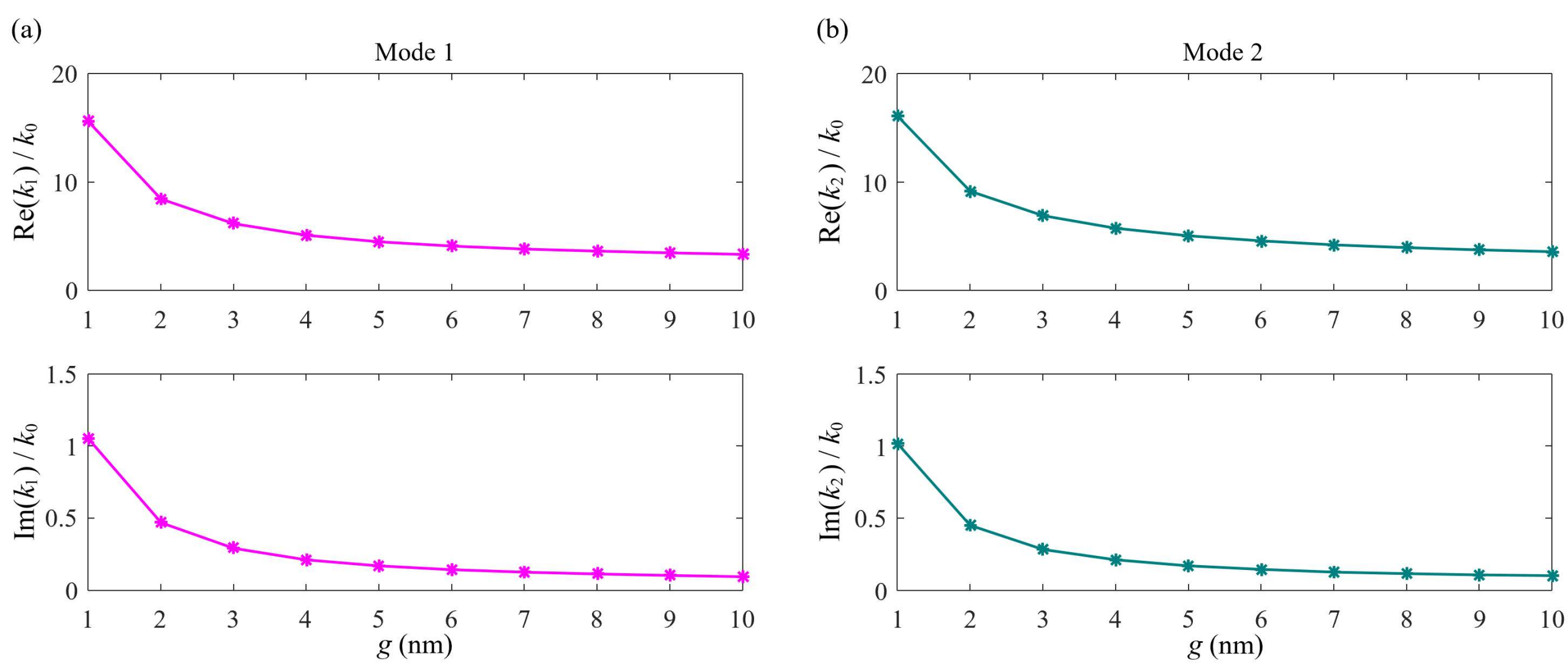

FIG. S10 Propagation constants $k_p$ [$p$=1, 2, corresponding to (a) and (b), respectively] of the two CWMs supported by the double-nanowire MPW as functions of the PMMA nanogap thickness $g$. The shown normalized quantity $k_p/k_0$ represents the complex effective index. These results are obtained via full-wave FMM numerical calculations.

For the double-nanowire MPW, the numerical results for its supported NWMs are presented below, which further verify the validity of the perturbation theory and exhibit conclusions analogous to those of the single-nanowire MPW, as detailed in the following.

In Fig. S11, with $d_{\perp} = d_{\perp}^{\text{Au-PMMA}}\delta$ and $d_{\parallel} = d_{\parallel}^{\text{Au-PMMA}}\delta$ [i.e., $d_{\perp}^{(1)} = d_{\perp}^{\text{Au-PMMA}}$ and $d_{\parallel}^{(1)} = d_{\parallel}^{\text{Au-PMMA}}$ in Eq. (4) in the main text] and $g$=5 nm, the propagation constants $\tilde{k}_p$ ($p$=1, 2) of the two NWMs are plotted as functions of the scaling factor $\delta$ of $d$-parameters. The normalized quantity $\tilde{k}_1/k_0$ represents the complex effective index. When $\delta$=0, there is $\tilde{k}_p = k_p$, which is the propagation constant of the CWM. Figure S11 shows that the predictions of the perturbation theory (orange dashed lines with pluses) are consistent with the tangent lines to the full-wave FMM numerical results at $\delta$=0 [i.e., the first two terms on the right-hand side of Eq. (5) in the main text calculated by the FMM, black dashed lines], confirming the validity of the perturbation theory. For $\delta$=1 (corresponding to the actual physical $d$-parameters), the predictions of the perturbation theory are quite close to the full-wave FMM numerical results (red solid lines with circles). The results show that $\text{Re}(\tilde{k}_p) < \text{Re}(k_p)$ and $\text{Im}(\tilde{k}_p) > \text{Im}(k_p)$, which are consistent with the theoretical predictions of Eq. (7) in the main text.

In Fig. S12, with $\delta$=1, $\text{Re}(k_p - \tilde{k}_p)$ and $\text{Im}(\tilde{k}_p - k_p)$ (both are positive according to the results) are plotted as functions of the PMMA nanogap thickness $g$. The results show that for all the values of $g$, the predictions of the perturbation theory (orange dashed lines with pluses) agree with the results of tangent lines to the full-wave FMM numerical results at $\delta$=0 (black dashed lines), further verifying the validity of the perturbation theory. Additionally, the predictions of the perturbation theory are quite close to the full-wave FMM numerical results (red solid lines with circles). Figure S12 reveals that the discrepancy between $\tilde{k}_p$ and $k_p$ gradually increases as $g$ decreases, which is consistent with the theoretical predictions of Eq. (7) in the main text (blue squares). This behavior originates from the enhanced confinement of the CWM electric field at the Au-PMMA interface within the nanogap.

Figure S12 also presents the predictions of the perturbation theory with $d_{\parallel} = 0$ (purple pluses), which are close to the predictions with $d_{\perp}d_{\parallel} \neq 0$, indicating that the contribution of $d_{\perp}$ is dominant. To further validate the contributions of $d_{\perp}$ and $d_{\parallel}$ in the perturbation theory, Fig. S13 depicts the variations of $\text{Re}(k_p - \tilde{k}_p)$ and $\text{Im}(\tilde{k}_p - k_p)$ ($p$=1, 2) as functions of the PMMA nanogap thickness $g$ when $d_{\parallel}$ is artificially magnified by a factor of 10 (i.e., $d_{\perp} = d_{\perp}^{\text{Au-PMMA}}$ and $d_{\parallel} = 10d_{\parallel}^{\text{Au-PMMA}}$). The results show that for all the values of $g$, the predictions of the perturbation theory (orange dashed lines with pluses) agree with the results of tangent lines to the full-wave FMM numerical results at $\delta$=0 (black dashed lines), which further confirms the validity of the perturbation theory. In addition, the predictions of the perturbation theory with $d_{\parallel} = 0$ (purple pluses) differ significantly from those with $d_{\perp}d_{\parallel} \neq 0$, indicating that both $d_{\perp}$ and $d_{\parallel}$ make significant contributions. Figures S13(a2) and (b2) reveal that for all the values of $g$, there is $\text{Re}(k_1 - \tilde{k}_1) > 0$ (which is consistent with Fig. S12); and for most values of $g$, there is $\text{Im}(\tilde{k}_1 - k_1) < 0$ when $d_{\perp}d_{\parallel} \neq 0$ (which differs from Fig. S12).

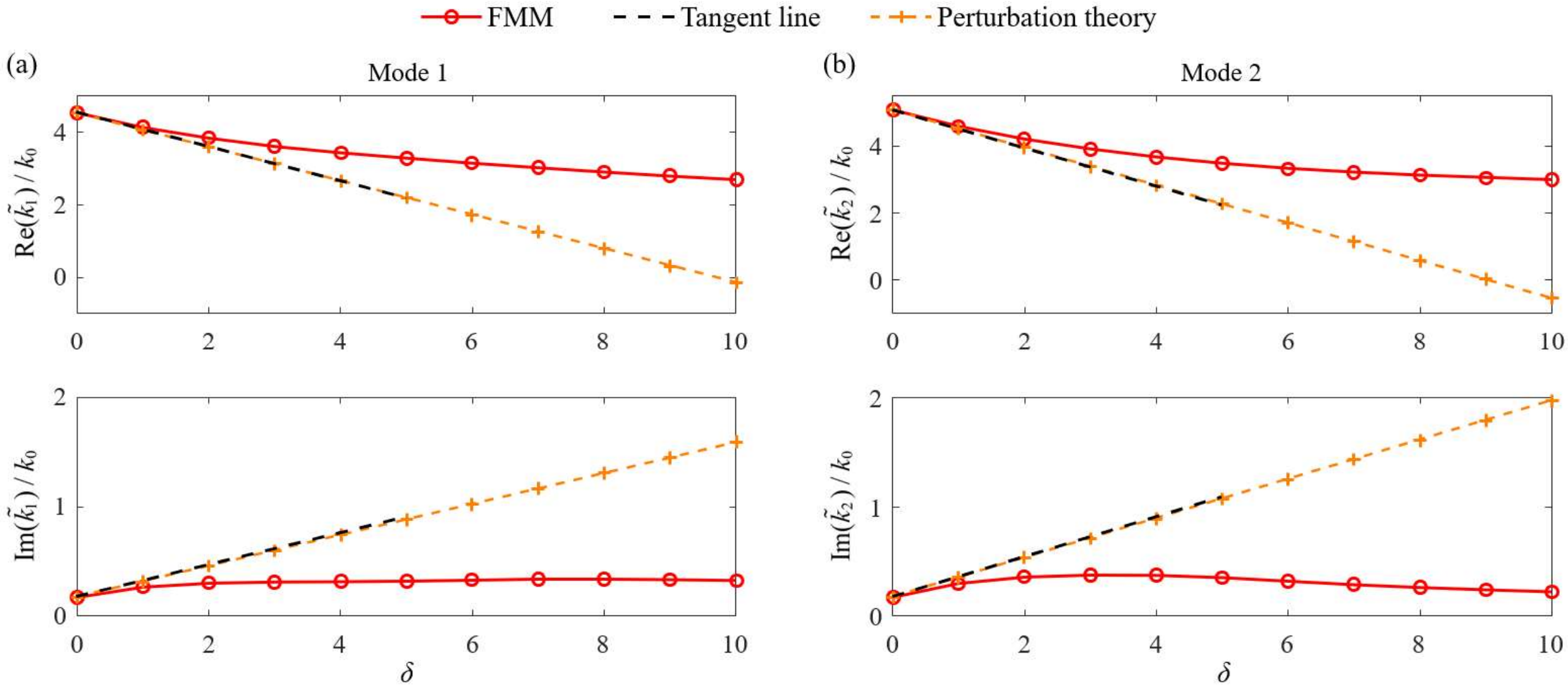

FIG. S11 Propagation constants $\tilde{k}_p$ [$p$=1, 2, corresponding to (a) and (b), respectively] of the two NWMs supported by the double-nanowire MPW as functions of the scaling factor $\delta$ of the Feibelman $d$-parameters, with the PMMA nanogap thickness $g$=5 nm. The shown normalized quantity $\tilde{k}_p / k_0$ represents the complex effective index. The red solid lines with circles show the full-wave FMM numerical results, the black dashed lines are the tangent lines to the FMM results at $\delta$=0, and the orange dashed lines with pluses represent the predictions of the perturbation theory.

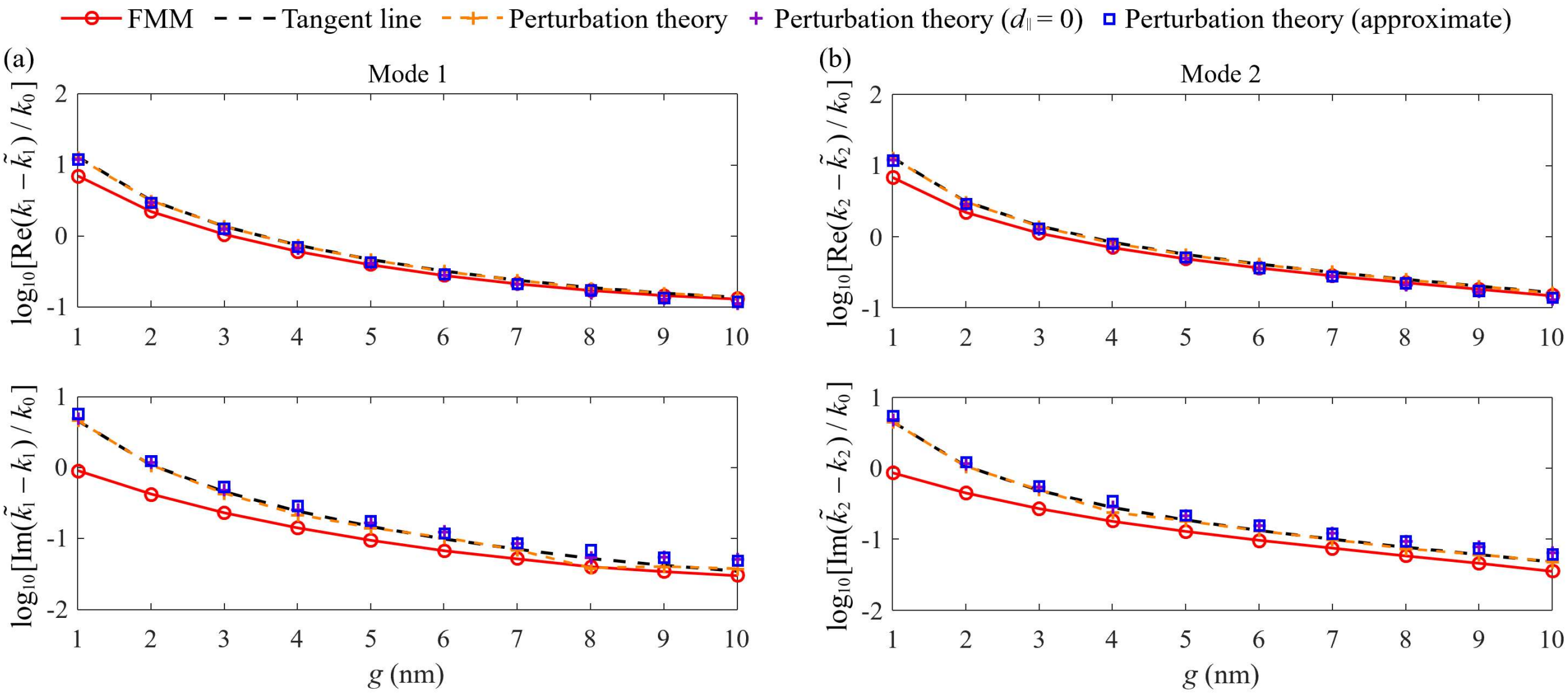

FIG. S12 Propagation constants $\tilde{k}_p$ [$p$=1, 2, corresponding to (a) and (b), respectively] of the two NWMs supported by the double-nanowire MPW as functions of the PMMA nanogap thickness $g$, with the $d$-parameters scaling factor $\delta$=1, i.e., $d_\perp = d_\perp^{\text{Au-PMMA}}$, $d_\parallel = d_\parallel^{\text{Au-PMMA}}$. The red solid lines with circles represent the full-wave FMM numerical results; the black dashed lines reperesent the results of tangent lines to the FMM results at $\delta$=0; the orange dashed lines with pluses show the predictions of the perturbation theory; the purple pluses show the predictions of the perturbation theory considering only the contribution of $d_\perp$ (i.e., $d_\parallel = 0$); and the blue squares show the results obtained from Eq. (7) in the main text.

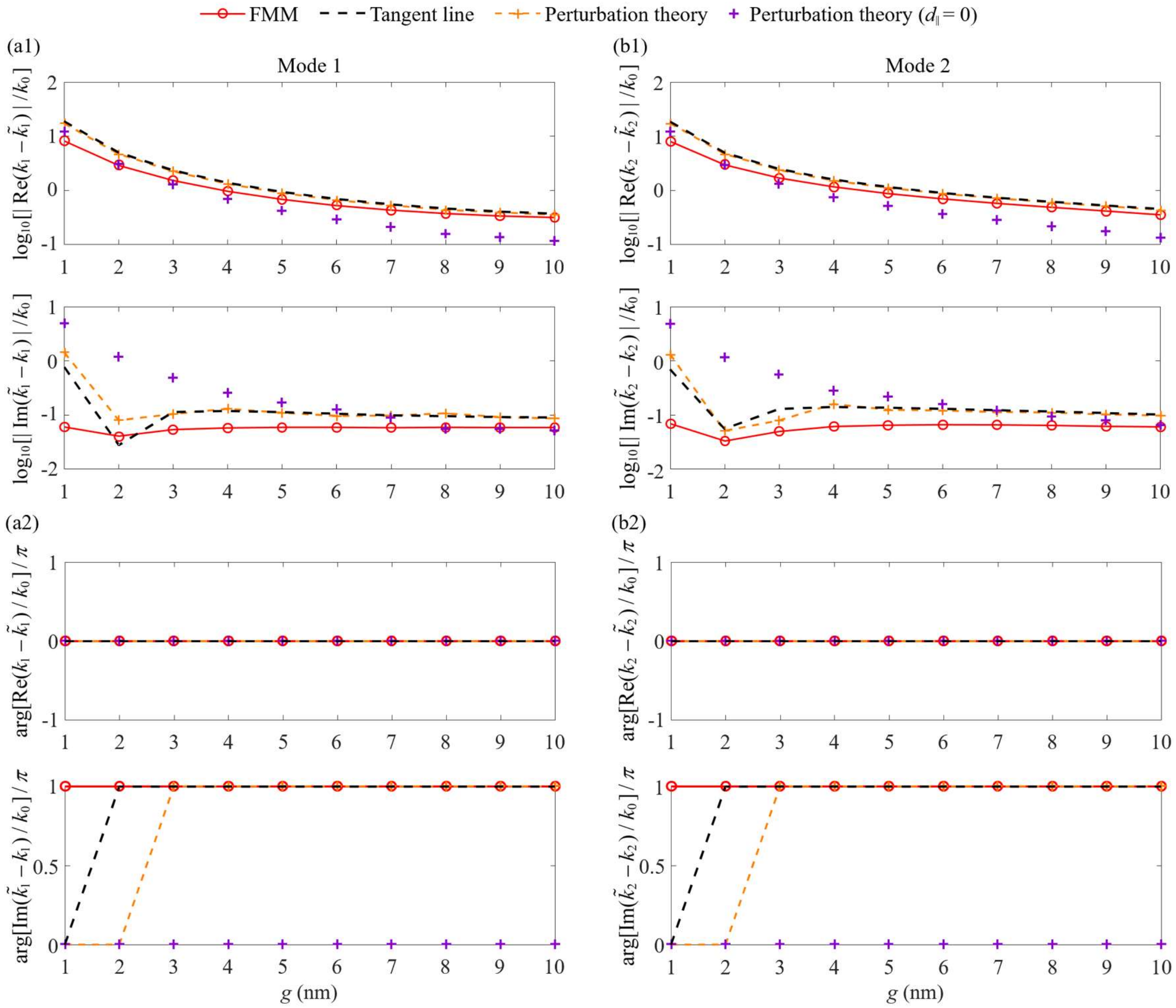

FIG. S13 Propagation constants $\tilde{k}_p$ [$p$=1, 2, corresponding to (a) and (b), respectively] of the two NWMs supported by the double-nanowire MPW as functions of the PMMA nanogap thickness $g$, with $d_\perp = d_\perp^{\text{Au-PMMA}}$, $d_\parallel = 10 d_\parallel^{\text{Au-PMMA}}$. In (a1) and (b1), the ordinates are $\log_{10}[|\mathrm{Re}(k_p-\tilde{k}_p)|/k_0]$ and $\log_{10}[|\mathrm{Im}(\tilde{k}_p-k_p)|/k_0]$. In (a2) and (b2), the ordinates are $\arg[\mathrm{Re}(k_p-\tilde{k}_p)/k_0]/\pi$ and $\arg[\mathrm{Im}(\tilde{k}_p-k_p)/k_0]/\pi$. In (a) and (b), the red solid lines with circles represent the full-wave FMM numerical results; the black dashed lines represent the results of tangent lines to the FMM results at $\delta$=0; the orange dashed lines with pluses show the predictions of the perturbation theory; and the purple pluses show the predictions of the perturbation theory considering only the contribution of $d_\perp$ (i.e., $d_\parallel = 0$).

Figure S12 shows that as the PMMA nanogap thickness $g$ decreases, which leads to enhanced nonclassical effects in the nanogap [6,10-12], the discrepancy between the predictions of the perturbation theory and the full-wave FMM results increases. This indicates that, in addition to the first-order contributions of the $d$-parameters considered in the perturbation theory, it is necessary to account for their higher-order contributions at this point [i.e., the $O(\delta^2)$ terms on the right-hand side of Eq. (6) in the main text]. To this end, we investigate whether the expansion theory [i.e., Eq. (2) in the main text, which is capable of incorporating the higher-order contributions of the $d$-parameters] can be employed to improve the accuracy of the perturbation theory.

In the expansion theory, considering all the bound and propagative CWMs supported by the double-nanowire MPW (including two forward- and two backward-propagating CWMs), Eq. (2) becomes,

$$\begin{bmatrix} k_1 + i\kappa_{1,1}^{-,+} & i\kappa_{1,2}^{-,+} & i\kappa_{1,1}^{-,-} & i\kappa_{1,2}^{-,-} \\ i\kappa_{2,1}^{-,+} & k_2 + i\kappa_{2,2}^{-,+} & i\kappa_{2,1}^{-,-} & i\kappa_{2,2}^{-,-} \\ -i\kappa_{1,1}^{+,+} & -i\kappa_{1,2}^{+,+} & -k_1 - i\kappa_{1,1}^{+,-} & -i\kappa_{1,2}^{+,-} \\ -i\kappa_{2,1}^{+,+} & -i\kappa_{2,2}^{+,+} & -i\kappa_{2,1}^{+,-} & -k_2 - i\kappa_{2,2}^{+,-} \end{bmatrix} \tilde{\mathbf{a}}_n = \tilde{k}_n \tilde{\mathbf{a}}_n . \tag{S5.7}$$

Solving the matrix eigenvalue problem (S5.7), the resulting eigenvalues $\tilde{k}_n$ ($n$=±1, ±2) and eigenvectors $\tilde{\mathbf{a}}_n$ yield the propagation constants and electromagnetic field distributions of the NWMs, respectively.

In Fig. S14, with the PMMA nanogap thickness $g$=5 nm and $d_\perp = d_\perp^{\text{Au-PMMA}}\delta$, $d_\parallel = d_\parallel^{\text{Au-PMMA}}\delta$, the propagation constants $\tilde{k}_p$ ($p$=1, 2) of the forward-propagating NWMs are plotted as functions of the scaling factor $\delta$ of $d$-parameters. When $\delta$=0, there is $\tilde{k}_p = k_p$, which is exactly the propagation constant of the CWM. In Fig. S15, with $\delta$=1, $\tilde{k}_p$ is plotted as a function of $g$, where both $\text{Re}(k_p - \tilde{k}_p)$ and $\text{Im}(\tilde{k}_p - k_p)$ are greater than 0. Figures S14 and S15 present the results of tangent lines to the full-wave FMM numerical results at $\delta$=0 (black dashed lines) and the results of tangent lines to the results of expansion theory at $\delta$=0 (black squares), which represent the first two terms on the right-hand side of Eq. (5) in the main text calculated via the FMM and the expansion theory, respectively. The consistency between these two results, as well as their agreement with the perturbation theory [see Eq. (6) in the main text and Figs. S11, S12], verifies the correctness of the implementation of the expansion theory. The results in Figs. S14 and S15 demonstrate that compared with the perturbation theory, the expansion theory (orange dashed lines with pluses) does not significantly improve the accuracy in predicting the full-wave FMM numerical results (red solid lines with circles), especially for large $\delta$ or small $g$. Therefore, to ensure the accuracy of the expansion theory so as to fully account for the higher-order contributions of the $d$-parameters, it is necessary to incorporate not only the bound and propagative CWMs but also a large number of unbound or non-propagative (i.e., radiative or evanescent) CWMs into the expansion theory, so that the considered CWMs can form a complete set of basis functions [2]. However, this comes at the cost of a drastic increase in computational cost and a degradation of physical insight.

In addition, Fig. S16 shows that the propagation constants of the forward- and backward-propagating NWMs predicted by the expansion theory, $\tilde{k}_p$ ($p$=1, 2, orange dashed lines) and $\tilde{k}_{-p}$ (orange diamonds), satisfy the symmetry relation $\tilde{k}_{-p} = -\tilde{k}_p$. This conclusion is consistent with the general conclusion of the expansion theory [see Eq. (S2.7)] and also agrees with the full-wave FMM numerical results (red solid lines and red crosses).

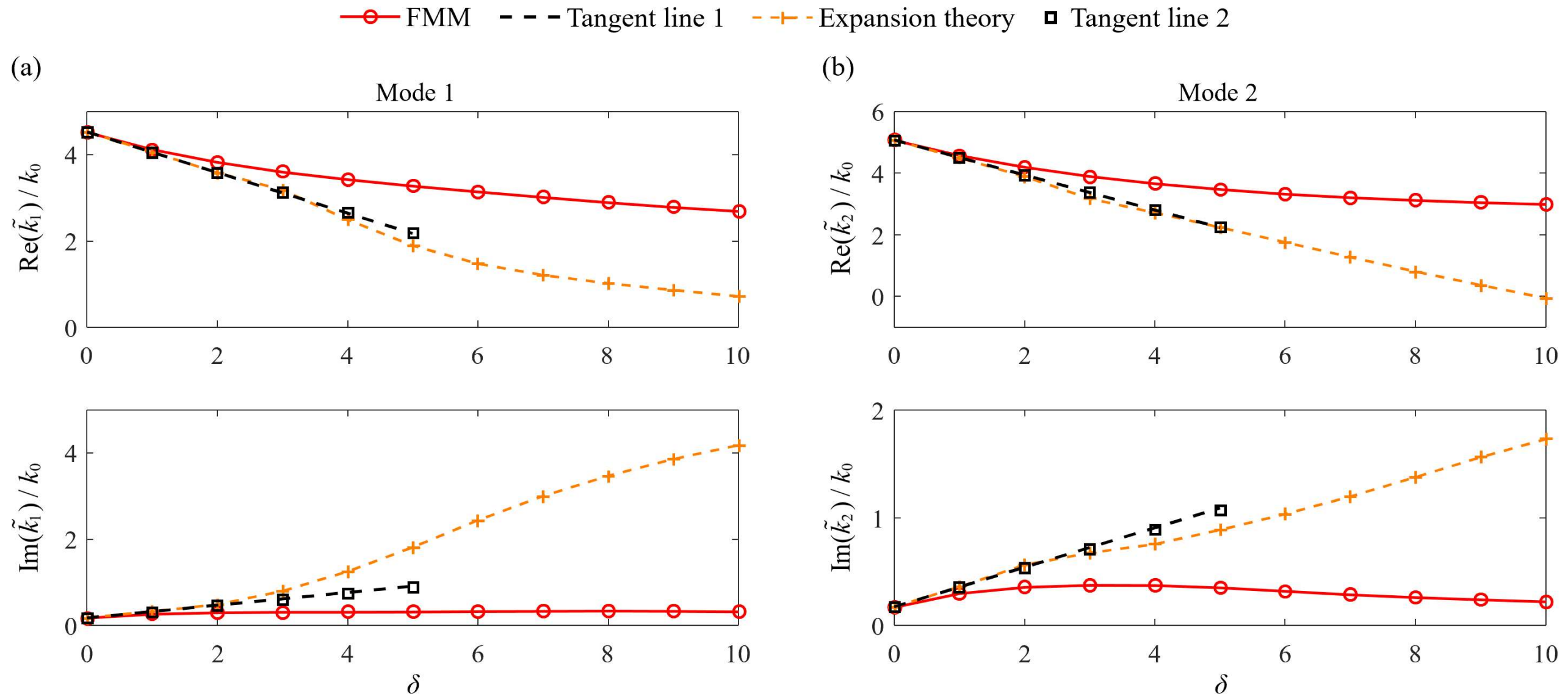

FIG. S14 Propagation constants $\tilde{k}_p$ [$p$=1, 2, corresponding to (a) and (b), respectively] of the two NWMs supported by the double-nanowire MPW as functions of the $d$-parameters scaling factor $\delta$, with the PMMA nanogap thickness $g$=5 nm. The shown normalized quantity $\tilde{k}_p / k_0$ represents the complex effective index. The red solid lines with circles represent the full-wave FMM numerical results; the orange dashed lines with pluses represent the results of the expansion theory; the black dashed lines and black squares are the tangent lines to the results of the FMM and the expansion theory at $\delta$=0, respectively.

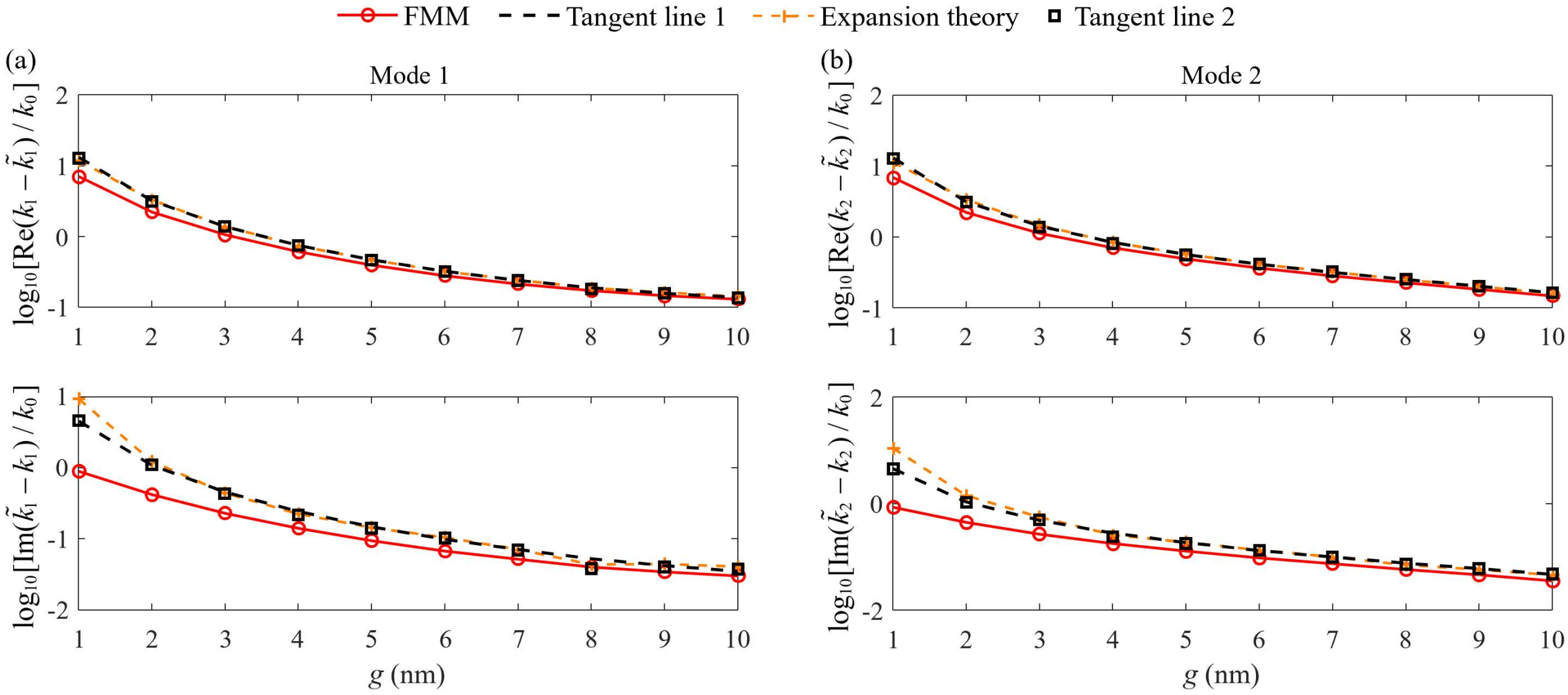

FIG. S15 Propagation constants $\tilde{k}_p$ [$p$=1, 2, corresponding to (a) and (b), respectively] of the two NWMs supported by the double-nanowire MPW as functions of the PMMA nanogap thickness $g$, with the scaling factor $\delta$=1 of the Feibelman $d$-parameters, i.e., $d_\perp = d_\perp^{\text{Au-PMMA}}$ and $d_\parallel = d_\parallel^{\text{Au-PMMA}}$. The red solid lines with circles represent the full-wave FMM numerical results; the orange dashed lines with pluses represent the results of the expansion theory; the black dashed lines and black squares show the results of tangent lines to the results of the FMM and the expansion theory at $\delta$=0, respectively.

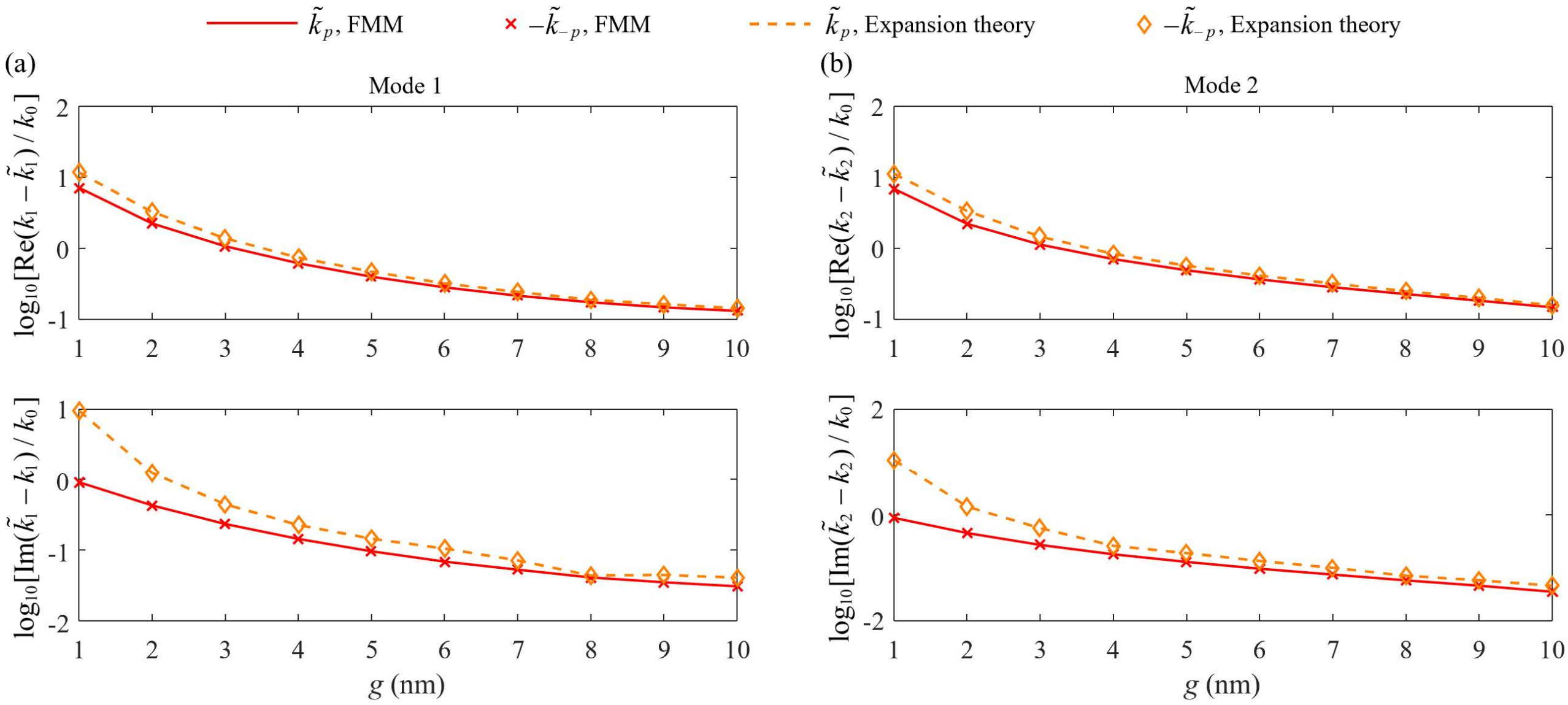

FIG. S16 Propagation constants $\tilde{k}_p$ and $\tilde{k}_{-p}$ [$p$=1, 2, corresponding to (a) and (b), respectively] of the forward- and backward-propagating NWMs supported by the double-nanowire MPW as functions of the PMMA nanogap thickness $g$. Here, $\text{Re}(k_p - \tilde{k}_p)$, $\text{Im}(\tilde{k}_p - k_p)$, $\text{Re}[k_p - (-\tilde{k}_{-p})]$, and $\text{Im}[(-\tilde{k}_{-p}) - k_p]$ are all greater than 0. The scaling factor of the Feibelman $d$-parameters is $\delta$=1, i.e., $d_\perp = d_\perp^{\text{Au-PMMA}}$ and $d_\parallel = d_\parallel^{\text{Au-PMMA}}$. The red solid lines and red crosses represent $\tilde{k}_p$ and $-\tilde{k}_{-p}$ calculated by the full-wave FMM, respectively, while the orange dashed lines and orange diamonds show the corresponding predictions of the expansion theory.

## S6. Numerical example of the NWM at a planar metal-dielectric interface

### A. Prediction of the perturbation theory of NWM under NEBC

For a metal-dielectric planar interface (at $x$=0, as shown in the inset of Fig. S17 for an Au-air planar interface), the electromagnetic field distribution $\boldsymbol{\psi}_p^+ = [\mathbf{E}_p^+, \mathbf{H}_p^+]^{\mathrm{T}}$ of the CWM propagating along the positive $z$-direction (with mode index $p$=1) in the dielectric region ($x$>0) is given by,

$$\mathbf{H}_p^+ = \mathbf{y}H_{y,p}^+ = \mathbf{y}\exp(-\kappa_d x)\exp(ik_p z), \tag{S6.1a}$$

$$\mathbf{E}_p^+ = \mathbf{x}E_{x,p}^+ + \mathbf{z}E_{z,p}^+ = \frac{1}{\omega\varepsilon_d}(\mathbf{x}k_p - \mathbf{z}i\kappa_d)\exp(-\kappa_d x)\exp(ik_p z), \tag{S6.1b}$$

and that in the metal region ($x$<0) is given by,

$$\mathbf{H}_p^+ = \mathbf{y}\exp(\kappa_m x)\exp(ik_p z), \tag{S6.2a}$$

$$\mathbf{E}_p^+ = \frac{1}{\omega\varepsilon_m}(\mathbf{x}k_p + \mathbf{z}i\kappa_m)\exp(\kappa_m x)\exp(ik_p z). \tag{S6.2b}$$

Here, $k_p$ is the propagation constant with an analytical expression,

$$k_p = k_0\sqrt{\frac{\varepsilon_{r,m}\varepsilon_{r,d}}{\varepsilon_{r,m}+\varepsilon_{r,d}}}, \tag{S6.3}$$

where $k_0 = \omega\sqrt{\varepsilon_0\mu_0} = 2\pi/\lambda$ is the vacuum wavenumber, $\omega$ is the angular frequency, $\varepsilon_0$ and $\mu_0$ are the vacuum permittivity and permeability, respectively, and $\lambda$ is the wavelength. $\varepsilon_j=\varepsilon_0\varepsilon_{r,j}$ is the permittivity of the dielectric ($j$=$d$) and metal ($j$=$m$), with $\varepsilon_{r,j}$ being the relative permittivity. The media are assumed to be nonmagnetic, i.e., with the vacuum permeability $\mu_0$. $\kappa_j = \sqrt{k_p^2 - k_j^2}$ [satisfying $\mathrm{Re}(\kappa_j)>0$], and $k_j = k_0\sqrt{\varepsilon_{r,j}}$ . For the CWM propagating along the negative $z$-direction, the electromagnetic field distribution $\boldsymbol{\psi}_p^- = [\mathbf{E}_p^-, \mathbf{H}_p^-]^{\mathrm{T}}$ in the dielectric region ($x$>0) is given by,

$$\mathbf{H}_p^- = \mathbf{y}H_{y,p}^- = -\mathbf{y}\exp(-\kappa_d x)\exp(-ik_p z), \tag{S6.4a}$$

$$\mathbf{E}_p^- = \mathbf{x}E_{x,p}^- + \mathbf{z}E_{z,p}^- = \frac{1}{\omega\varepsilon_d}(\mathbf{x}k_p + \mathbf{z}i\kappa_d)\exp(-\kappa_d x)\exp(-ik_p z), \tag{S6.4b}$$

and that in the metal region ($x$<0) is given by,

$$\mathbf{H}_p^- = -\mathbf{y}\exp(\kappa_m x)\exp(-ik_p z), \tag{S6.5a}$$

$$\mathbf{E}_p^- = \frac{1}{\omega\varepsilon_m}(\mathbf{x}k_p - \mathbf{z}i\kappa_m)\exp(\kappa_m x)\exp(-ik_p z). \tag{S6.5b}$$

For the metal-dielectric planar interface, the perturbation theory of the NWM under the NEBC [Eq. (6) in the main text] provides the first-order asymptotic expansion of the propagation constant $\tilde{k}_p$ of the NWM,

$$\tilde{k}_p = k_p + i\kappa_{p,p}^{-,+} + O(\delta^2), \tag{S6.6}$$

where

$$\begin{aligned}\kappa_{p,p}^{-,+} &= \frac{1}{F_p}\oint_L \left(i\omega d_\parallel [\![\varepsilon]\!]\mathbf{E}_{p,\parallel}^- \cdot \mathbf{E}_{p,\parallel}^+ - i\omega d_\perp [\![\varepsilon E_{p,\perp}^- E_{p,\perp}^+]\!]\right)_{z=0} dl \\ &= \frac{1}{F_p} i\omega\left(d_\parallel [\![\varepsilon]\!]\mathbf{E}_{p,\parallel}^- \cdot \mathbf{E}_{p,\parallel}^+ - d_\perp [\![\varepsilon E_{p,\perp}^- E_{p,\perp}^+]\!]\right)_{z=0},\end{aligned} \tag{S6.7}$$

$$F_p = \int_{-\infty}^{\infty} \mathbf{z}\cdot(\mathbf{E}_p^- \times \mathbf{H}_p^+ - \mathbf{E}_p^+ \times \mathbf{H}_p^-)_{z=0}\, dx = \frac{k_p}{\omega}\left(\frac{1}{\varepsilon_m\kappa_m} + \frac{1}{\varepsilon_d\kappa_d}\right). \tag{S6.8}$$

Substituting Eqs. (S6.1), (S6.2), (S6.4), (S6.5), and (S6.8) into Eq. (S6.7) yields,

$$\kappa_{p,p}^{-,+} = i\frac{\varepsilon_{r,m}\varepsilon_{r,d}\sqrt{-\varepsilon_{r,m}\varepsilon_{r,d}}\,(\varepsilon_{r,m}-\varepsilon_{r,d})}{(\varepsilon_{r,m}^2+\varepsilon_{r,d}^2)(\varepsilon_{r,m}+\varepsilon_{r,d})}k_0^2(d_\perp - d_\parallel). \tag{S6.9}$$

Substituting Eqs. (S6.3) and (S6.9) into Eq. (S6.6) yields the first-order asymptotic expansion of the propagation constant $\tilde{k}_p$ of the NWM.

### B. Prediction of the asymptotic expansion method

For the metal-dielectric planar interface under the NEBC, the propagation constant $\tilde{k}_p$ ($p$=1) of the supported NWM satisfies the transcendental equation (see Eq. (S2.9) in Supplement 1 of Ref. [13]),

$$\varepsilon_{r,m}\tilde{\kappa}_d + \varepsilon_{r,d}\tilde{\kappa}_m - (\varepsilon_{r,m} - \varepsilon_{r,d})(d_\perp \tilde{k}_p^2 - d_\parallel \tilde{\kappa}_d \tilde{\kappa}_m) - d_\perp d_\parallel \tilde{k}_p^2 (\varepsilon_{r,m}\tilde{\kappa}_m + \varepsilon_{r,d}\tilde{\kappa}_d) = 0, \tag{S6.10}$$

where $\tilde{\kappa}_j = \sqrt{\tilde{k}_p^2 - k_j^2}$ [satisfying $\mathrm{Re}(\tilde{\kappa}_j) > 0$ ], $k_j = k_0\sqrt{\varepsilon_{r,j}}$ with $\varepsilon_{r,j}$ denoting the relative permittivity of the dielectric ($j$=$d$) and metal ($j$=$m$), $k_0$=$2\pi/\lambda$ is the vacuum wavenumber, and $\lambda$ is the wavelength. Note that, in contrast to Ref. [14], where the higher-order $d_\perp d_\parallel$ terms are neglected, Eq. (S6.10) does not neglect any higher-order terms of the $d$-parameters. When $d_\perp = d_\parallel = 0$ , Eq. (S6.10) becomes,

$$\varepsilon_{r,m}\tilde{\kappa}_d + \varepsilon_{r,d}\tilde{\kappa}_m = 0, \tag{S6.11}$$

whose solution is exactly the analytical expression of the propagation constant of the CWM under the CEBC, given by Eq. (S6.3).

When $d_\perp d_\parallel \neq 0$ , let [i.e., Eq. (4) in the main text],

$$d_\perp = d_\perp^{(1)}\delta,\ d_\parallel = d_\parallel^{(1)}\delta. \tag{S6.12}$$

Since $\tilde{k}_p$ is a function of the $d$-parameters, i.e., $\tilde{k}_p = \tilde{k}_p(d_\perp, d_\parallel)$ , it can be expressed as an asymptotic expansion in terms of $\delta$ [i.e., Eq. (5) in the main text],

$$\tilde{k}_p = \tilde{k}_p(d_\perp^{(1)}\delta, d_\parallel^{(1)}\delta) = k^{(0)} + k^{(1)}\delta + O(\delta^2). \tag{S6.13}$$

In the following, we employ the asymptotic expansion method to solve for $k^{(0)}$ and $k^{(1)}$ in Eq. (S6.13), and compare them with the perturbation theory of the NWM [i.e., Eq. (S6.6)] so as to verify the validity of the latter. There is,

$$\tilde{\kappa}_j = \sqrt{\tilde{k}_p^2 - k_j^2} = f(\delta), \tag{S6.14}$$

which is a function of $\delta$ and thus can be expressed as an asymptotic expansion,

$$\begin{aligned}\tilde{\kappa}_j &= f(0) + f'(0)\delta + O(\delta^2)\\ &= \sqrt{(k^{(0)})^2 - k_j^2} + \frac{k^{(0)}k^{(1)}}{\sqrt{(k^{(0)})^2 - k_j^2}}\delta + O(\delta^2)\\ &= \kappa_j^{(0)} + \kappa_j^{(1)}\delta + O(\delta^2),\end{aligned} \tag{S6.15}$$

where the definitions are,

$$\kappa_j^{(0)} = \sqrt{(k^{(0)})^2 - k_j^2},\ \kappa_j^{(1)} = \frac{k^{(0)}k^{(1)}}{\sqrt{(k^{(0)})^2 - k_j^2}}. \tag{S6.16}$$

Substituting Eqs. (S6.12), (S6.13), and (S6.15) into Eq. (S6.10) yields,

$$\begin{aligned}&\varepsilon_{r,m}(\kappa_d^{(0)} + \kappa_d^{(1)}\delta) + \varepsilon_{r,d}(\kappa_m^{(0)} + \kappa_m^{(1)}\delta) - (\varepsilon_{r,m} - \varepsilon_{r,d})[d_\perp^{(1)}\delta(k^{(0)})^2 - d_\parallel^{(1)}\delta\kappa_d^{(0)}\kappa_m^{(0)}] + O(\delta^2)\\ &= \varepsilon_{r,m}\kappa_d^{(0)} + \varepsilon_{r,d}\kappa_m^{(0)} + \{\varepsilon_{r,m}\kappa_d^{(1)} + \varepsilon_{r,d}\kappa_m^{(1)} - (\varepsilon_{r,m} - \varepsilon_{r,d})[d_\perp^{(1)}(k^{(0)})^2 - d_\parallel^{(1)}\kappa_d^{(0)}\kappa_m^{(0)}]\}\delta + O(\delta^2) = 0.\end{aligned} \tag{S6.17}$$

From Eq. (S6.17), we obtain,

$$\varepsilon_{r,m}\kappa_d^{(0)} + \varepsilon_{r,d}\kappa_m^{(0)} = 0, \tag{S6.18a}$$

$$\varepsilon_{r,m}\kappa_d^{(1)} + \varepsilon_{r,d}\kappa_m^{(1)} - (\varepsilon_{r,m} - \varepsilon_{r,d})[d_\perp^{(1)}(k^{(0)})^2 - d_\parallel^{(1)}\kappa_d^{(0)}\kappa_m^{(0)}] = 0. \tag{S6.18b}$$

Substituting Eq. (S6.16) into Eq. (S6.18) yields a system of equations for $k^{(0)}$ and $k^{(1)}$, solving which then determines $k^{(0)}$ and $k^{(1)}$, as detailed below.

Equation (S6.18a) is identical to Eq. (S6.11), solving which yields,

$$k^{(0)} = k_0\sqrt{\frac{\varepsilon_{r,m}\varepsilon_{r,d}}{\varepsilon_{r,m} + \varepsilon_{r,d}}}, \tag{S6.19}$$

which is exactly the propagation constant $k_p$ of the CWM given by Eq. (S6.3). Substituting Eq. (S6.19) into Eq. (S6.18b) and solving the resulting equation yields,

$$k^{(1)} = -\frac{\varepsilon_{r,m}\varepsilon_{r,d}\sqrt{-\varepsilon_{r,m}\varepsilon_{r,d}}\,(\varepsilon_{r,m} - \varepsilon_{r,d})}{(\varepsilon_{r,m}^2 + \varepsilon_{r,d}^2)(\varepsilon_{r,m} + \varepsilon_{r,d})}k_0^2(d_\perp^{(1)} - d_\parallel^{(1)}). \tag{S6.20}$$

Substituting Eqs. (S6.19) and (S6.20) into Eq. (S6.13), and using Eq. (S6.12), we obtain the first-order asymptotic expansion of the propagation constant $\tilde{k}_p$ of the NWM. It is seen that this result is consistent with the perturbation theory of the NWM [i.e., Eq. (S6.6)], thereby confirming the validity of the latter.

**C. Numerical test of theoretical predictions**

By employing the perturbation theory and the asymptotic expansion method, respectively, Sections S6A and S6B have provided analytical expressions for the first-order asymptotic expansion of the propagation constant of the NWM supported by the planar metal-dielectric interface, leading to consistent results. These results will be compared with rigorous numerical results in this section to further verify their validity. Considering an Au-air planar interface, according to the Feibelman *d*-parameters being proportional to the permittivity of the dielectric [6], and given that $d_{\perp}^{\text{Au-PMMA}} = (-0.4+0.2i)$ nm (see the *Verification* Section of the main text), we set $d_{\perp}^{\text{Au-Air}} = (-0.2+0.1i)$ nm. Due to the same considerations as in setting $d_{\parallel}^{\text{Au-PMMA}} = (0.4+0.2i)$ nm (see the *Verification* Section of the main text), we set $d_{\parallel}^{\text{Au-Air}} = (0.2+0.1i)$ nm. The refractive indices of Au and air are set consistently with those of the single-nanowire MPW (see the *Verification* Section of the main text). Consistent with the main text, a wavelength of $\lambda$=1 μm is adopted in the calculations of this section.

In Fig. S17, with $d_{\perp} = d_{\perp}^{\text{Au-Air}}\delta$ and $d_{\parallel} = d_{\parallel}^{\text{Au-Air}}\delta$, the propagation constant $\tilde{k}_1$ of the NWM supported by the planar Au-air interface is plotted as a function of the scaling factor $\delta$ of the Feibelman *d*-parameters. The results show that the predictions of the perturbation theory and the asymptotic expansion method (which yield identical results, represented by the orange dashed lines with pluses) are consistent with the tangent lines to the full-wave FMM numerical results at $\delta$=0 [i.e., the first two terms on the right-hand side of Eq. (S6.13) calculated using the FMM, black dashed lines], which verifies the validity of the former. In addition, for $\delta$=1 (corresponding to the actual physical *d*-parameters), the predictions of the perturbation theory and the asymptotic expansion method are quite close to the full-wave FMM numerical results (red solid lines with circles), while the latter are consistent with the results obtained by a rigorous numerical solving of the transcendental equation (S6.10) (green asterisks), even for large values of $\delta$.

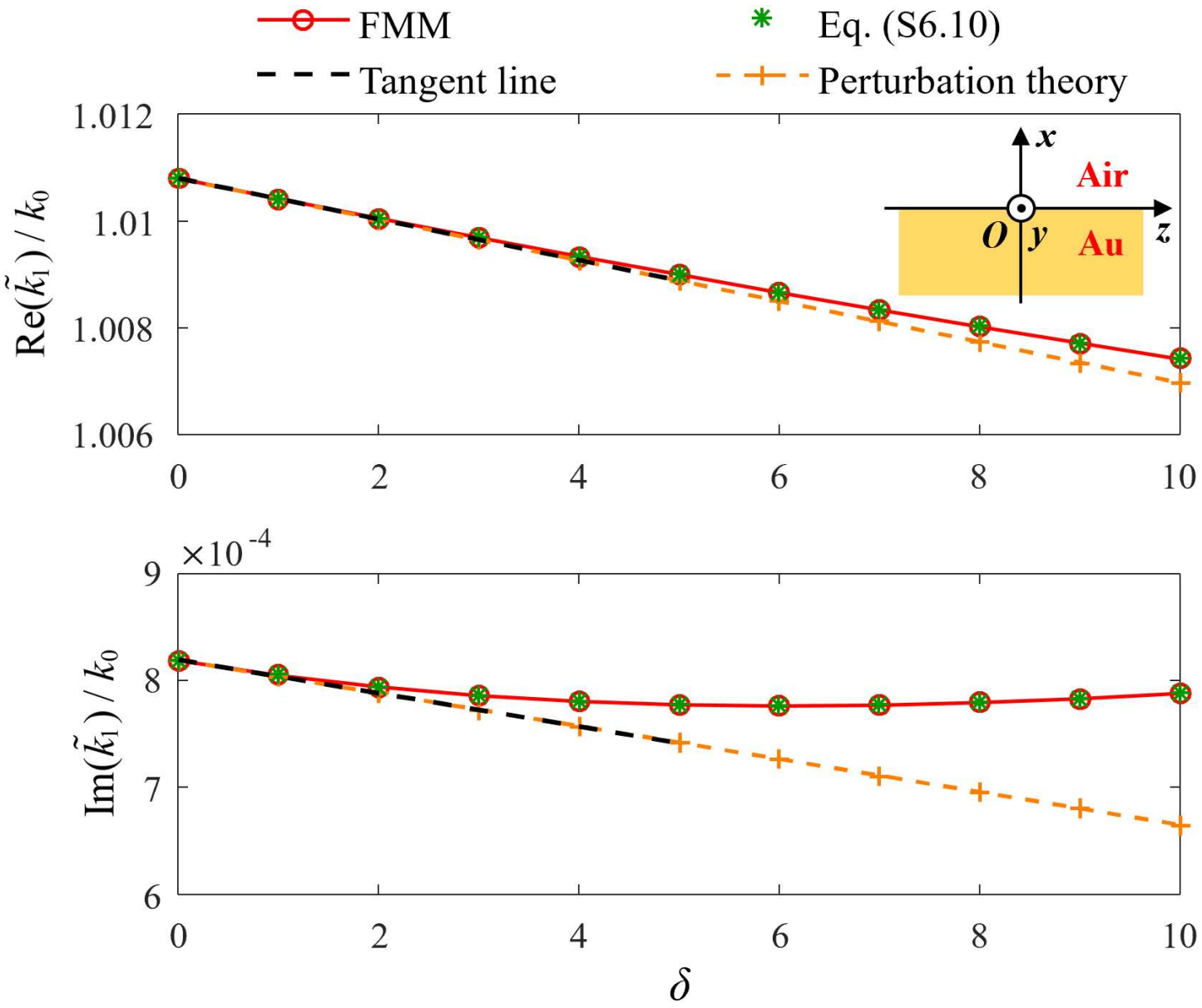

FIG. S17 Propagation constant $\tilde{k}_1$ of the NWM supported by a planar Au-air interface (as illustrated in the inset) as a function of the scaling factor $\delta$ of the Feibelman *d*-parameters. The shown normalized quantity $\tilde{k}_1 / k_0$ represents the complex effective index. The red solid lines with circles and the black dashed lines represent the full-wave FMM numerical results and their tangent lines at $\delta$=0, respectively; the green asterisks represent the results obtained by a rigorous numerical solving of the transcendental equation (S6.10); and the orange dashed lines with pluses show the predictions of both the perturbation theory and the asymptotic expansion method (which yield identical results).